\documentclass[]{tADP2e}
\usepackage{dcolumn}
\usepackage{bm}
\usepackage{graphicx}
\usepackage{amsmath}
\usepackage{amssymb}

\begin{document}
\newcommand{\rhoH}{\ensuremath{\rho_{\text H}}}
\newcommand{\RH}{\ensuremath{R_{\text H}}}
\newcommand{\TRH}{\ensuremath{\tilde{R}_{\text H}}}
\newcommand{\ten}[1]{\ensuremath{\uline{#1}}}      
\newcommand{\YRS}{YbRh$_2$Si$_2$}
\newcommand{\LRS}{LuRh$_2$Si$_2$}
\newcommand{\TN}{\ensuremath{T_{\text N}}}
\jvol{61} 
\jyear{2012} \jmonth{October}

\vspace*{3cm}This is an Author's Accepted Manuscript of an article
to be published in Advances in Physics [copyright Taylor \&
Francis], available after publication online at:
http://www.tandfonline.com/

\title{Hall effect in heavy-fermion metals}
\author{Sunil Nair$^{\rm a}$$^{\dagger}$\thanks{$^{\dagger}$Present
address: Indian Institute of Science Education and Research
(IISER), Dr Homi Bhabha Road, Pune 411008, India}, S. Wirth$^{\rm
a}$$^{\ast}$ \thanks{$^{\ast}$Corresponding author. Email:
wirth@cpfs.mpg.de}, S. Friedemann$^{\rm a,b}$, F. Steglich$^{\rm
a}$, Q. Si$^{\rm c}$ and A.~J.~Schofield$^{\rm d}$ \\
\vspace{12pt} $^{\rm a}${\em{Max Planck Institute for Chemical
Physics of Solids, N\"othnitzer Str. 40, 01187 Dresden, Germany}}\\
$^{\rm b}${\em{Cavendish Laboratory, JJ Thomson Avenue, Cambridge
CB3 0HE, United Kingdom}}\\
$^{\rm c}${\em{Department of Physics and Astronomy, Rice University,
Houston, Texas 77005, USA}}\\
$^{\rm d}${\em{School of Physics and Astronomy, University of
Birmingham, Birmingham B15 2TT, United Kingdom}}} \maketitle

\begin{abstract}
The heavy fermion systems present a unique platform in which
strong electronic correlations give rise to a host of novel, and
often competing, electronic and magnetic ground states. Amongst a
number of potential experimental tools at our disposal,
measurements of the Hall effect have emerged as a particularly
important one in discerning the nature and evolution of the Fermi
surfaces of these enigmatic metals. In this article, we present a
comprehensive review of Hall effect measurements in the
heavy-fermion materials, and examine the success it has had in
contributing to our current understanding of strongly correlated
matter. Particular emphasis is placed on its utility in the
investigation of quantum critical phenomena which are thought to
drive many of the exotic electronic ground states in these
systems. This is achieved by the description of measurements of
the Hall effect across the putative zero-temperature instability
in the archetypal heavy-fermion metal YbRh$_2$Si$_2$. Using the
Ce$M$In$_5$ (with $M =$ Co, Ir) family of systems as a paradigm,
the influence of (antiferro-)magnetic fluctuations on the Hall
effect is also illustrated. This is compared to prior Hall effect
measurements in the cuprates and other strongly correlated systems
to emphasize on the generality of the unusual magnetotransport in
materials with non-Fermi liquid behavior.
\end{abstract}
\begin{keywords}heavy fermion metals; Hall effect; quantum
criticality; magnetotransport
\end{keywords}

\noindent\begin{tabular}{p{0.2cm}p{11.4cm}r} & \centering{\bf
Contents} & \small{\textsc{Page}}\\[0.1cm]
1. & Introduction & \pageref{sec:intro}\\
 & \hspace*{0.3cm} 1.1. \enspace Significance of Hall effect
   to heavy-fermion systems & \pageref{sec:pert}\\
 & \hspace*{0.3cm} 1.2. \enspace Contemporary issues in
   heavy-fermion systems & \pageref{sec:motiv}\\
 & \hspace*{0.3cm} 1.3. \enspace Classification of quantum critical
   points via Hall effect & \pageref{sec:class}\\
 & \hspace*{0.3cm} 1.4. \enspace Outline and scope &
   \pageref{sec:outline}\\[0.2cm]
2. & Basics of Hall effect and heavy-fermion systems &
   \pageref{sec:basics}\\
 & \hspace*{0.3cm} 2.1. \enspace History of the Hall effect &
 \pageref{sec:hist}\\
 & \hspace*{0.3cm} 2.2. \enspace The influence of magnetism:
 Anomalous Hall effect & \pageref{sec:anom}\\
  & \hspace*{0.3cm} 2.3. \enspace Mechanisms contributing to the
 anomalous Hall effect & \pageref{sec:Hallmech}\\
 & \hspace*{1.37cm}2.3.1. \enspace Skew scattering &
     \pageref{sec:skew}\\
 & \hspace*{1.37cm}2.3.2. \enspace Side-jump mechanism &
     \pageref{sec:side}\\
 & \hspace*{1.37cm}2.3.3. \enspace Berry phase contributions &
     \pageref{sec:berry}\\
\end{tabular}

\noindent\begin{tabular}{p{0.2cm}p{11.4cm}r} & &
   \small{\textsc{Page}}\\[0.1cm]
 & \hspace*{0.3cm} 2.4. \enspace Hall effect and Fermi surface &
    \pageref{sec:HallFS}\\
 & \hspace*{0.3cm} 2.5. \enspace Basic remarks on heavy-fermion
 systems & \pageref{sec:HF}\\
 & \hspace*{0.3cm} 2.6. \enspace Anomalous Hall effect in
 heavy-fermion systems & \pageref{sec:AHE-HF}\\[0.2cm]
3. & Theoretical work on the Hall effect & \pageref{sec:theory}\\[0.2cm]
 & \hspace*{0.3cm} 3.1. \enspace Theoretical Overview &
   \pageref{sec:theory-overview}\\
 & \hspace*{0.3cm} 3.2. \enspace Key results from Boltzmann theory
   & \pageref{sec:Boltzmann}\\
 & \hspace*{0.3cm} 3.3. \enspace Hall effect within Fermi liquid
   theory & \pageref{sec:Hall-flt}\\
 & \hspace*{0.3cm} 3.4. \enspace Quantum critical heavy fermion
   metals & \pageref{sec:beyond-flt}\\
 & \hspace*{1.37cm}3.4.1. \enspace ``Conventional'' Hertz-Moriya-Millis
   quantum & \\
  & \hspace*{2.62cm}criticality & \pageref{sec:HM-qc}\\
 & \hspace*{1.37cm}3.4.2. \enspace Spin-fluctuation theory &
   \pageref{sec:spinfluct}\\
 & \hspace*{0.3cm} 3.5. \enspace Hall effect across Kondo breakdown
   quantum critical point & \pageref{sec:Kondobreak}\\
 & \hspace*{1.37cm}3.5.1. \enspace Quantum criticality in heavy
     fermion metals and & \\
 & \hspace*{2.62cm}jump of Hall coefficient & \pageref{sec:Halljump}\\
 & \hspace*{1.37cm}3.5.2. \enspace Crossover and scaling of the
     Hall coefficient at $T \neq 0$ & \pageref{sec:crosstheo}\\[0.2cm]
4. & Experimental aspects of Hall effect measurements in metals&
 \pageref{sec:exp}\\
 & \hspace*{0.3cm} 4.1. \enspace Measurement techniques &
   \pageref{sec:techn}\\
 & \hspace*{0.3cm} 4.2. \enspace Advanced aspects of Hall effect
   measurements & \pageref{sec:advexp}\\
 & \hspace*{1.37cm}4.2.1. \enspace Single-field Hall experiments &
   \pageref{sec:single}\\
 & \hspace*{1.37cm}4.2.2. \enspace Crossed-field Hall experiments &
   \pageref{sec:cross}\\
 & \hspace*{1.37cm}4.2.3. \enspace Realization of crossed-field
   Hall experiments & \pageref{sec:crossexp}\\
 & \hspace*{1.37cm}4.2.4. \enspace Comparing single and crossed-field
   Hall experiments & \pageref{sec:comcross}\\[0.2cm]
5. & Hall effect and Kondo-breakdown quantum criticality
 & \pageref{sec:qpt}\\
 & \hspace*{0.3cm} 5.1. \enspace Hall effect evolution at the QCP
    in \YRS\ & \pageref{sec:YRS} \\
 & \hspace*{0.3cm} 5.2. \enspace Comparison to other candidates for
    Kondo breakdown & \pageref{sec:HFcomp} \\
 & \hspace*{0.3cm} 5.3. \enspace Hall effect and scaling behavior &
   \pageref{sec:scal} \\[0.2cm]
6. & Hall effect in systems with 115 type of structure &
   \pageref{sec:fluc}\\
 & \hspace*{0.3cm} 6.1. \enspace Influence of magnetic fluctuations on
      superconductivity & \pageref{sec:flucsup}\\
 & \hspace*{0.3cm} 6.2. \enspace Interplay magnetism and superconductivity
   in Ce$M$In$_5$ & \pageref{sec:fluc115}\\
 & \hspace*{0.3cm} 6.3. \enspace Hall effect measurements on Ce$M$In$_5$
   systems & \pageref{sec:Hall115}\\
 & \hspace*{1.37cm}6.3.1. \enspace Scaling relations in Hall
   effect & \pageref{sec:comp}\\
 & \hspace*{1.37cm}6.3.2. \enspace CeCoIn$_5$ & \pageref{sec:Co115}\\
 & \hspace*{1.37cm}6.3.3. \enspace CeIrIn$_5$ & \pageref{sec:Ir115}\\
 & \hspace*{1.37cm}6.3.4. \enspace Comparative remarks &
     \pageref{sec:comp115}\\[0.2cm]
7. & Comparison to Hall effect of other correlated materials
   & \pageref{sec:sces}\\
 & \hspace*{0.3cm} 7.1. \enspace Copper oxide superconductors and
      related systems & \pageref{sec:copper}\\
 & \hspace*{1.37cm}7.1.1. \enspace Cuprates & \pageref{sec:cup}\\
 & \hspace*{1.37cm}7.1.2. \enspace Comparison of cuprates and
     heavy-fermion systems & \pageref{sec:cupHF}\\
 & \hspace*{1.37cm}7.1.3. \enspace Hall effect in oxy-pnictides
     and related systems & \pageref{sec:pnic}\\
  & \hspace*{0.3cm} 7.2. \enspace Other systems of related interest &
     \pageref{sec:relat}\\
 & \hspace*{0.3cm} 7.3. \enspace Colossal magnetoresistive manganites
 &   \pageref{sec:cmr}\\[0.2cm]
8. & Summary & \pageref{sec:sum}\\[0.2cm]
9. & Appendix & \pageref{sec:append}\\[0.3cm]
   & Acknowledgement & \pageref{sec:ackn}\\
   & References & \pageref{sec:bib}\\
\end{tabular}\pagebreak

\section{Introduction} \label{sec:intro}
\subsection{Significance of Hall effect in heavy-fermion systems}
\label{sec:pert} The heavy fermion metals represent an enigmatic
class of strongly correlated electron systems, in which the strong
hybridization between the conduction electrons and localized $f$
moments results in a Landau Fermi liquid (LFL) at low temperatures
with heavily renormalized quasiparticle properties \cite{gre91}. A
dramatic manifestation of the strong many body effects in these
systems was uncovered in 1979 with the discovery of
superconductivity in CeCu$_2$Si$_2$ \cite{ste79}. Predating the
flurry of activity in the high temperature superconducting
cuprates, this represented the first observation of
superconductivity in an inherently magnetic environment. It was
followed by the observation of superconductivity in many heavy
fermion metals \cite{ott83}. It has also motivated the search for
electronic (more specifically, magnetic) mechanisms for
superconducting pairing. The resurgence of research into the heavy
fermion metals is primarily due to current interest in continuous
quantum phase transitions---a zero-temperature instability which
can be tuned by the use of a non-thermal control parameter
\cite{her76,hvl07,geg08}. The presence of such a zero-temperature
instability is often manifested in a large region of the
experimentally accessible phase space, as has been clearly
demonstrated in many systems. The celebrated Landau-Fermi-liquid
theory of conventional metals breaks down in the vicinity of such
instabilities \cite{hvl07}, and anomalous experimental behavior
like a linear temperature dependence of the electrical resistivity
\cite{cus03}, a non-saturating specific heat coefficient
\cite{loh94} and an apparent violation of the Wiedemann-Franz law
\cite{tan07,pfau12} have been observed. This instability also
appears to be linked to the emergence of novel states of
matter---for example superconductivity (as mentioned above) is
often observed in the vicinity of a quantum phase transition
\cite{mat98}. A related development has been the growing
realization that the physics of the heavy-fermion systems and the
high-temperature superconducting cuprates have much in common.
What makes this so fascinating is that unlike the heavy fermions
metals, the parent high-transition-temperature (high-$T_{\rm c}$)
superconductors are Mott insulators. In this context it is
surprising that the superconducting and normal state properties of
some heavy fermion systems are so similar to the high-$T_{\rm c}$
cuprates \cite{nak07}.

A microscopic explanation of the physics of the heavy-fermion
systems calls for an understanding of how the Fermi surfaces of
these complex systems evolve as a function of various experimental
control parameters. Measurements of the de Haas--van Alphen (dHvA)
effect and of angle resolved photo electron spectroscopy (ARPES)
are amongst the most powerful tools for this purpose. In
particular, dHvA effect measurements demonstrated \cite{tai88} the
applicability of a description by strongly renormalized
quasiparticles in UPt$_3$. In general, however, extreme
sensitivity to disorder along with prerequisites of very low
temperatures and high fields have limited the application of these
techniques to the heavy-fermion systems. It is in this context
that measurements of the Hall effect \cite{hur72}---a relatively
simpler experimental technique---has proven illuminating. This is
in spite of the fact that the Hall effect itself is a complex
quantity, especially in systems like the heavy-fermion systems
which usually have more than one band crossing the Fermi level.
Moreover, the contribution from the anomalous Hall effect which
results from skew scattering \cite{smi55}, is an important factor
in materials comprised of an array of localized magnetic moments.
However, it has been shown that at low temperatures the
contribution from skew scattering is negligible in the
heavy-fermion systems \cite{fer87}, and thus the experimentally
measured Hall voltage predominantly arises from the normal part of
the Hall effect. Consequently, at these temperatures the Hall
effect can provide a crucial---albeit indirect---measure of the
Fermi surface volume.

Measurements of the Hall effect have also been extremely useful in
revealing the normal-state properties of heavy-fermion
superconductors. For instance, the possible presence of a
pseudogap-like precursor state to superconductivity in Ce$M$In$_5$
system ($M$ = Co, Ir) has been proposed. Comparison with prior
results on the Hall effect in the superconducting cuprates reveals
many striking similarities, indicating that the non-Fermi-liquid
physics probably affect the Fermi surfaces of these diverse
materials in a rather similar fashion. An anisotropic
reconstruction of the Fermi surface by the formation of ``hot''
and ``cold'' regions with different scattering rates has been
suggested, which then manifests itself in the form of two distinct
scattering times that influence the resistivity and Hall effect in
these materials in disparate manners \cite{xu99,nai09a}. Recent
measurements of the Hall effect in cuprates have reinforced its
utility in uncovering the relation between quantum magnetism and
superconductivity in these complex materials
\cite{bal03,leb07,bal09}.

The theory of the Hall effect in strongly correlated systems
remains an exceedingly difficult problem, in part due to the acute
sensitivity of the Hall response to the constitution and topology
of the Fermi surface. Earlier calculations on metals relied on
approximating the Fermi surface to simple geometric constructions
without taking into account the specific band structure of the
materials under consideration \cite{all70,ban78}. Unusual Hall
behavior in the cuprates \cite{chi91} triggered intense activity
in this area, and besides a more generic geometric treatment
\cite{ong91}, models like the nearly antiferromagnetic Fermi
liquid \cite{sto97} and the spin-charge separation \cite{and91}
scenario, and others were employed to account for the experimental
observations. For the heavy-fermion metals, the Hall constant was
predicted to allow for a qualitative distinction between different
scenarios for quantum criticality \cite{col01}. The Hall effect
and the resistivity have also been calculated using current vertex
corrections (CVC) on a Fermi liquid in the presence of
antiferromagnetic fluctuations, in an attempt to unify
observations in the cuprates and the heavy-fermion metals
\cite{kon08}.

\subsection{Contemporary issues in heavy-fermion systems}
\label{sec:motiv} The recent interest in heavy-fermion systems
stems from their model character in the investigation of quantum
critical points. Although occurring at zero temperature, a quantum
critical point (QCP) may lead to unusual properties up to
surprisingly high temperatures. We shall see that the Hall effect
is a sensitive tool to explore the nature of a QCP.

\begin{figure}[tb]
\centering \includegraphics[width=12.8cm]{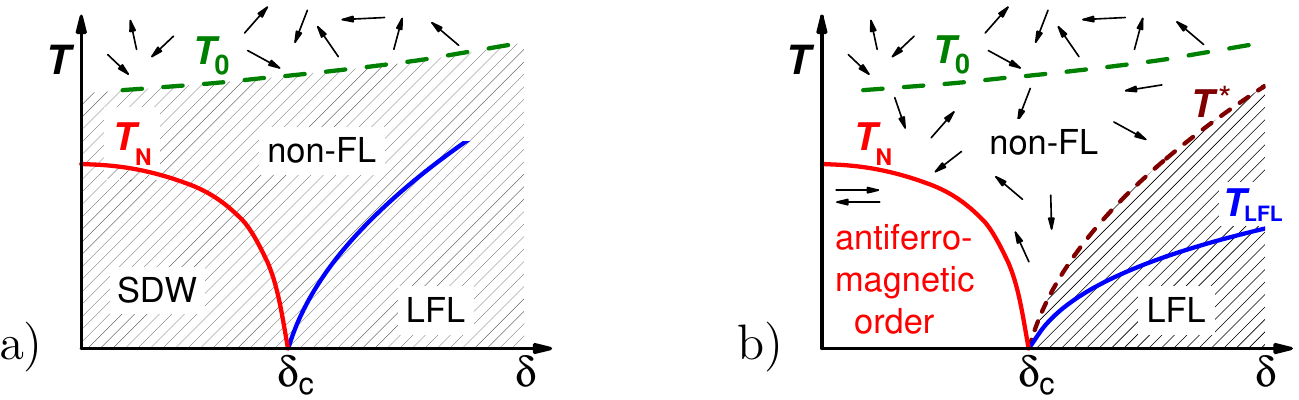}
\caption{Comparison of quantum criticality in a) the spin-density
wave (SDW) and b) the locally critical scenario. $\delta$ refers
to an external control parameter for tuning the material to its
QCP at $\delta_c$. The hatched areas indicate phase space regions
within which the heavy quasi-particles are formed, whereas the
arrows illustrate the existence of local magnetic moments. In the
local scenario (b) the quasi-particles themselves break apart at
$\delta_c$. At finite temperatures the quasiparticles disintegrate
within a crossover range indicated by the $T^*$ line.
$T_{\mathrm{LFL}}$ marks the temperature below which LFL behavior
is found, whereas $T_0$ denotes the onset temperature for initial
Kondo screening.} \label{sdwlocal}
\end{figure}

Two main approaches are available to describe QCPs in heavy
fermion metals ({\it cf.} Fig.~\ref{sdwlocal}). Conventionally,
the order parameter notation is generalized by incorporating
quantum corrections. For the heavy-fermion systems magnetism is
treated itinerantly giving rise to a spin-density wave (SDW)
\cite{her76,mor85,mil93}. This approach relies on the fact that
the composite quasiparticles formed by the Kondo effect stay
intact at the quantum critical point, Fig.~\ref{sdwlocal}(a).

By contrast, more recent studies additionally incorporate quantum
modes to become critical \cite{qsi,sen03}. In the case of the
heavy-fermion systems the critical quantum modes are associated
with the Kondo effect. When the Kondo effect becomes critical, the
quasiparticles disintegrate at the QCP, Fig.~\ref{sdwlocal}(b). As
a consequence, the Fermi surface is expected to jump at such an
unconventional QCP. This is to be contrasted with the conventional
QCP for which a continuous evolution of the Fermi surface is
expected. Hence, the evolution of the Fermi surface is a
fundamental feature distinguishing the Kondo breakdown from the
SDW scenario. In fact, Hall effect measurements were suggested to
characterize QCPs \cite{col01}. The second fundamental difference
between the two scenarios is their scaling behavior. For the 3D
SDW QCP an $E/T^{3/2}$ scaling is predicted for the spin dynamics,
whereas an $E/T$ scaling is expected at a Kondo breakdown QCP. The
first example for an unconventional QCP was identified in
CeCu$_{6-x}$Au$_x$ by observation of an $E/T$ scaling in inelastic
neutron scattering measurements \cite{loe-na}. This Kondo
breakdown scenario will be further discussed in section
\ref{sec:crosstheo}. In the following we shall see how Hall effect
measurements may shed light on the dynamics and on the scaling
behavior.

\subsection{Classification of quantum critical points via Hall
effect} \label{sec:class} Hall effect measurements currently
provide the best probe to study the Fermi surface evolution at
QCPs. This is mainly due to the experimental difficulties
associated with other probes: Photo electron spectroscopy on the
one hand is---in the context discussed here---limited to
relatively high temperatures and moderate energy resolution
($\Delta E$ in the order of meV). Quantum oscillation
measurements, on the other hand, can only be conducted in high
magnetic fields which are often beyond interesting energy scales,
and require clean samples which often excludes the investigation
of composition-driven QCPs.

In order to distinguish whether the Fermi surface evolves
continuously or discontinuously one needs to have sufficient
resolution of the control parameter used to access the QCP. Widely
employed tuning parameters in heavy fermion systems are pressure,
composition, and magnetic field. So far, however, only magnetic
field-tuned QCPs appear to allow the high resolution required.

The QCP in CeCu$_{6-x}$Au$_x$ is tuned either by variation of gold
content or by the application of pressure. Probing a pressure- or
composition-driven QCP with sufficient resolution was not yet able
to show an abrupt Fermi surface change in this system by means of
Hall effect measurements. The critical gold concentration of
$x_{\mathrm c}=0.1$ implies that a considerable amount of
scatterers are present which change the Hall effect dramatically
\cite{loe06,fuk07}. The latter appear to lead to a dominance of
the anomalous contributions to the Hall resistivity for all
non-stoichiometric samples. Consequently, the normal contributions
cannot be determined reliably and hence, information on the Fermi
surface evolution is difficult to extract. This is also seen from
the fact that no signatures are seen in the Hall resistivity when
probing the field-induced QCP in samples on the magnetically
ordered side of the QCP \cite{rog07}.

Pressure reverses the effect of Au substitution in
CeCu$_{6-x}$Au$_x$. Consequently, the pressure-driven QCP is only
accessible for samples with finite Au content. A Hall effect study
on such a sample would presumably suffer from the strong anomalous
contributions \cite{Bartolf2005,fuk07}. Moreover, Hall effect
studies under pressure are highly demanding. The fine tuning of
pressure required to resolve a rapid change in the Fermi surface
is rarely achieved. This is seen for instance in the heavy fermion
system CeRh$_2$Si$_2$ where the Hall coefficient was extracted as
a function of pressure \cite{bou06}. The few data points obtained
in the pressure range up to 1.3\,GPa provide only limited insight.
Here, the Hall coefficient seems to be sensitive to a change in
the magnetic structure at 0.6\,GPa. At the low temperature
magnetic transition, however, the data appear to evolve smoothly
despite the fact that de Haas--van Alphen measurements indicate a
strong Fermi surface reconstruction \cite{ara02}. It would be
necessary to collect further data to draw a firm conclusion about
the evolution of the Fermi surface. We note that the sign change
of \RH\ observed in proximity to the critical pressure cannot
directly be related to a severe Fermi surface change since
CeRh$_2$Si$_2$ features multiple bands \cite{ara02}, {\it cf.}
discussion in section \ref{sec:HallFS}.

For the material CeRhIn$_5$ a drastic change of the Fermi surface
at a critical pressure of about 2.3 GPa was observed by de
Haas--van Alphen measurements \cite{shi05} which has been ascribed
to a breakdown of the Kondo effect at an underlying QCP
\cite{par06}. It is to be expected that the Hall coefficient will
correspondingly show a drastic change across the critical
pressure. Existing Hall measurements within the zero-field limit
and at relatively high temperatures (above 2~K) have already
revealed a rapid change of the Hall coefficient across the
critical pressure; this rapid change is likely associated with
both the Kondo physics and the QCP, because it is absent in both
LaRhIn$_5$ and CeCoIn$_5$ \cite{nak07}. In order to determine the
precise nature of the evolution of the Hall effect across the
critical pressure, it is important to carry out measurements at
low temperatures (0.4~K or below, where the electrical resistivity
shows a $T^2$ dependence \cite{kne08}) and at magnetic fields
where superconductivity is suppressed. It should also be stressed
that the precise evolution of the Hall coefficient across the QCP
needs to be studied with care, given the discrete steps in
pressure used as the control parameter.

\begin{figure}
\centering\includegraphics[width=7.8cm]{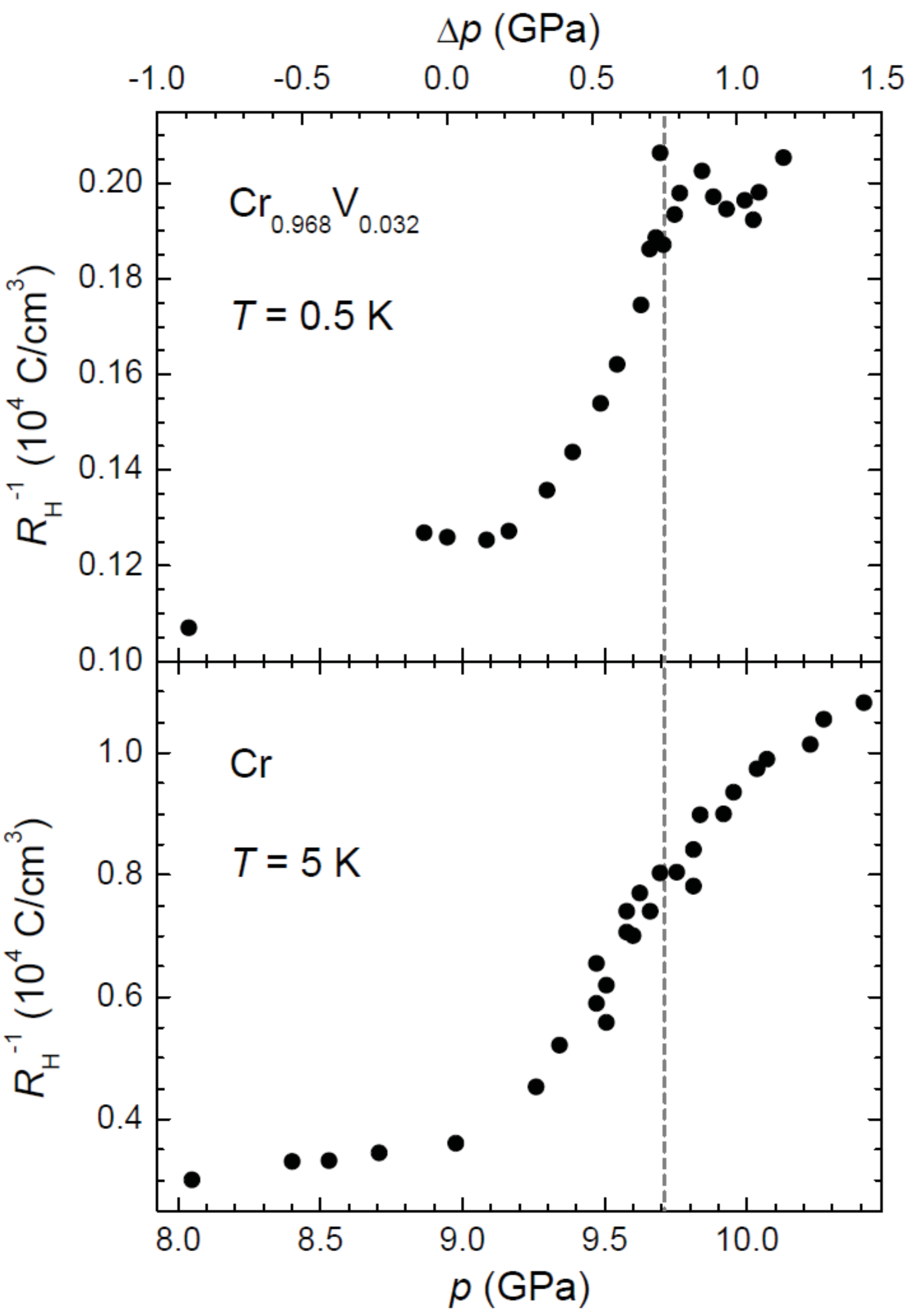}
\caption{\label{fig:CrV} Inverse Hall coefficient of pure Cr
(lower panel) and Cr$_{0.968}$V$_{0.032}$ (upper panel) as a
function of pressure. In the upper panel, $\Delta p$ denotes the
external pressure for Cr$_{0.968}$V$_{0.032}$, or an effective
pressure for samples in which the V-concentration is different
from 0.032 \cite{Lee2004}. In the lower panel, the result is shown
only in the vicinity of the critical pressure \cite{Jaramillo2010}
so that the two panels cover the same pressure extent. The dashed
lines label the critical pressure. Reproduced with permission from
S. Friedemann {\it et al.}, Journal of Physics: Condensed Matter
{\bf 23} (2011) 094216 \cite{Friedemann2010e}. Copyright (2011)
IOP Publishing Ltd.}
\end{figure}

A canonical SDW QCP is realized in pure and V-doped Cr
\cite{yeh02}. Hall effect measurements under pressure conducted on
these materials represent the rare examples where the Hall
coefficient was measured for a large series of pressure points
\cite{Jaramillo2010,Lee2004}, see Fig.\ \ref{fig:CrV}. These
measurements were conducted at 5 and 0.5\,K, respectively, {\it
i.e.} at temperatures which are small compared to the natural
temperature scales of these systems. The measurements reveal a
smooth crossover of the Fermi surface when the magnetic order is
suppressed, a behavior which is expected theoretically
\cite{col01,Norman2003,Bazaliy2004}. The Fermi surface of a SDW
state is reconstructed from that of the paramagnetic state through
a band folding, which is more pronounced for systems like Cr whose
Fermi surfaces are nested. However, if the SDW order parameter is
adiabatically switched off the folded Fermi surface is smoothly
connected to the paramagnetic one. As a result, the Hall
coefficient does not show a jump as long as the nesting is not
perfect \cite{Bazaliy2004}. So far, no information was extracted
regarding how this crossover evolves with temperature; it would be
interesting to see if the finite crossover width persists in the
extrapolation to zero temperature as expected for a SDW scenario.
However, we wish to point out that these pressure measurements
were very extensive and it would be an even larger effort to
implement the techniques used to the low temperatures needed to
study heavy fermion systems.

The remaining common tuning parameter, magnetic field, appears to
be the obvious choice to achieve high resolution. In case of a
magnetic-field driven QCP, however, the non-zero critical field
can give rise to a small discontinuity in the magnetotransport
coefficients even in the case of an SDW QCP \cite{Fenton2005}.
Approaching the QCP, the energy scale associated with the
vanishing order parameter becomes smaller than the scale
associated with the Lorentz force, leading to a non-linearity in
the system's response to the Lorentz force. This is reflected in a
breakdown of the weak-field magnetotransport. The linear field
dependence of the magnetoresistance in Ca$_3$Ru$_2$O$_7$ was taken
as indication thereof \cite{kik10}. Moreover, the breakdown of
weak-field magnetotransport may lead to a jump in the Hall
coefficient which, however, is likely to be very small. In fact,
for the case of the cuprates mean field theory predicts that the
anomaly in the Hall response at the critical doping level is
negligibly small even though the SDW gap is large \cite{mil05}.
Disorder may smear the jump of the Hall coefficient into a smooth
crossover.

For the above-mentioned pressure-driven SDW QCP of pure and
V-doped Cr \cite{Lee2004,Jaramillo2010}, the Hall coefficient (and
the resistivity) exhibits a smooth crossover near the QCP the
width of which does not track the strength of disorder
(Fig.~\ref{fig:CrV}). This behaviour is in contrast to the
scenario of a breakdown of the weak-field limit. For the
field-driven QCP of YbRh$_2$Si$_2$, the effect of non-linear
response to the Lorentz force is expected to be negligible, given
that at the critical field, $\omega_c \tau$ as defined in eq.\
(\ref{HallAngle}) is very small (on the order of 0.01 and 0.002
for the samples described in Ref.~\cite{Friedemann2010b}).
Furthermore, it is completely avoided by the crossed-field Hall
setup, see section \ref{sec:cross}.

\subsection{Outline and Scope} \label{sec:outline}
In this article, we give an overview of experimental and
theoretical work on the Hall effect in the heavy fermion metals.
We start with a brief description of the basics of the Hall
effect, and an introduction to the anomalous Hall effect in
section \ref{sec:basics}. Subsequently, we introduce the
heavy-fermion metals and describe the ingredients which go into
the formation of the heavy-fermion fluid. A description of the
anomalous Hall effect is carried over from the earlier section,
and a more microscopic description of this effect is provided. The
relevance of Hall effect measurements in the investigation of
these systems is stated, and the use of the Hall effect in
identifying some transitions and crossovers observed in the
low-temperature phase diagrams of the heavy-fermion systems is
described. Section \ref{sec:theory} provides an account of
theoretical work on the Hall effect in strongly correlated
electron systems. This incorporates models ranging from the
earlier geometric methods to contemporary ones where the details
of band structure are explicitly considered. A brief description
of the experimental aspects of Hall effect measurements are
provided in section \ref{sec:exp}. Special mention is made of the
crossed-field Hall experiments, where a tuning field is introduced
in addition to the magnetic field used for generating the Hall
response. The influence of strong magnetic fluctuations on the
physical properties of the heavy-fermion systems is well
documented. Section \ref{sec:fluc} describes the signatures of
these magnetic fluctuations as seen in Hall effect measurements.
Special mention will be made of the Ce$M$In$_5$ ($M$ = Co, Ir or
Rh) family of compounds where a complex interplay between
superconductivity and magnetism has been observed. Section
\ref{sec:qpt} deals with the investigation of quantum critical
phenomena -- a topic of contemporary interest. The utility of Hall
effect measurements in the investigation of quantum criticality
will be outlined, and experimental work in this area is reviewed.
In section \ref{sec:sces}, we compare the Hall effect in the
heavy-fermion systems with observations in other strongly
correlated electron systems, with emphasis on the manganites,
doped Mott insulators and the copper oxide superconductors.
Section \ref{sec:sum} sets out some open questions.

One cannot hope to be comprehensive in this vast subject, so we
have omitted several issues entirely. These include the
heavy-fermion transition metal oxides as well as the mixed state
of the heavy-fermion superconductors. Since only a very brief
description on early work in metals is included in section
\ref{sec:basics}, we refer the reader to earlier reviews
\cite{faw64,hur72,hur74} which deal exclusively with the Hall
effect in metals and alloys.

\section{Basics of Hall effect and heavy-fermion systems}
\label{sec:basics}
\subsection{History of the Hall effect}
\label{sec:hist} Unconvinced by a passage from Maxwell's treatise
on {\em Electricity and Magnetism} which stated that ``the path of
a current through a conductor is not permanently altered by a
magnetic field'', Edwin H. Hall, in 1879, set about investigating
the action which a magnetic field would have on such a current.
Using experiments on a gold leaf held between the poles of an
electromagnet, he observed that when a magnetic field is applied
perpendicular to the direction of a current flow, an electric
field is generated perpendicular to both, the direction of the
current, and the direction of the magnetic field \cite{hal79}.
This resultant electric field could then be detected using a
sensitive galvanometer. Hall also realized that the product of the
current through the specimen and the strength of the magnetic
field, when divided by the current through the galvanometer, was
reasonably constant. In current notation, this implies that
$V_{\rm H} \propto B I$, where $B$ and $I$ refer to the magnetic
flux density and the current flowing through the specimen,
respectively, and $V_{\rm H}$ denotes the transverse (Hall)
voltage.

\begin{figure}
\centering\includegraphics[width=7.5cm,clip]{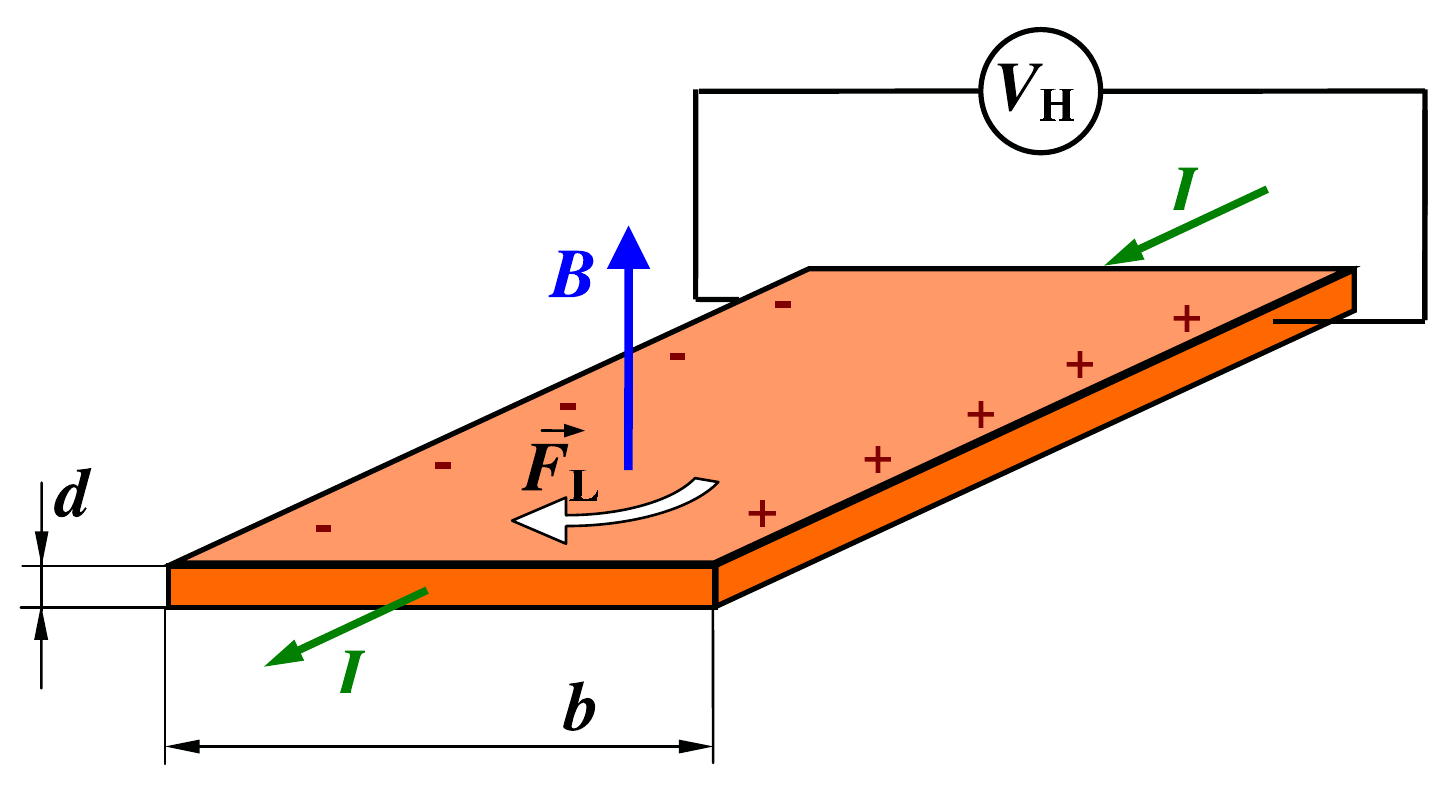}
\caption{\label{fig:Hall} Schematic setup for the measurement of
the Hall effect.}
\end{figure}

Intuitively, this can be seen as a direct consequence of the
Lorentz force acting on the electron current flowing through the
solid. Consider a rectangular specimen of width $b$ and height $d$
as shown in Fig.\ \ref{fig:Hall} and assume a current $I_x$ of
charge carriers $q$ with density $n$ flowing uniformly along its
length with velocity $v_x$. When $B_z$ is applied perpendicular to
the direction of the current, the electrons are deflected by the
Lorentz force towards the edge of the specimen. This transient
motion builds up a charge at (and thus an electric field $E_y
\equiv E_{\rm H}$ between) the edges of the sample, which in turn
impedes further electrons from accumulating at the edges. A
stationary state is achieved when the deflection of charge
carriers due to the Lorentz force is balanced by the force
resulting from the transverse electric field: $E_{\rm H} -(v_x
B_z) = 0$. With $I_x/bd = nqv_x$ one finds for the Hall voltage
$V_y = E_{\rm H} /b$ within the simple free electron model:
\begin{equation}
V_y = \frac{1}{nq} \frac{I_x}{d} B_z = R_{\rm H} \frac{I_x}{d} B_z
\label{Hallvolt}
\end{equation}
The constant of proportionality $R_{\rm H} = 1/nq$ is the
so-called Hall coefficient. Obviously, Hall measurements can
reveal information on the {\em type} of charge carriers. This was
soon realized with the observation of a positive $R_H$ in some
metals. Eq.\ (\ref{Hallvolt}), however, is strictly valid only if
the material can be described by a simple one-band picture. In
more complex materials, more then one band may exist at the Fermi
energy $E_F$ and hence, contributes to the conductivity. In these
cases, the charge carrier concentration as determined from Hall
measurements has to be considered an effective one, $n_{\rm eff}$,
and the interpretation of the results can be considerably
complicated, see {\it e.g.} \cite{hur72} and section
\ref{sec:HallFS}.

\subsection{The influence of magnetism: Anomalous Hall effect}
\label{sec:anom} The Hall effect in metals is a direct consequence
of the breaking of time reversal symmetry, in this case by the
application of an external magnetic field $H$. It was realized
very early on that the Hall effect could possibly arise even in
the absence of an external field, as long as time reversal
symmetry is broken. Early experimental insight was provided by
measurements on ferromagnetic elements and alloys, where unlike
that observed in simple diamagnetic metals, a pronounced
nonlinearity in the field dependence of the Hall signal was
observed \cite{smi10,smi16,smi21} (paramagnetic materials may
exhibit small nonlinearities). The Hall voltage was not only seen
to be proportional to the magnetization $M$, but also appeared to
reproduce the irreversibilities observed in the magnetization
curves \cite{smi29,pug30,pug32}. This additional contribution to
the ordinary, or normal, Hall effect was termed the extraordinary,
or anomalous, Hall effect (AHE). At fields larger than that
required to drive the magnetization of the material into
saturation, the Hall effect continued to increase, albeit at a
much slower rate.

\begin{figure}
\centering \includegraphics[width=6.4cm,clip]{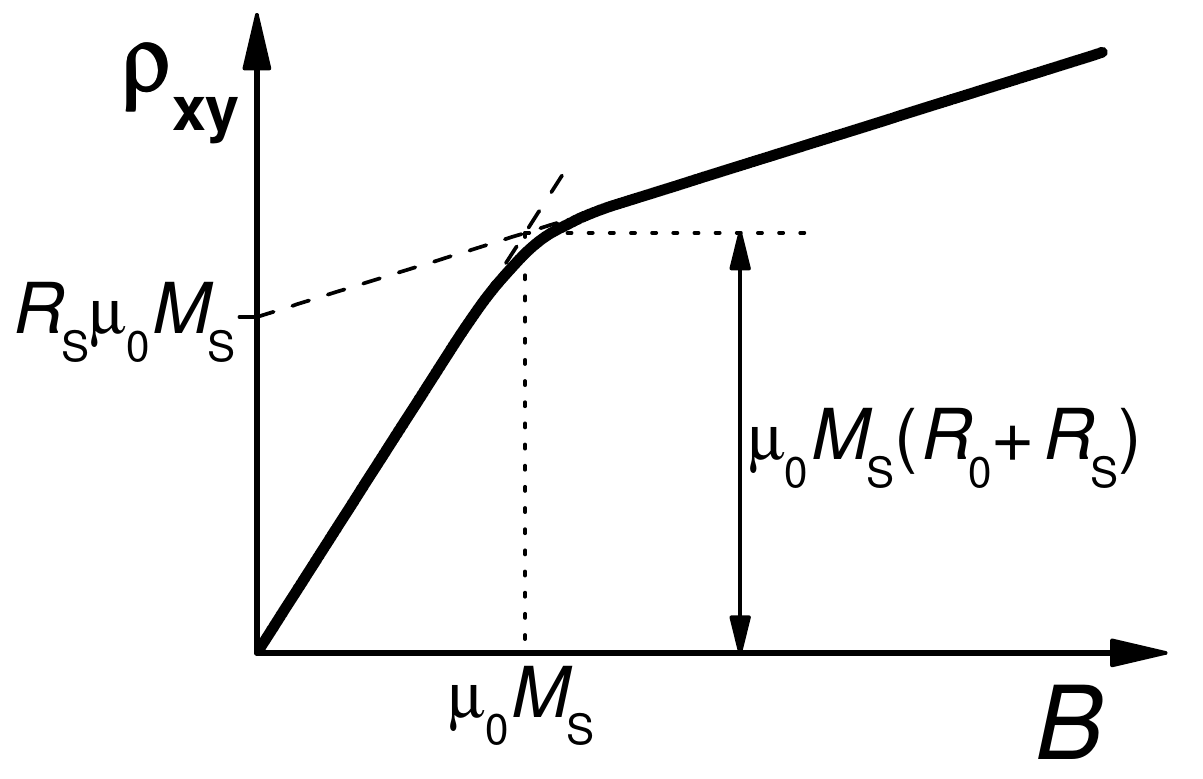}
\caption{\label{fig:AHE} Schematic of the typical behavior of the
Hall resistivity $\varrho_{xy}$ in a magnetic field.}
\end{figure}

Empirically, the measured Hall resistivity in ferromagnets could
thus be written as
\begin{equation}
\varrho_{xy} = R_{\rm 0} B + R_{\rm S} \mu_0 M \label{rhoAHE}
\end{equation}
where $R_0$ and $R_S$ refer to the normal and the anomalous Hall
effect contributions, respectively. Fig.\ \ref{fig:AHE} shows a
schematic of the typical behavior of the Hall resistivity in a
metal with appreciable magnetization as a function of applied
field. Two distinct, linear regimes with different slopes are
clearly discernible. In the first one, the slope equals $R_0 +
R_{\rm S}$. The second regime lies above the technical saturation
field, and here the smaller slope corresponds to $R_0$ only. The
straight extrapolations of these two regimes intersect at $B =
\mu_0 M_S$ where $M_S$ is the spontaneous magnetization. Though
the empirical formula (\ref{rhoAHE}) assumes that the anomalous
Hall effect is directly proportional to the magnetization, the
observed phenomenon could not be simply explained in terms of the
internal field of the ferromagnetic system. A non-trivial
temperature dependence and even a change of sign of the Hall
signal at low temperature in some systems, warranted the use of
more sophisticated models to explain the experimental data (see,
{\it e.g.}, Ref.\ \cite{jiang10}). It is now understood that the
extent and dependencies of the AHE arise as a consequence of the
spin-orbit coupling, although some of the microscopic details
continue to remain investigated. In what follows three basic
mechanisms which have been used to account for the anomalous Hall
effect, namely ``skew scattering'', the ``side-jump'' mechanism
and the ``Berry phase'' induced Hall signal, are described in more
detail. For early reviews of the anomalous Hall effect, the reader
is referred to Refs.~\cite{hur72,pug53} whereas more contemporary
reviews on the anomalous Hall effect are provided in
Refs.~\cite{nag06,wol06,nag10}.

\subsection{Mechanisms contributing to the anomalous Hall effect}
\label{sec:Hallmech}
\subsubsection{Skew scattering} \label{sec:skew}
First postulated by Smit, skew scattering refers to the phenomenon
of electrons traversing in a plane perpendicular to the applied
magnetic field being scattered {\em asymmetrically} from a
magnetic impurity potential \cite{smi55}. Considering a stream of
electrons traversing along a plane, Smit argued that in a
perfectly periodic lattice, the transverse polarization resulting
from the spin-orbit interaction is fully compensated by the
periodic electrostatic forces. However, in a real crystal where
periodicity is imperfect, the compensation near this impurity
potential is incomplete. This can result in a finite force which
accelerates electrons perpendicular to the direction of electron
flow. It was also suggested that for the skew scattering mechanism
the Hall resistivity should vary linearly with the conventional
electrical resistivity ($\varrho_{xy} \propto \varrho_{xx}$).
Since the early days of investigations on Kondo lattice systems,
this mechanism has been widely cited to be primarily responsible
for the anomalous Hall contribution observed in these systems.
Here, the magnetic impurities which are immersed in the sea of
conduction electrons act as impurity potentials for electron
scattering. These magnetic impurities can be polarized by applying
an external magnetic field, thus deflecting the charge carriers
preferentially in one direction giving rise to a large anomalous
contribution to the Hall effect.

\subsubsection{Side-jump mechanism} \label{sec:side}
The observation of the Hall resistivity varying quadratically as a
function of the resistivity ($\varrho_{xy} \propto
\varrho_{xx}^2$) in iron \cite{koo54} and some alloys \cite{sch55}
is inconsistent with the predictions of the simple skew scattering
mechanism, where a linear relation is expected. This led Berger to
propose an alternative explanation for the anomalous Hall effect
which was termed the side-jump mechanism \cite{ber70}. Berger
argued that when an electron is scattered by an impurity potential
(or a phonon), the trajectory of the electron is shifted sideways
by the action of the spin-orbit coupling. This abrupt jump takes
place in a direction perpendicular to both the initial electron
trajectory as well as the magnetization. Thus, unlike the skew
scattering mechanism in which the electron can propagate in a
direction perpendicular to the original direction after
scattering, here the electron is simply displaced sideways by a
small distance. This mechanism which was suggested to be
independent of the density of impurities could be used to account
for the observed $\varrho_{xy} \propto \varrho_{xx}^2$ behavior.
Since this displacement is characteristically smaller than the
electronic mean free path, the contribution arising due to the
side-jump mechanism is also typically much smaller than that
stemming from skew scattering. However, if the mean free path
becomes of the order of this displacement, then this mechanism
could dominate. Therefore, consequences of the side-jump mechanism
are more probable to be observed at relatively higher temperatures
or in samples with appreciable degree of disorder.

Recent calculations using density functional theory (DFT) showed
that the side-jump contribution to the anomalous Hall effect can
directly be computed from the electronic structure of a pristine
crystal, and good agreement with experimental data was observed in
the cases of some ferromagnetic elements and alloys \cite{wei11}.

\subsubsection{Berry phase contributions} \label{sec:berry}
Unlike the skew scattering and side-jump mechanism which deal with
scattering from impurity potentials and thus, are clearly
extrinsic in nature, the Berry phase contribution to the AHE is an
intrinsic geometric contribution. The nomenclature is due to M.
Berry, who first pointed out that a quantum system which is
adiabatically transported along a closed loop acquires a phase
which depends purely on the geometry of the loop \cite{ber84}.
Interestingly, though the realization of the existence of this
geometric phase and its importance is relatively recent, Karplus
and Luttinger had already in 1954 discovered the probably earliest
instance of a Berry phase in a solid in an attempt to explain the
origin of the anomalous Hall effect \cite{kar54,lut58}.  In
calculating the electron transport in a system with broken time
reversal symmetry, they uncovered a term which mimicked the
influence of a magnetic field, thus giving rise to a
dissipationless transverse current. The Berry phase contribution
has now been observed in some systems such as spinels \cite{lee04}
and pyrochlores \cite{tag01} and continues to be a field of
intense activity especially in the field of spintronics
\cite{jun02}. However, its relevance---if any---to the
heavy-fermion systems remains to be evaluated.

We note that the side-jump mechanism could also be considered as a
consequence of the Berry phase contribution in samples with
moderate impurity concentration. This leads to the same
dependencies on resistivity, $\varrho_{xy} \propto
\varrho_{xx}^2$, in case of both mechanisms \cite{nag10}.

\subsection{Hall effect and Fermi surface}
\label{sec:HallFS} Eq.~(\ref{Hallvolt}) and all considerations
based on it (including $\RH\ = 1/nq$) rely on Drude's model of a
free electron gas. However, already in simple metals like Al its
assumptions no longer hold and, {\it e.g.,} \RH\ depends strongly
on magnetic field \cite{lue66}. In general, \RH\ in metals
strongly depends on the actual band structure and the particular
shape of the Fermi surface onto which the electrons traverse.

Before continuing we shall introduce a useful quantity, the
so-called Hall angle $\theta_{\rm H}$. It basically describes the
angle between the total electric field and the current
\begin{equation}
\tan \theta_{\rm H} = \frac{\varrho_{xy}}{\varrho_{xx}} \; \propto
\; \omega_c \tau = \mu B\:\: , \label{HallAngle}
\end{equation}
where $\omega_c$ is the cyclotron frequency, $\tau$ is the
relaxation time and $\mu$ the mobility. For small magnetic fields,
the Hall angle can be considered as the angular deviation of the
electron's motion within $\tau$. In high magnetic field and within
the simple Drude picture, $\theta_{\rm H}$ approaches $\pi /2$.
According to eq.~(\ref{HallAngle}) the Hall angle provides a
direct measure of $\omega_c \tau$ and hence, the mobility $\mu$.
Note that for large values of $\omega_c \tau = eB \tau / m_{\rm
eff}$ not only high magnetic fields but also low temperatures and
single crystals of sufficiently good quality are required.

If all occupied electronic levels fall on closed orbits the high
field limit of the Hall coefficient again simplifies to $\RH\ =
1/nq$ (a similar consideration holds for holes). This is no longer
valid if one (or more) orbit is open in $\pmb k$-space. Depending
on direction of this orbit in real space the high field limit of
$\theta_{\rm H}$ can drastically deviate from $\pi /2$ and, more
generally spoken, the Hall effect can become anisotropic with
respect to the sample's crystallographic orientation. Of course,
in the small-field limit ($B \rightarrow 0$) the precise nature of
the orbits is insignificant.

In the majority of metals there is more than one band crossing the
Fermi energy $E_{\rm F}$. In consequence, all these bands,
electron- or hole-like, contribute to the resulting Hall effect.
Although in principal the voltages generated by the individual
bands simply sum up, the resulting expressions quickly become
cumbersome (in general, the Hall resistivity is a matrix element
obtained by inverting the conductivity tensor, see below). For two
bands (the expression can easily be generalized to more than two
bands) one finds in the low-field limit \cite{hur72}
\begin{equation}
\RH\ = \frac{1}{e \, \sigma^2_t} \left[ \frac{\sigma_1^2}{n_1} +
\frac{\sigma_2^2}{n_2} \right] \qquad \textrm{where} \qquad
\sigma_t = \sigma_1 + \sigma_2 \label{twoband}
\end{equation}
and the indices refer to the individual bands. If an effective
carrier concentration is introduced as
\begin{equation}
\frac{1}{n_{\rm eff}} = \frac{1}{\sigma^2_t} \left[
\frac{\sigma_1^2}{n_1} + \frac{\sigma_2^2}{n_2} \right]
\end{equation}
than the expressions for the single-band case are recovered. Each
band is characterized by two parameters: its carrier concentration
$n_i$ and its mobility $\mu_i$ (or related quantities). Therefore,
an analysis of measured Hall effect data even for two-band
conductors is complicated as has clearly been demonstrated for
CrO$_2$ \cite{watts}. For more than two bands any quantitative
analysis is severely hampered by the numerous free parameters.
Only in the high-field limit, {\it i.e.} for $B \rightarrow
\infty$, is a simple relation within the two-band model recovered,
\RH\ = $1 / e (n_1 + n_2)$. Experimentally, however, it is
difficult to ascertain that this condition is met, and so for all
bands involved.

A peculiar case arises for two-band conductors in which an
electron and a hole band exhibit equal carrier concentrations,
so-called \emph{compensated} metals. In this case, there is a
dominating contribution to $\varrho_{xy}(B)$ which goes as $B^2$
\cite{ree64,ber69}. The latter has been demonstrated to hold by
high-field experiments on the heavy fermion metal UPt$_3$
\cite{Kambe1999}. This aspect will be important for the discussion
of the Hall effect on Ce$M$In$_5$ systems, sections
\ref{sec:Co115} and \ref{sec:Ir115}.

\subsection{Basic remarks on heavy-fermion systems} \label{sec:HF}
The assumption of non-correlated, {\it i.e.} essentially
non-interacting, electrons is well established throughout many
areas of solid state physics. An excellent example for this is
semiconductor physics which is typically understood in terms of
non-interacting electrons. However, strong correlations between
the electrons could offer new concepts and applications. Interest
is fueled by the fact that the properties of the whole ensemble of
interacting entities (in our case the electrons) may lead to new
organizational principles or low-lying excitations neither related
to nor expected from the properties of the individual
constituents. The occurrence of such new principles are referred
to as {\em emergent behavior}. Typical examples here range from
superconductivity, colossal magnetoresistance, to the fractional
quantum Hall effect.

Originally formulated in an effort to explain the bulk properties
of $^{3}$He, Landau's Fermi liquid theory \cite{lan57} has found
remarkable success in explaining the low temperature properties
also of many materials that show strong electronic correlations. A
key ingredient here is the notion of ``quasiparticles'', which
refer to low-lying excitations that have a one-to-one
correspondence in terms of their quantum numbers with the original
particles, which in the case of metals are the conduction
electrons. Moreover, the Landau Fermi liquid theory could also
predict the temperature dependencies of experimentally measurable
quantities at low temperatures. For instance, the contribution of
electron-electron scattering to the electrical resistivity
$\varrho$ varies as $T^2$, the spin susceptibility $\chi$ is
temperature independent, and the specific heat divided by
temperature, $C/T$, is a constant. Based on this concept, the
occurrence of low temperature superconductivity in simple metals
can well be accounted for by the BCS theory \cite{bcs}, where
phonons provide the glue that holds together electrons into
Cooper-pairs.

The heavy fermion metals are a remarkable example of the robust
nature of the Fermi liquid state. This is especially the case if
these materials contain certain elements with partially filled $f$
electron shells, like Ce, Yb or U. At relatively high
temperatures, these incompletely filled $f$ shells give rise to
localized atomic moments which interact only weakly with the sea
of conduction electrons that engulf them: the conduction electrons
experience a weak energetic preference to align their spins
antiparallel to the total spin of the open $f$-shell. However,
when the temperature is reduced, this antiferromagnetic
interaction between the localized $f$-spins and the conduction
electron spins becomes continuously stronger, and this radically
influences the low-energy ground states of the system. These
strong correlations are manifested in the form of anomalous
contributions to various thermodynamic and transport properties.
Moreover, the $f$ electron moments in this low temperature state
are seen to be appreciably smaller than their high temperature
values and, eventually, disappear in static properties for $T \ll
T_{\rm K}$, the so-called Kondo temperature---a consequence of the
``Kondo effect''. The resistivity in the high temperature limit is
large as a consequence of spin flip scattering, and a dramatic
reduction is seen as one enters the low temperature coherent
regime. If not interrupted by the onset of magnetic order, the
heavy fermion metal can then settle into a LFL regime with
strongly renormalized quasiparticles having effective masses
$m_{\rm eff}$ up to three orders of magnitude larger than the bare
free electron mass.

The low temperature electronic ground states of these systems are
thus dictated by the intricate interplay between two competing
phenomena: the Kondo effect \cite{kon64}, which screens the local
moments and promotes the formation of a nonmagnetic ground state,
and an indirect exchange interaction, the so-called
Ruderman-Kittel-Kasuya-Yosida (RKKY) interaction
\cite{rud54,kas56,yos57}, which couples the (partially
compensated) $f$-spins via the conduction electrons. The Doniach
phase diagram \cite{don77} provides a useful illustration of how
the electronic ground states of these systems evolve as a function
of the coupling (or the hybridization) between the $f$ electrons
and the conducting electrons, as is
\begin{figure}[t]
\centering \includegraphics[width=8.4cm,clip]{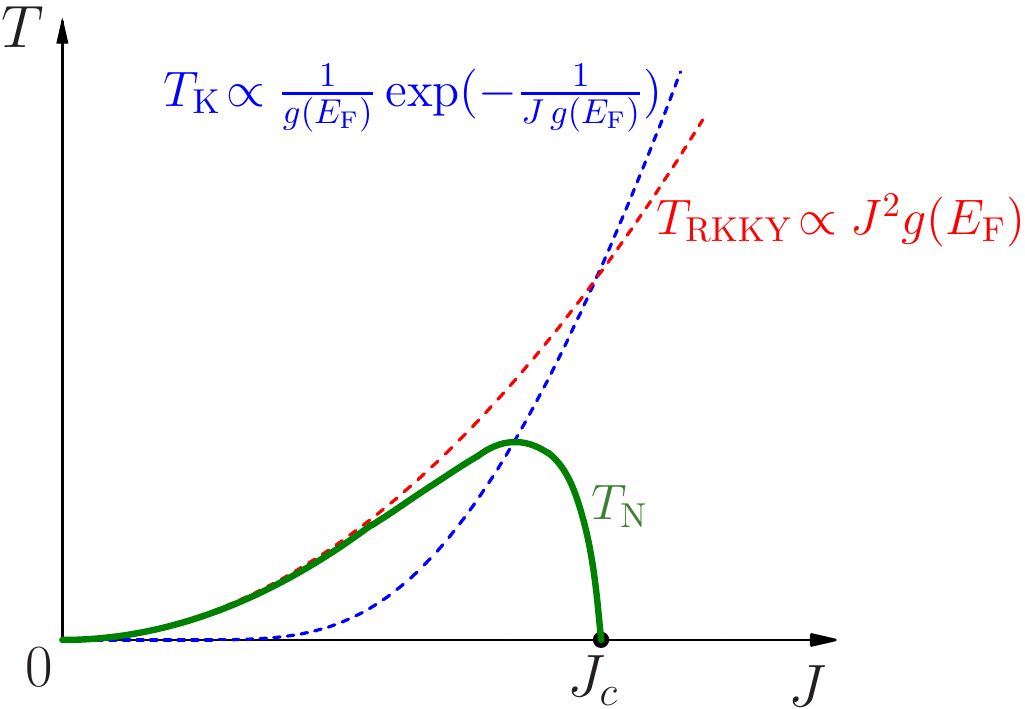}
\caption{\label{fig:doniach} Phase diagram of heavy fermion metals
as inspired by Doniach \cite{don77}. Here, $J$ refers to the Kondo
exchange coupling between the spins of the $f$ and conduction
electrons and $g(E_{\rm F})$ to the density of states (DOS) at
$E_{\rm F}$.} \label{doniach}
\end{figure}
schematically shown in Fig.~\ref{doniach}. At low
antiferromagnetic Kondo coupling ($J$) values, the RKKY
interaction (characterized by the energy scale $T_{\rm RKKY}
\propto J^2$) dominates the Kondo screening temperature ($T_{\rm
K} \propto e^{-1/J}$) and the system orders magnetically. At high
values of $J$, on the other hand, the Kondo screening prevails and
the system condenses into a sea of highly interacting
quasiparticles with LFL characteristics. The resurgence in the
field of heavy fermions is predominantly due to current interest
in quantum critical behavior. This behavior is observed if the
material undergoes a {\em continuous} phase transition at absolute
zero temperature; in the scenario just discussed such a transition
can take place from a magnetically ordered phase into a heavy LFL
one, with the latter representing an ordered state in $\pmb
k$-space. Within the framework of Doniach's description of the
Kondo lattice, systems for which such phenomena are likely to
occur are those where the Kondo effect and the RKKY interaction
have comparable energies, {\it i.e.,} for intermediate values of
$J$. A quantum phase transition can then be brought about by a
well-directed change of a non-thermal experimental (``control'')
parameter such as chemical doping, pressure or magnetic field
\cite{her76}. While the former two can directly influence the
value of $J$ ({\it e.g.} via changed lattice parameters) the
latter may suppress antiferromagnetic order and, in particular for
transverse field tuning, even ferromagnetic order. Even though
such a quantum phase transition at $T = 0$ is not directly
accessible to experiment it affects the finite temperature
properties of the material if investigated sufficiently close to
the QCP at which the continuous phase transition takes place in
the temperature--control parameter diagram \cite{hvl07}. The
Landau Fermi liquid theory of conventional metals breaks down in
the vicinity of such instabilities, and anomalous experimental
behavior like a power-law temperature dependence of the electrical
resistivity ($\varrho(T) - \varrho(T\! =\! 0) \propto
T^{\epsilon}$ where $1 \le \epsilon < 2$) \cite{cus03}, a
diverging specific heat coefficient \cite{loh94} and even an
apparent violation of the Wiedemann-Franz law \cite{pfau12} have
now been observed in heavy fermion systems.

Experimentally, quantum critical behavior and unconventional
superconductivity are often found in close vicinity in phase space
in this class of materials. In addition to the compound
CeCu$_2$Si$_2$ \cite{ste79}, the compounds in which the phenomenon
has been observed include other Ce \cite{pet01a,pet01b}, but also
U \cite{ott83,ste84,pal85,schl86,gei91} and more recently a Yb
\cite{nak08} based system. The observation of superconductivity in
these systems which are comprised of a dense array of magnetic
atoms is striking since the conventional BCS theory predicts that
the presence of even very small amounts of magnetic impurities is
highly detrimental to the formation of the superconducting
condensate. The fact that superconductivity can exist in an
inherently magnetic environment has forced researchers to think
beyond the conventional scenario of phonon-mediated
superconductivity. The driving force for Cooper pair formation in
these heavy fermion metals is believed to be electronic (or more
specifically, magnetic) in origin
\cite{miy86,sca86,mor03,mon07,geg08}. In case of the compound
CeCu$_2$Si$_2$, an inelastic neutron scattering study of the
magnetic excitation spectrum provided indications for
superconducting pairing resulting from antiferromagnetic
excitations in this prototypical heavy-fermion compound
\cite{stockert11}. An earlier dramatic manifestation of this
aspect is the superconductivity observed when a continuous
magnetic phase transition is suppressed to absolute zero
temperatures, in other words in the vicinity of a QCP
\cite{mat98}. This has also reinforced the connection of heavy
fermion systems with the high transition temperature
superconducting cuprates. For the latter, one of the competing
scenarios is that a QCP lies beneath the superconducting dome and
is responsible for both superconductivity as well as the strange
metallic (non-Fermi-liquid like) behavior observed in an
appreciable region of the experimentally accessible phase space
\cite{bro08}.

The formation of the heavy fermion state and superconductivity in
these materials has been a subject of extensive investigations in
recent years, the details of which are  beyond the scope of this
article. \nocite{ste01} We refer the reader to some comprehensive
reviews devoted to both the experimental and theoretical aspects
of these systems \cite{gre91,ste01,tha05,hvl07,pfl09}.

\subsection{Anomalous Hall effect in heavy-fermion systems}
\label{sec:AHE-HF} Though the skew scattering mechanism has been
used extensively in trying to account for the observed Hall effect
in heavy-fermion systems, a few caveats need to be borne in mind.
Firstly, the skew scattering mechanism was proposed for systems
with ferromagnetic order, {\it i.e.,} where the time reversal
symmetry is irrevocably broken. This criterion is not strictly met
in the heavy-fermion systems, although a lack of time reversal
symmetry can be induced by the application of a magnetic field.
Secondly, these scattering theories considered the simple case in
which the same type of electrons are responsible for both
magnetism and electrical conduction. In the heavy-fermion systems
it is clear that the magnetism arises from localized (partially
screened) $f$ electrons, and are thus of a different origin from
the conduction $s, p$ or $d$ electrons. A more relevant model here
is that by Kondo \cite{kon62}, who considered a situation in which
$d$ electrons were localized in a sea of conduction ($s$ shell)
electrons. These localized spins can be disordered by the
influence of thermal fluctuations and, in turn, can then act as
scattering potentials. Thus, the non-periodicity originates in
this case from thermal fluctuations of the local spins, and not
primarily from the introduction of impurities. Though this $s\! -
\!d$ spin-spin interaction itself did not give rise to skew
scattering, including the spin-orbit interaction of the $d$
electrons within the magnetic ions did result in skew scattering,
and thus an anomalous contribution to the Hall effect. A related
treatment of the Kondo model was presented by Maranzana
\cite{mar67} who considered a ($d$-spin--$s$-orbit) interaction
which described the force acting on moving electrons as a
consequence of the magnetic field produced by the magnetic $d$
ions. This was used to reproduce the temperature dependence of the
anomalous Hall effect in some ferromagnets as well as the abrupt
decrease in the Hall coefficient at the antiferromagnetic
transition temperature ($T_N$), though the calculated values were
smaller than the experimentally measured ones. A similar model was
also applied by Giovannini to explain the experimentally observed
anomalous Hall effect in some dilute magnetic alloys \cite{gio73}.

\begin{figure}
\centering \includegraphics[width=7.8cm,clip]{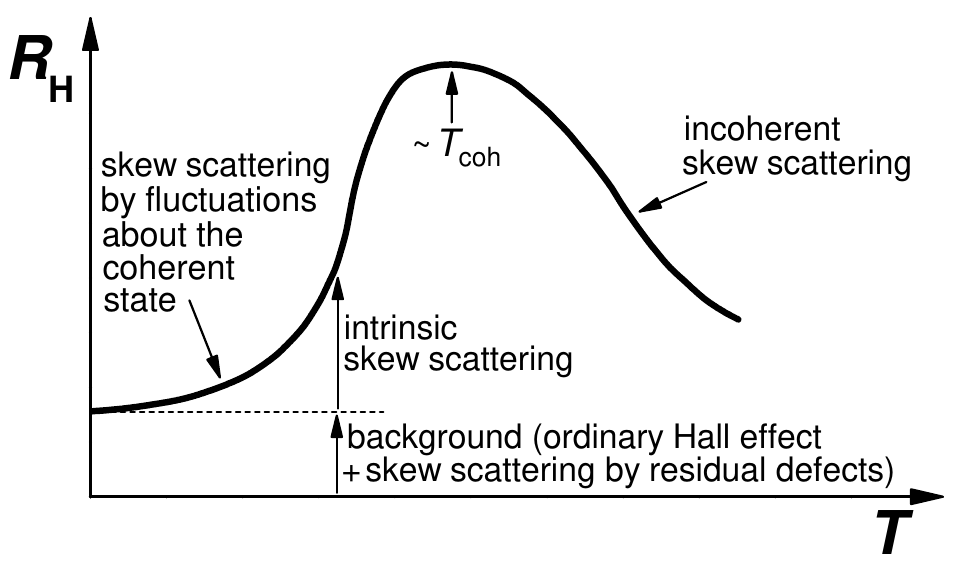}
\caption{\label{fig:fert} Modeling of the Hall effect in
heavy-fermion compounds. Reprinted figure with permission from A.
Fert and P.M. Levy, Physical Review B {\bf 36}, 1907 (1987)
\cite{fer87}. Copyright \copyright\,(1987) by the American
Physical Society.}
\end{figure}

Considerable theoretical insight into the Hall effect in the
heavy-fermion systems originates from the early work by Fert
\cite{fer73} who considered the skew scattering in alloys
containing Ce impurities. Using a Coqblin-Schrieffer interaction
\cite{coq86}, which considers a 4$f^1$ configuration of Ce and its
interaction with the conduction ($s$) electrons, a reasonable
agreement with the low-field experimental data of some La-Ce
alloys was established. Subsequently, a model pertaining to a
Kondo lattice system was used \cite{col85,ram85} in an effort to
explain the early experimental signatures observed in the
heavy-fermion metals. Here, the ground state was modeled by a
degenerate effective $f$ resonance level with an unquenched
orbital moment. Application of a magnetic field lifts the
degeneracy of the resonance level due to Zeeman splitting, and
gives rise to skew scattering of the band electrons. It is to be
noted that both these models were effectively meant to be applied
to the incoherent Kondo regime ($T \gg T_K$) where the Kondo
ions scatter independent of each other; they were not strictly
valid for the low temperature regime ($T \ll T_K$), where a
coherent heavy-fermion band condenses out of the local moment
landscape. Nevertheless, their results could---at least
qualitatively---be extended down to the coherent regime. Combining
both these approaches, a generic interpretation of the Hall effect
in the heavy-fermion systems was put forward \cite{fer87} which is
schematically shown in Fig.\ \ref{fig:fert}. It was concluded that
the skew scattering contribution increases as the temperature is
lowered and attains a maximum at the temperature $T_{\rm coh}$
where spatial coherence sets in. At even lower temperatures this
contribution rapidly falls down to zero. In the high temperature
regime, the Hall effect arising due to skew scattering is
proportional to the product of the magnetic susceptibility
($\chi$) and the magnetic contribution to the resistivity
($\varrho_M$). Well below $T_{\rm coh}$, the Hall effect is
primarily composed of the ordinary Hall effect, in addition to a
small contribution from defects or impurities. This inference is
striking, for it suggests that at very low temperatures the
measured Hall effect is an intrinsic quantity and thus can be
effectively used to monitor the evolution of the Fermi surface.
This has now been exploited in the investigation of quantum
critical phenomena, an aspect which will be dealt in detail in
section \ref{sec:qpt}.

\begin{figure}
\centering \includegraphics[width=8.0cm,clip]{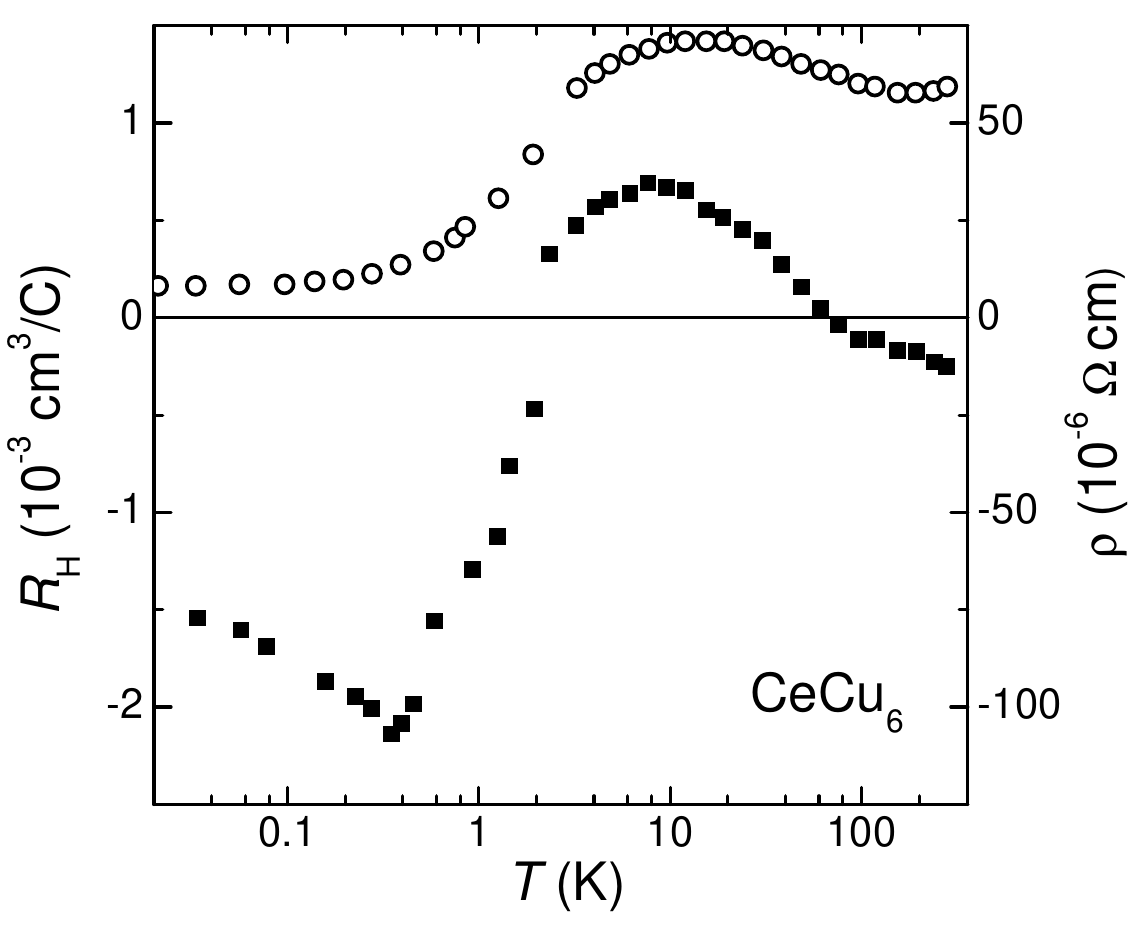}
\caption{\label{fig:compar} Comparison of the resistivity
$\varrho$ ($\circ$ and right scale) and the Hall coefficient \RH\
($\scriptstyle \blacksquare$ and left scale) at the onset of the
coherent state in CeCu$_6$. Reprinted figure with permission from
T. Penney {\it et al.}, Physical Review B {\bf 34}, 5959(R) (1986)
\cite{pen86b}. Copyright \copyright\,(1986) by the American
Physical Society.}
\end{figure}

Since the anomalous Hall effect exhibits the just described
maximum at the onset of the Kondo coherence, it is not surprising
that a large positive maximum followed by a precipitous drop was
observed in the measured Hall signal of a number of Ce and U based
heavy-fermion systems
\cite{cat85,pen86a,had86,sch86,onu87,lap87,ham88}. In these
reports, this drop in the measured Hall response was usually
ascribed to the formation of a coherent band which drastically
reduces electron scattering. In some systems, such as CePd$_3$,
CeBe$_{13}$ and CeCu$_6$, the drop in the Hall coefficient was
accompanied by a change of its sign. This change of sign of \RH\
was accounted for by the models described above in spite of their
limited validity in the low temperature ($T \ll T_K$) regime. For
instance, Ramakrishnan and coworkers had predicted \cite{ram85}
that the Hall coefficient can be written as
\begin{equation}
R_{\rm H} = R_0 + g \mu_B |\alpha| \, \varrho \sin(\phi + \delta_2)
/ \sin \delta_2 \; .\label{rama}
\end{equation}
Here $g$ is the gyromagnetic ratio of the $f$ electrons, $\mu_B$
is the Bohr magneton and $\delta_2$ refers to the phase shift due
to scattering in the $l = 2$ channel. In addition, $\alpha$ is
proportional to $\chi_1 (1 - \chi_1 T)$ where $\chi_1$ refers to
the reduced susceptibility. If the resonant scattering in the $l =
3$ channel at low temperatures gives rise to a phase shift
$\delta_3$, then $\phi$ changes from $-\pi$ to $-2 \delta_3$ as
one traverses from the high temperature ($T \gg T_K$) regime to
the low temperature ($T \ll T_K$) one. According to this model, in
the high temperature regime, $\phi = - \pi$, and thus the
anomalous Hall contribution is always positive. In the low
temperature regime one finds $\phi = -2 \delta_3$
and---interestingly---here the Hall effect can have either sign.
However, based on measurements on the Ce$_{1-x}$Y$_x$Pd$_3$
alloys, it was suggested that this change of sign in $R_H$ is
solely associated with the onset of coherence \cite{fer85}. This
was demonstrated by the fact that in compounds with finite Y
substitution, in which no low-temperature coherent state is
attained, the Hall coefficient \RH\ decreased monotonically as a
function of decreasing temperature. Qualitatively similar behavior
was also observed in the system Ce(Pd$_{1-x}$Ag$_x$)$_3$
\cite{cat85}. These results are in contrast to those observed for
the undoped compound CePd$_3$, where the formation of a
low-temperature coherent state is accompanied by a sign change in
\RH . The experimental signature of the onset of coherence
appeared to be much sharper in the Hall effect if compared to that
typically seen in resistivity measurements \cite{pen86b}. This
difference is exemplary demonstrated for the system CeCu$_6$ in
Fig.\ \ref{fig:compar}.

In many heavy-fermion systems, the onset of the low-temperature
coherent state is interrupted by the emergence of long-range
(antiferro-)magnetic order. In the system U$_2$Zn$_{17}$, it was
shown that $R_H$ increases abruptly at the onset of the
antiferromagnetic order \cite{sie86}. It was suggested that this
change reflects the opening of a gap at the Fermi surface due to
magnetic ordering. In the system CePtSi, this increase was seen to
be more dramatic, with the Hall constant increasing by a factor of
15 on entering the antiferromagnetically ordered regime
\cite{ham88}. Moreover, below the N{\'e}el transition temperature
$T_N$, the Hall resistivity was highly nonlinear, and also
exhibited signatures of a possible metamagnetic transition.
However, the resistivity was almost constant around $T_N$, thus
prompting the authors to conclude that the observed features
cannot be accounted for on the basis of a Fermi surface
modification alone, which would have affected the resistivity and
Hall measurements in similar fashions. Thus, irrespective of the
interpretation of the observed Hall coefficient across the
antiferromagnetic transition, the sensitivity of this measurement
in tracking such transitions is beyond doubt. Moreover, it appears
to be larger than that commonly observed in resistivity
measurements. This has been reinforced by measurements on the
heavy-fermion system URu$_2$Si$_2$ where a transition at 17.5~K
has been the focus of extensive investigations. The challenge here
is to reconcile the large entropy release and the accompanying gap
in the magnetic excitation spectrum with the anomalously small
value ($\sim 0.03\; \mu_B / U$) of the magnetic moment, giving
rise to suggestions that a ``hidden order'' coexists and probably
competes with the antiferromagnetic order in this system
\cite{tha11}. Early investigations in URu$_2$Si$_2$ revealed that
the Hall coefficient \RH\ exhibits a sharp anomaly at this
transition \cite{sch87}. An analysis based on the model by Fert
and Levy \cite{fer87} has been used to extract a carrier density
of about 0.04 holes/U above the transition, but is an order of
magnitude smaller below the transition \cite{daw89}. In
combination with Nernst effect measurements, it was thus inferred
that this transition is accompanied by a reconstruction of the
Fermi surface into small high-mobility pockets, resulting in an
abrupt increase in the entropy per itinerant charge carrier and a
decrease of the scattering rate \cite{bel04b}. Recently, Hall
effect measurements have been extended up to magnetic fields of
the order of 45 T and revealed a cascade of field induced
transitions accompanied by changes in the Fermi surface topology
\cite{oh07,lev09}.

\section{Theoretical Work on the Hall effect} \label{sec:theory}
\subsection{Theoretical Overview}
\label{sec:theory-overview} While the conceptual basis of the Hall
effect is accessible enough for it to be captured in a simple
sentence---a manifestation of the perpendicular Lorentz force on
moving charges---the interpretation of Hall measurements can be
considerably more challenging. Spectroscopies are usually the more
straightforward experimental quantities to connect to the
underlying theory because they often directly measure a response
function. In transport theory it is the elements of the
appropriate conductivity tensor that are related to correlation
functions~\cite{abrikosov1963a}---current-current for the
electrical conductivity. However the Hall resistivity (which is
the quantity typically measured in experiment) is not related to a
single correlation. Rather it is a ratio of conductivities through
the inverse of the conductivity tensor. One is helped to some
extent by the Onsager relation for the off diagonal elements of
the conductivity ({\it e.g.} $\sigma_{xy} = - \sigma_{yx}$).
Nevertheless the measured quantities can look complex:
\begin{equation}
\rho_{xy}=
-\frac{\sigma_{xz}\sigma_{yz} -
  \sigma_{xy}\sigma_{zz}}{\sigma_{xz}^2 \sigma_{yy} +\sigma_{yz}^2
  \sigma_{xx} + \sigma_{xy}^2 \sigma_{zz} + \sigma_{xx} \sigma_{yy}
  \sigma_{zz}}
\xrightarrow[\rm symmetry]{\rm tetragonal}-\frac{\sigma_{xy}}
  {\sigma_{xy}^2 + \sigma_{xx}^2} \; .
\end{equation}
Fortunately for materials with high symmetry and in the weak field
limit, this ratio can lead to cancellations between numerator and
denominator. This can, in turn, give direct insight into
physically measurable quantities, of which carrier concentration
in doped semiconductors is one elementary example. It is
remarkable that the heavy fermion systems may equally allow quite
simple interpretations of the Hall effect as we will show later.

Although the Hall resistivity may not be directly related to a
response function, it has been shown that the related
quantity---the tangent of the Hall angle $\theta_{\rm H}$ as
defined by eq.~(\ref{HallAngle})---is~\cite{drew1997a}. Physically
$\theta_H$ is the angle of deflection a free current would
experience in a transverse magnetic field. Its role as a response
function means that the frequency dependent Hall angle can be
integrated to provide a sum rule independent of the underlying
microscopic physics~\cite{drew1997a}. Optical Hall angle
measurements have been undertaken in the
cuprates~\cite{kaplan1996a}, but not yet in the heavy fermion
metals.

\subsection{Key results from Boltzmann theory}
\label{sec:Boltzmann} The most elementary theoretical treatment of
the Hall effect is via a semi-classical Boltzmann equation in the
relaxation time approximation which is well described in standard
texts~\cite{abrikosov1988a}. It assumes fermionic excitations with
a Fermi surface determined by a given dispersion relation and that
the collision processes can be approximated by a decay time to
equilibrium. In section~\ref{sec:beyond-flt} we consider the
limitations of this approach. Nevertheless, this is likely to be
valid in the low temperature limit of the Fermi liquid state when
impurity scattering dominates. The simple results of such an
approach already illustrate a number of important features of
magnetotransport in general.

Within the Jones-Zener expansion the conductivities are found
order by order in magnetic field~\cite{ziman1960a}, and are
related to surface integrals around the Fermi surface. The
conductivities are defined by
\begin{equation}
j_\alpha = \sigma_{\alpha\beta} E_\beta +
\sigma^{(1)}_{\alpha\beta\gamma} E_{\beta} B_\gamma +
\sigma^{(2)}_{\alpha\beta\gamma\delta} E_{\beta} B_\gamma B_\delta
+ \cdots \; ,
\end{equation}
where
\begin{eqnarray}
\sigma_{\alpha\beta} &=& \int e^2 \tau \frac{\partial
f^0}{\partial \varepsilon} v_\alpha v_\beta \; d^3 k \; , \\
\sigma^{(1)}_{\alpha\beta\gamma} &=& \epsilon_{\gamma\delta\sigma}
\int e^3 \tau^2 \frac{\partial f^0}{\partial \varepsilon}
v_\alpha v_\sigma M^{-1}_{\beta\delta} \; d^3 k\; , \\
\sigma^{(2)}_{\alpha\beta\gamma\delta} &=&
\epsilon_{\rho\gamma\nu} \epsilon_{\mu\delta\sigma} \int e^4
\tau^3 \frac{\partial f^0}{\partial \varepsilon} v_\alpha v_\rho
\left\{ M^{-1}_{\sigma\beta} M^{-1}_{\mu\nu} + \frac{v_\mu}{\hbar}
\frac{\partial}{\partial k_\beta}\left( M^{-1}_{\sigma \nu}
 \right) \right\} d^3 k \; ,
\end{eqnarray}
$\epsilon_{\alpha\beta\delta}$ is the antisymmetric tensor, and
the inverse effective mass is
\begin{equation}
M^{-1}_{\alpha\beta} = \frac{1}{\hbar^2} \frac{\partial^2
\varepsilon}{\partial k_\alpha \partial k_\beta} \; .
\end{equation}
As can be seen, each conductivity is an integral over the Fermi
surface (since the energy derivative of the Fermi function, $f^0$,
is approximately a delta function). Furthermore, higher order
dependence on the external magnetic field is related to a higher
order derivative of the dispersion ($\varepsilon_k$) around the
Fermi surface. This suggests that magnetotransport is, with each
increasing order of field, a sensitive probe of the details of the
Fermi surface.

These expressions reduce to the ``pre-quantum" forms of Drude for
the case of a parabolic dispersion ($\sigma_{xx} = ne^2\tau/m$,
$\sigma_{xy} = \omega_c \tau \sigma_{xx}$) so that the Hall
coefficient $R_H=\rho_{xy}/B = 1/ne$. However, they now account
for the otherwise mysterious observation that the Hall effect can
change sign. This was first recognized by Peierls who showed how
the almost full band of electrons produces the same Hall effect as
an almost empty band of positively charged carriers because of the
sign of the curvature of the Fermi surface. In the case of a
multiband system then conductivities arising from each band must
be added together before the resultant conducivity tensor is
inverted. This of course spoils the simple relationship between
carrier density and Hall constant even for a parabolic band. A
peculiar case of note emerges for compensated metals (where the
Hall conductivity of the individual bands in, say, a two band
system are equal and opposite and so cancel). Under these
circumstances the leading contribution to the measured Hall effect
occurs at higher order so $\rho_{xy} \sim B^2$~\cite{ree64,ber69}.
The latter has been demonstrated to hold by high-field experiments
on the heavy fermion metal UPt$_3$ \cite{Kambe1999}. This aspect
will be important for the discussion of the Hall effect on
Ce$M$In$_5$ systems, sections \ref{sec:Co115} and \ref{sec:Ir115}.

Some rather beautiful results have been proved by Ong within
Boltzmann transport theory in the 2D limit which give geometric
interpretation to magnetotransport quantities~\cite{ong91}. The
Hall conductivity can be related to the flux through the area
swept out by the mean free path as it traced around the Fermi
surface. The magnetoresistance is related to the variance of the
local Hall angle around the Fermi surface~\cite{har95}.

The Boltzmann equation can also be solved in the limit of high
magnetic field as an expansion in $1/B$ (though of course Landau
level quantization means that this limit is never strictly
obtained). Under these conditions the Hall effect becomes
insensitive to Fermi surface shape and can be directly related to
the volume of the Fermi surface. In this limit even in a multiband
system one can use the Hall effect to determine the carrier
concentration independent of band structure details $R_H = 1/e
(\sum_i n_i)$. It is arguable as to whether this limit is ever
physically relevant.

Finally, while the results quoted above relate to solutions
obtained via expansion in magnetic field (either low field or high
field), it is not necessary to solve the Boltzmann equation by
expansion. The conductivity can be obtained to all orders in the
field at the expense of an additional surface integral using the
so-called Chambers formula~\cite{chambers1969}. Such an approach
is necessary when dealing with quantities which vary rapidly
around the Fermi surface as the Jones-Zener expansion can
breakdown. This will be important at a density wave transition for
example (see later).

These results from Boltzmann theory are often very useful as a
starting point to understand magnetotransport phenomena in the
heavy fermion metals. In fact, sometimes they are all one can do.
However, to make further progress we proceed to consider the
foundations of this approach within Fermi liquid theory and thence
the physics beyond.

\subsection{Hall effect within Fermi liquid theory}
\label{sec:Hall-flt} A more formal justification for the Boltzmann
approach (and thereby revealing its limitations) is obtained
through the Kubo formula and its diagrammatic representation.
Direct methods of calculating typically involve treating the
magnetic field perturbatively. Given we are looking for the
current response to an electric and magnetic field in the lowest
order, the Hall conductance involves a three current correlation.
This issue of maintaining gauge invariance in these calculations
requires that Ward identities be preserved~\cite{fukuyama1969}. In
essence this means that the treatment of vertex corrections must
be consistent with the self-energy ({\it i.e.} that every process
included in the self-energy has a corresponding vertex correction
obtained by including a current vertex at any point within the
self-energy diagram).

The simplest question concerns the recovery of the results of
Boltzmann theory in the relaxation time approximation. If one
assumes non-interacting electrons and dilute impurities then these
relaxation time approximation results  above can be obtained if
one assumes $s$-wave potential scattering~\cite{john1992}. The
isotropic scattering ensures that the calculation is insensitive
to vertex corrections. It also suggests that in the case of
strongly momentum dependent scattering which might be expected
near a QCP then reliance on the relaxation time approximation may
be suspect.

Treating the problem of interacting fermions is considerably more
difficult. However, a remarkable simplification ensues if one is
limited to systems where the low energy physics can be described
within Fermi liquid theory~\cite{lan57}. In Landau Fermi liquid
theory the low lying excitations of the interacting system are
adiabatically connected to a non-interacting Fermi gas (a
pedagogical account being found in
Refs.~\cite{anderson1984,schofield1999}). Landau considered a
neutral Fermi liquid in deriving the transport equation, but the
extension of this equation to the charged case is straightforward
and dictated by minimal coupling. The derivation assumes that the
Fermi liquid is characterized by a coarse-grained distribution of
quasiparticles $n(\bm p, \bm r, t)$ ({\it i.e.} the $\bm r$
dependence is slowly varying on the length scale of the Fermi
wavelength). Under these conditions the local quasiparticle energy
can be considered to be the effective classical Hamiltonian of the
quasiparticle
\begin{equation}
\epsilon(\bm p, \bm r) = \epsilon^0(\bm p) + \sum_{\bm p'} f_{\bm
p \bm p'} \delta n(\bm p',\bm r) \; ,
\end{equation}
where the interaction is assumed to be local on the scale of
spatial coarse graining. Minimal coupling introduces the electric
and magnetic fields through the vector potential $\bm p
\rightarrow \bm p - e\bm A$. The time evolution of the
distribution is then determined by the total time derivative
\begin{equation}
\frac{d n}{d t} = \frac{\partial n}{\partial t} + {\bm
\nabla}_{\bm r}n \cdot \frac{\partial \bm r}{\partial t} + {\bm
\nabla}_{\bm p}n \cdot \frac{\partial \bm p}{\partial t} = I \; ,
\end{equation}
where $I$ is the collision integral. Using Hamilton's equation of
motion and expressing the result in terms of the deviation from
local equilibrium, $\delta n(\bm p, \bm r) = n(\bm p, \bm r) -
n_0(\epsilon_{\bm p}(\bm r))$, we find for steady state,
translationally invariant solutions
\begin{equation}
e \bm E \cdot \bm \nabla_{\bm p} n_0 + e \bm v \times \bm B \cdot
\bm \nabla_{\bm p} \delta n = I \; .
\end{equation}
This is exactly the result one would write down for simple
Boltzmann transport yet it includes the electron-electron
interaction within the Fermi liquid formalism.

Establishing that this result is rigorous within a diagrammatic
approach was achieved by Betbeder-Matibet and
Nozi\`eres~\cite{betbeder1966} and by Khodas and
Finkel'stein~\cite {khodas2003}. Thus the quasiparticle
renormalization, $Z$, and other interaction corrections all cancel
in the Hall coefficient at least to zeroth order in $1/\tau E_F$.
Thus we conclude that if a Fermi liquid state emerges at low
temperatures in the heavy fermion system, then transport can be
described within a Boltzmann-like formalism.

\subsection{Quantum critical heavy fermion metals}
\label{sec:beyond-flt} All the results of the previous discussion
require the existence of Fermi liquid like quasiparticles.
Understanding the Hall effect without quasiparticles is more
challenging~\cite{taras1997}. Here we focus on what can be learnt
from the Hall coefficient and higher order magnetotransport
coefficients in the heavy fermion systems and will argue that the
most useful diagnosis is obtained in the $T \rightarrow 0$ limit
where a Fermi liquid like picture should emerge. The heavy fermion
metals hold particular interest here since their inherently small
energy scales mean that relatively weak perturbations like
pressure or magnetic field are able to tune magnetic transition
temperatures to absolute zero.

At zero temperature the QCP can be viewed a transition between
Fermi liquids but the nature of that transition is a matter of
very active interest. Two scenarios are emerging. On the one-hand,
the nature of the QCP could be dictated by the fluctuations
associated with a magnetic transition driven or tuned to zero
temperature. This QCP might refer to as a conventional
one---though as we will show the theoretical description remains
challenging in a number of physically relevant cases. In this case
the Fermi liquids on either side of the quantum critical point
would be both adiabatically connected to the {\em same}
non-interacting Fermi gas (albeit evolving continuously as the
density wave gap opens at the transition point).

An alternative view is that the essence of the QCP in heavy
fermion metals is associated with the zero temperature breakdown
of the Kondo effect. In that scenario the magnetic instability
plays a secondary role as a mechanism to quench the spin entropy
of the local moments. In this scenario while the QCP would divide
the zero temperature axis into two Fermi liquids, these two states
would be adiabatically connected to {\em entirely different}
non-interacting systems. A central tenant of this review is that
the low temperature Hall data can play a key diagnostic role in
discriminating between these two scenarios.

\subsubsection{``Conventional'' Hertz-Moriya-Millis quantum criticality}
\label{sec:HM-qc} We begin by considering the Hall effect at a
``conventional'' SDW QCP. The effective action for quantum
criticality in a metal has been argued to be the so called
Hertz-Moriya-Millis action~\cite{her76,mor85,mil93}
\begin{equation}
S=\frac{1}{\beta} \! \int \! d^D q \sum_n \phi_{-q, -\omega_n} E_0
\! \left[ r_0 + \xi_0^2 q^2 + \frac{|\omega_n|}{\Gamma_q} \right]
\phi_{q, \omega_n}  + u \!\int \!\! d^D {\pmb r} \int_0^\beta \!\!
d \tau  |\phi({\pmb r},\tau)|^4 + \cdots \; .
\end{equation}
In its simplest terms this action can be viewed as the quantum
extension of the Ginsburg-Landau free energy functional in the
vicinity of a continuous phase transition where the form of the
expansion is deduced by symmetry and analyticity. The frequency
sum of bosonic Matsubara frequencies is a consequence of the
quantum nature of the transition ({\it i.e.} that the order
parameter is not an eigenstate of the Hamiltonian and so must be
summed over in imaginary time). This action can be obtained from a
random phase approximation (RPA)-like saddle-point treatment of
the interacting fermion problem~\cite{her76}. The non-analytic
term in $\omega_n$ arises from the damping induced by the electron
fluid on the order parameter. The form of the damping rate depends
on the conservation laws applied to the order parameter:
\begin{equation}
\Gamma_q \sim q^{z-2} = \left\{
\begin{array}{ll}
\Gamma_0 & \quad (z=2) \quad \text{antiferromagnetic}, \\
v_F q & \quad (z=3) \quad \text{clean ferromagnet} ,\\
Dq^2 & \quad (z=4) \quad \text{dirty ferromagnet} \: .\\
\end{array}
\right.
\end{equation}
The experimental consequences of this action have been well
studied and so too have been the conditions for its validity.
These issues have been reviewed~\cite{hvl07} but in brief the low
energy particle-hole excitations of the metallic state cannot
generally be safely integrated out to yield the action
above~\cite{belitz2005}. The problem is particularly acute for
clean ferromagnetic instabilities where the soft particle-hole
excitations generate non-local terms in the effective action for
the order parameter and would seem to generically drive the
transition first order~\cite{belitz1999}. The case of two
dimensional antiferromagnets is also problematic from a
theoretical standpoint~\cite{abanov2003}.

Because of the theoretical questions concerning the precise nature
of the magnetic QCP in itinerant systems, for the purpose of this
review we consider only the antiferromagnetic case and that at two
levels. The first is the finite temperature effect of
antiferromagnetic spin fluctuations on transport and the Hall
effect. The second is the zero temperature limit.

\subsubsection{Spin-fluctuation theory}
\label{sec:spinfluct} At a phenomenological level one can
introduce a form of the spin-fluctuations first introduced in the
context of the high temperature cuprate
materials~\cite{millis1990a}:
\begin{equation}
\chi^s_{\bm q}(\omega) = \sum_{\bm Q} \frac{\chi_{\bm Q}}{1 +
\xi^2(\bm q - \bm Q)^2 - i \omega/ \omega_{\rm sf}} \; .
\end{equation}
Here $\omega_{\rm sf}$ and $\chi_{\bm Q}$ scale with the square of
the magnetic correlation length, $\xi^2$, and $\bm Q$ are the
Fourier modes of the incipient magnetic order. Transport is then
dictated by quasiparticle scattering from spin fluctuations. Two
effects control the resulting transport which can be considered
within Boltzmann transport.

Firstly, the strong momentum dependence of the fluctuations
strongly scatters particles on the Fermi surface connected by the
ordering wave vectors ${\bm Q}$, while leaving other parts of the
Fermi surface relatively unscathed. The Fermi surface divides into
"hot-spots" where scattering is strong and ``cold regions'' where
it is not~\cite{stojkovic1996}. One issue within this scenario is
the observation that the cold parts should short-circuit the
transport and so render the metal relatively insensitive to the
presence of the magnetic fluctuations~\cite{hlubina1995}. Within
such a picture a relaxation time approximation has been
invoked~\cite{ioffe1998}. This produced a number of criticisms of
the spin-fluctuation model as it applied to the cuprates. The
origin of these criticisms is the observation made previously that
magnetotransport is very sensitive to anisotropy around the Fermi
surface and so a hot-spot--cold-region model generates a large
magnetoresistance \cite{har95,sandeman2001} (much larger than that
seen in the cuprates).

The second effect is that the fluctuation performing the
scattering is itself a particle-hole excitation of the fluid and
this modifies the current. This has been emphasized by Kontani who
recognized this effect within a tour-de-force diagrammatic
approach to high order transport coefficients in a magnetic field
(for a recent review see Ref.~\cite{kon08}). In diagramatics this
appears as a current vertex correction but its effect can be
modelled within Boltzmann transport provided one does not make the
relaxation time approximation but considers the collision integral
in detail.

Taking both of these effects together, Kontani's analysis of the
spin fluctuation model claims that one can obtain a strongly
temperature dependent Hall coefficient in the presence of
antiferromagnetic fluctuations: $R_{\rm H} \sim \xi^2 \sim 1/T$.
Moreover, higher order terms in magnetotransport do not get
anomalously large as they would in the relaxation time
approximation but adopt a temperature dependence associated with
the Hall effect. One consequence is that the magnetoresistance
$\Delta \rho/\rho_0$, which in the relaxation time approximation
would be expected to scale like $\rho_0^{-2}$ (Kohler's rule,
Ref.\ \cite{kohler38}), does not. Rather it would behave like
$\xi^4/\rho_0^2 \sim T^{-4}$ leading to a modified Kohler's rule
$\Delta \rho/\rho_0 \sim (R_{\rm H}/ \rho_0)^2$.

The observations above are, at face value, very reminiscent of the
behaviour of the normal state of the high temperature cuprates
superconductors. There, Anderson \cite{and91} and Ong \cite{chi91}
noticed that $\cot \theta_H$---which should, within the relaxation
time approximation, be proportional to the scattering rate as
measured in the resistivity ({\it i.e.} $\sim T$)---here behaved
with a distinct $T^2$ temperature dependence ({\it cf.} Fig.\
\ref{chien} and related discussion in section \ref{sec:cup}; see
also the discussion on results obtained on Ce$M$In$_5$ materials,
section \ref{sec:comp115}). Crucial to their observation was the
disorder dependence on the introduction of Zn impurities into the
cuprates. They identified $\rho$ and $\cot \theta_H$ as measuring
entirely separate scattering rates because while both had very
different temperature dependencies, they both also showed
Matthiesen's rule type behavior as a function of impurity
concentration: Impurity scattering behaved additively to both the
scattering rates. In the context of the cuprates this is a strong
constraint on the underlying mechanism for the temperature
dependence of the Hall coefficient.

Adding disorder to a quantum critical antiferromagnetic metal was
considered by Rosch~\cite{rosch1999a}. He showed that there is a
strong interplay between even very small amounts of disorder and
antiferromagnetic fluctuations. The disorder smears out the
hot-spots on the Fermi surface thereby considerably weakening the
short-circuiting argument of Hlubina and Rice. Moreover, the
expected temperature dependencies of the resistivity are also
modified from those of the spin-fluctuation model. Rosch also
showed that very small amounts of disorder are sufficient to
render the Hall coefficient relatively weakly dependent on
temperature~\cite{rosch1999b}. This is somewhat at odds with the
Kontani calculation which claims that the $\cot \theta_H$ behavior
remains distinct from the resistivity albeit with a weaker than
$T^2$ power law.

Given that there remains some uncertainty about the precise
temperature/disorder dependence of the Hall coefficient when it is
dominated by quantum critical fluctuations, we argue that the
$T=0$ limit provides a more reliable domain for the Hall effect's
interpretation. At temperatures low enough that the inelastic
scattering is a relatively small fraction of the overall
resistivity, one is dominated by elastic impurity scattering.
Under these circumstances a relaxation time approximation is valid
and one can use the magneto-transport data to characterize changes
in the Fermi surface.

Initial studies suggested that at a continuous density wave
transition all transport quantities should vary smoothly
\cite{col01,Bazaliy2004}. However, these considerations were based
on a weak field expansion. Crucially near a QCP there is an order
of limits question as to whether the field scale goes to zero
before the density wave gap \cite{Fenton2005,mil05}. A weak field
expansion is valid only if
\begin{equation}
B \tau < \frac{\Delta}{e v_F^2} \; ,
\end{equation}
where $\tau$ is the relaxation time and $\Delta$ the density wave
gap. Near enough to the QCP this expansion will breakdown for any
finite field experiment in the ordered phase. The consequences are
the Hall effect and resistivity develop non-analytic terms in
field $\omega_c \tau$. These arise whenever the Fermi surface
crosses the density wave Brillouin zone so Bragg reflection causes
the Fermi surface to develop a sharp corner. Under these
circumstances the Hall conductivity develops an additional term
that goes like $|\omega_c \tau|^2$ and the longitudinal
conductivity a term that behaves like $|\omega_c \tau|$. These
terms are absent in the paramagnetic phase so this implies that
there is a discontinuity in the Hall effect and in the resistivity
of order $|\omega_c \tau|^2$ and $|\omega_c \tau|$ respectively at
the transition point. These discontinuities are rounded by
magnetic breakdown effects and are suppressed by disorder.
Nevertheless, the observation of a magnetoresistance linear in
magnetic field is often indicative of a sharp feature (point of
small radius of curvature) on the Fermi surface.

In summary, the theory of the Hall effect in a conventional
density wave type QCP is least ambiguous in the zero temperature
limit. Here we expect to see rather smooth evolution of the weak
field Hall constant through the transition (unless the system is
unusually clean when it may be possible to see discontinuities due
to the reconstruction of the Fermi surface). At finite
temperature, Kontani predicts that antiferromagnetic fluctuations
can give rise to a temperature-dependent Hall coefficient though
Rosch argues that this temperature dependence is lost with
relatively small quantities of disorder.

\subsection{Hall effect across Kondo breakdown quantum critical
point} \label{sec:Kondobreak}
\subsubsection{Quantum criticality in heavy fermion metals and jump
of Hall coefficient} \label{sec:Halljump} Theoretical studies of
quantum phase transitions in heavy fermion metals depart from the
Kondo lattice Hamiltonian:
\begin{eqnarray}
{\cal H} = \frac{1}{2} \sum_{ ij} I_{ij} ~{\bf S}_{i} \cdot {\bf
S}_{j} ~+~\sum_{\bf k \sigma} \epsilon_{\bf k} c_{{\bf
k}\sigma}^{\dagger} c_{{\bf k}\sigma} +~ \sum_i J_K ~{\bf S}_{i}
\cdot {\bf s}_{c,i} \;. \label{kondo-lattice}
\end{eqnarray}
Here, a lattice of spin-$1/2$ local moments interact with each
other with an exchange interaction $I_{ij}$. To specify the
typical strength of the exchange interaction, we use $I$ to label
the nearest-neighbor interaction, and we will focus on the case
that it is antiferromagnetic. The model also contains a
conduction-electron band, $c_{{\bf k} \sigma}$, with a band
dispersion $\epsilon_{\bf k}$ and bandwidth $W$. At each site $i$,
the spin of the conduction electrons, ${\bf s}_{i,c} = (1/2)
c_{i}^{\dagger} {\pmb \tau} c_i$, where ${\pmb \tau}$ are the
Pauli matrices, is coupled to a spin-$1/2$ local moment, $\bf
{S}_i$, via an antiferromagnetic Kondo exchange interaction $J_K$.

When the Kondo coupling dominates over the RKKY interaction, the
ground state is a Kondo singlet ({\it cf.} section \ref{sec:HF}).
The Kondo screening effect leads to Kondo resonances, which are
charge-$e$ and spin-$1/2$ excitations. There is one such Kondo
resonance per site, and these excitations induce a ``large'' Fermi
surface \cite{Hewson,Allen,Auerbach86,Millis87,Oshikawa00}.
Consider that the conduction electron band is filled with $x$
electrons per site; for concreteness, we take $0<x<1$. The
conduction electron band and the Kondo resonances will be
hybridized, resulting in a count of $1+x$ electron per site. The
Fermi surface would therefore have to expand to a size that
encloses all these $1+x$ electrons. This defines the large Fermi
surface.

Consider the conduction electron Green's function:
\begin{eqnarray}
G_c({\bf k},\omega) \equiv F.T.[-<T_{\tau} c_{{\bf k},\sigma}(\tau)
c_{{\bf k},\sigma}^{\dagger}(0)>] \;,
\label{gc-definition}
\end{eqnarray}
where the Fourier-transform ($F.T.$) is taken with respect to
$\tau$. This Green's function is related to a self-energy,
$\Sigma({\bf k},\omega)$, via the standard Dyson equation:
\begin{eqnarray}
G_c({\bf k},\omega) = \frac{1}{\omega-\epsilon_{\bf k} -
\Sigma({\bf k},\omega)} \;.
\label{gc-Dyson-equation}
\end{eqnarray}
In the heavy Fermi liquid state, $\Sigma({\bf k},\omega)$ is
non-analytic and contains a pole in the energy space
\cite{Hewson,Auerbach86,Millis87}:
\begin{eqnarray}
\Sigma({\bf k},\omega) = \frac{(b^*)^2}{\omega-\epsilon_f^*} \;.
\label{sigma-pole}
\end{eqnarray}
Inserting eq.~(\ref{sigma-pole}) into
eq.~(\ref{gc-Dyson-equation}), we end up with two poles in the
Green's function:
\begin{eqnarray}
G_c({\bf k},\omega) = \frac{u_{\bf k}^2}{\omega-E_{1,{\bf k}}} +
\frac{v_{\bf k}^2}{\omega-E_{2,{\bf k}}} \;. \label{gc-two-poles}
\end{eqnarray}
Here,
\begin{eqnarray}
E_{1,{\bf k}} &=& (1/2)\left [\epsilon_{\bf k}+\epsilon_f^*
-\sqrt{(\epsilon_{\bf k}-\epsilon_f^*)^2+4(b^*)^2} \right ] \;,
\nonumber \\
E_{2,{\bf k}} &=& (1/2) \left [ \epsilon_{\bf k}+\epsilon_f^*
+\sqrt{(\epsilon_{\bf k}-\epsilon_f^*)^2+4(b^*)^2} \; \right ]
\label{hf-bands}
\end{eqnarray}
describe the dispersion of the two heavy-fermion bands. These
bands must accommodate $1+x$ electrons, so the new Fermi energy
has to lie in a relatively flat portion of the dispersion, leading
to a small Fermi velocity and a large quasiparticle mass $m^*$.

The self-energy contains only two parameters, the pole strength
({\it i.e.}, the residue), $(b^*)^2$, and the pole location,
$\epsilon_f^*$. eq.~(\ref{sigma-pole}) describe only the coherent
part with the well-defined pole; an incoherent part will specify
damping of the quasiparticle excitations.

In the opposite limit, when the RKKY interaction dominates over
the Kondo coupling, the system is antiferromagnetically ordered.
For any spatial dimension $d > 1$, it follows from a
renormalization group study \cite{Yamamoto07} that the Kondo
singlet breaks down. Correspondingly, there is no Kondo resonance,
and the Fermi surface is ``small", {\em i.e.} does not incorporate
the $f$-electrons. The conduction-electron self-energy in a
large-$N$ limit takes the form
\begin{eqnarray}
\Sigma({\bf k},\omega) = a \omega - i (b |\omega|^d + c \omega^2)
\; {\rm sgn} \, \omega \label{sigma-af}
\end{eqnarray}
From a comparison with eq.~(\ref{sigma-pole}), we see that the
strength of the pole in the conduction-electron self-energy has
vanished. Inserting eq.~(\ref{sigma-af}) into
eq.~(\ref{gc-Dyson-equation}) gives rise to a Fermi surface that
is entirely specified by the conduction electron dispersion.

Quantum phase transition in the Kondo lattice arises by tuning the
parameter  $\delta \equiv T_K^0 / I$, where $T_K^0 \approx
\rho_0^{-1} \exp (-1/\rho_0 J_K)$ with $\rho_0$ being the density
of states of the conduction electrons at the Fermi energy
\cite{don77,var89a} ({\it cf.} also Fig.\ \ref{doniach}). In
recent years, several theoretical approaches have  been undertaken
to study the quantum phase transition. The extended dynamical mean
field theory (EDMFT) \cite{SmithSi00,Chitra00} focuses on the
destruction of the Kondo effect by the antiferromagnetic
fluctuations, leading to two classes of solutions. In one, the
Kondo breakdown occurs at the onset of the antiferromagnetic
order; this is referred to as local quantum criticality. In
another class of solution, the Kondo breakdown takes place inside
the antiferromagnetic order. The QCP separating the
antiferromagnetic and paramagnetic states has the SDW form, but a
Kondo breakdown gives rise to another quantum phase transition
inside the antiferromagnetic part of the phase diagram.

The evolution of the Hall coefficient across the local QCP can be
most clearly seen in the zero-temperature limit. At $\delta >
\delta_c$, the ground state is a Fermi liquid, and the
conduction-electron self-energy has the form of
eq.~(\ref{sigma-pole}). The associated quasiparticles, with the
dispersion given by eq.~(\ref{hf-bands}), are located near the
large Fermi surface. At $\delta < \delta_c$, the ground state is
also a Fermi liquid, but now the conduction-electron self-energy,
having the form of eq.~(\ref{sigma-af}), no longer has the pole
and the quasiparticles are located near the small Fermi surface.
As discussed earlier in this section, the quasiparticle residue
for each Fermi liquid will cancel in its Hall coefficient. Since
the underlying non-interacting system for the two Fermi liquids
corresponds respectively the fermions with $f$-interactions
itinerant or localized, the Hall coefficient has different values
in the two Fermi liquids \cite{Si03}. In other words, the Hall
coefficient at zero temperature jumps as the control parameter
$\delta$ is tuned through $\delta_c$, the QCP.

Another approach to the Kondo breakdown is based on a large-$N$
formulation in conjunction with slave-particle representations of
the spin operator. The fermionic representation of the spin
provides a means to access a spin liquid when the Kondo amplitude
$(b^*)^2$ is suppressed \cite{Senthil04,Paul07}, where fermionic
spinons and the bosonic $b$ fields are coupled to a gauge field.
The Kondo destruction of $b^*=0$ corresponds to a metallic state
(as opposed to a Bose-Einstein condensate) of the bosonic
component. Calculation of the Hall coefficient and longitudinal
resistivity in this approach shows a jump of both quantities at
the Kondo-breakdown transition at zero temperature
\cite{Coleman05}.

\subsubsection{Crossover and scaling of the Hall coefficient at
non-zero temperatures} \label{sec:crosstheo} The $T$-$\delta$
phase diagram for the Kondo-destruction local QCP was already
illustrated in Fig.~\ref{sdwlocal}(b). The energy scale
$E_{\text{\small loc}}^*$ specifies a corresponding $T^*$ line,
which separates the phase diagram at low temperatures into two
parts. To the right of the $T^*$ line, the system flows (in the
renormalization-group sense, as energy is lowered) towards a
ground state with complete Kondo screening and therefore large
Fermi surface; the $T_{\text{\small LFL}}$ scale describes the
renormalized Fermi energy for this Fermi liquid state. To the left
of the $T^*$ line, however, the Kondo screening process is
incomplete; in the ground state, the static Kondo screening is
absent and the Fermi surface is small. Here, the N\'{e}el
temperature, $T_N$, marks the onset of the antiferromagnetic
order. In the zero-temperature limit, the end of the $T^*(\delta)$
line, which is at $\delta_c$, marks a genuine $f$-electron
delocalization-localization phase transition.

Compared to the zero temperature case, the behavior of the Hall
coefficient at non-zero temperatures across a Kondo-destruction
local QCP is much more difficult to analyze and can only be
considered qualitatively. The single-electron Green's function,
$G({\bf k}, \omega)$, on either side of the zero-temperature
transition can be written as
\begin{eqnarray}
G({\bf k}, \omega) = G_{\rm coh} ({\bf k}, \omega)
+G_{\rm inc} ({\bf k}, \omega) \: .
\label{G-decomposition}
\end{eqnarray}
This decomposition is an immediate consequence of the fact that
the phases separated by the QCP are Fermi liquids. The coherent
part is given by
\begin{eqnarray}
G_{\rm coh} ({\bf k}, \omega)
= \frac{z_{\bf k}}{\omega - E({\bf k}) + i \Gamma_{\bf k} (T)}
\label{G-coh}
\end{eqnarray}
describing a quasiparticle, and $G_{\rm inc} ({\bf k}, \omega)$ is
a background contribution. The quasiparticle residue, $ z_{\bf
k}$, is non-zero in either phase, but $ z_{\bf k}$ vanishes as the
QCP is approached: $ z_{\bf k} \rightarrow 0$ as $\delta
\rightarrow \delta_c$. At $T=0$, the quasiparticle damping
$\Gamma_{\bf k}$ vanishes at the small Fermi-momenta ${\bf k}_F^S$
for $\delta < \delta_c$, and at the large Fermi-momenta ${\bf
k}_F^L$ for $\delta > \delta_c$; at these respective
Fermi-momenta, the quasiparticles become infinitely-sharp
excitations at zero temperature. The coherent part of $ G({\bf k},
\omega)$ is therefore the diagnostic feature on either side of the
transition, and it jumps at the QCP in accordance with the sudden
change of the Fermi surface. As already described, this jump is
manifested in the Hall measurement (see, {\it e.g.},
section~\ref{sec:YRS} where results obtained on YbRh$_2$Si$_2$ are
presented), because the Hall coefficient is independent of the
quasiparticle residue.

At non-zero temperatures, the quasiparticle relaxation rate at
either ${\bf k}_F^S$  for $\delta < \delta_c$, or ${\bf k}_F^L$ no
longer vanishes. In fact, inside the Fermi-liquid phase (with
either large or small Fermi surface), the temperature dependence
of $\Gamma_{{\bf k}_F}$ has to be quadratic in $T$. However, the
Fermi surface remains well defined in these regimes. The change
from one  Fermi surface to the other is restricted to the
intermediate quantum critical regime. Because of the absence of a
phase transition at any non-zero temperature, the sharp
reconstruction of the Fermi surface at $T=0$ is turned into a
Fermi-surface crossover across the $T^*(\delta)$ line. This
implies that the relaxation rate determines the parameter range
corresponding the Fermi surface change. From the relation of the
Hall coefficient to the Fermi surface we can  associate the width
of the Hall crossover with this broadening and consequently with
the relaxation rate $\Gamma$ of the single-electron Green's
function.

Consider a general scaling form for the single-electron Green's
function at the Fermi momentum in the quantum critical region:
\begin{equation}
G({\bf k}_F,\omega) =\frac{1}{T^\alpha} g \left(
{\bf k}_F,\frac{\omega}{T^x} \right) \, .
\end{equation}
The $\omega/T$ scaling at an interacting fixed point implies
$x=1$. The associated relaxation rate, defined in the quantum
relaxational regime (${\hbar \omega} \ll k_B T$) according to
\begin{eqnarray}
\Gamma({\bf k}_{\mathrm F},T) \equiv \left [-i \partial \ln G
({\bf k}_{\mathrm F},{\omega},T) / \partial {\hbar \omega } \right
]_{{\omega }=0}^{-1} \; , \label{Gamma-def}
\end{eqnarray}
is linear in temperature: $\Gamma({\bf k}_{\mathrm F},T) = c T$,
where $c$ is a universal constant. Correspondingly, the crossover
width is expected to be linear in temperature. This is indeed
experimentally observed for YbRh$_2$Si$_2$, see
Fig.~\ref{fig:FWHM_YRS} in section \ref{sec:YRS}.

\section{Experimental aspects of Hall effect measurements in metals}
\label{sec:exp}
\subsection{Measurement techniques}
\label{sec:techn} For the following considerations we rely on a
discussion of eq.~(\ref{Hallvolt}) for simplicity. Because of the
typically large density of free charge carriers in a metal the
Hall coefficient \RH\ is small. As an example, for aluminum \RH\
is about $-3.5 \cdot 10^{-11}$ m$^3$C$^{-1}$ at room temperature
\cite{hur72}. Consequently, even for favorable experimental
conditions only a very small Hall voltage is generated. For Al and
$I_x = 100$ mA, $B_z = 1$ T and $d =$ 1 mm, the Hall voltage to be
measured is as small as $V_y = 3.5$ nV. The requirement to measure
such small voltages calls for some preconditions concerning the
experimental measurement setup as well as sample preparation. We
note that these preconditions are specific for the investigation
of metals. Semiconductors can exhibit values of \RH\ in the order
of 100 m$^3$C$^{-1}$ rendering Hall effect measurements
effortless. Therefore, all device applications based on the Hall
effect use a semiconductor rather than a metallic sample.

\begin{figure}
  \begin{center}
    \includegraphics[width=6.4cm,clip]{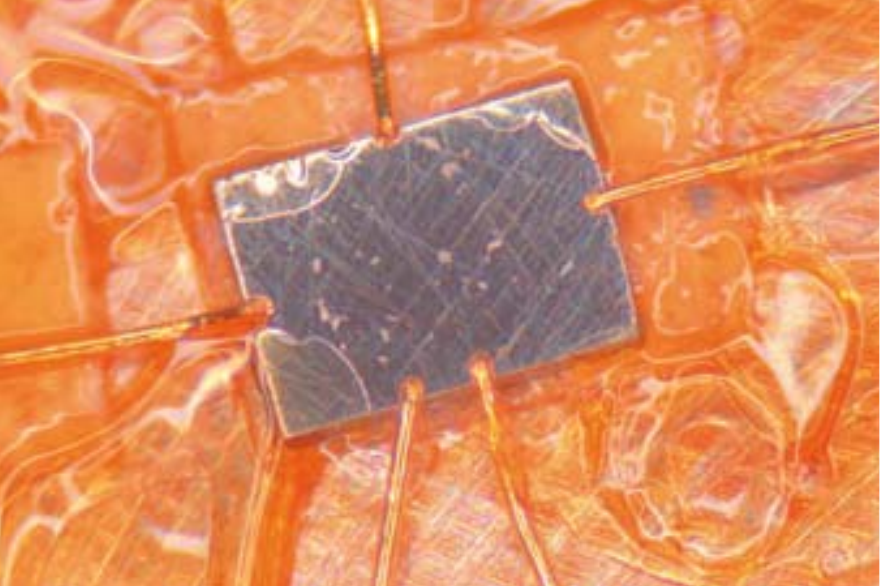}
    \caption{Photograph of a single crystal YbRh$_2$Si$_2$ contacted
    for magnetotransport measurements. The sample size is about $0.87
    \times 0.6$ mm$^2$, with a thickness of about 65 $\mu$m.}
    \label{fig:YRSsamp}%
  \end{center}
\end{figure}

Obviously, one way of bringing up the Hall voltage independently
of all measurement equipment is to thin the sample down. This is
typically achieved by polishing or, in the final step, lapping. A
single crystal YbRh$_2$Si$_2$ prepared in this way and
investigated in section \ref{sec:YRS} is shown in
Fig.~\ref{fig:YRSsamp}. The sample thickness is about 65 $\mu$m.
The contacts were spot-welded, with the left and right ones used
as current leads and the top and bottom ones for voltage
measurements. Having two contacts along a line parallel to the
current $I_x$ (lower sample side) allows for an additional
magnetoresistance measurement \emph{simultaneous} with the Hall
measurement. The sample was embedded in varnish for good thermal
contact and fixation. This sample was thinned down along the
crystallographic $c$-direction. According to Fig.~\ref{fig:Hall}
this implies $B_z \parallel c$. More generally spoken: Once the
sample geometry is optimized the direction with respect to the
sample in which the magnetic field is to be applied is fixed.
Consequently, a comparison of the Hall effect along different
crystallographic directions of a single crystal typically requires
the preparation of one sample for each envisioned direction of the
applied magnetic field.

In principle, thin samples could also be prepared by thin film
deposition techniques. Such films ({\it e.g.} of the manganites
discussed in section \ref{sec:cmr}) are well suited for Hall
measurements if they are patterned. However, thin film deposition
of heavy fermion metals has met limited success so far
\cite{jour,jou04,iza07,shi10}.

If the Hall measurements are to be conducted at cryogenic
temperatures, self-heating effects of the sample due to the
driving current $I_x$ have to be prevented. This holds
specifically true for the investigation of heavy fermion metals
which develop their most fascinating properties at lowest
temperatures such that experiments are often run in the mK
temperature range. For the experiments presented in chapters
\ref{sec:fluc} and \ref{sec:qpt}, ac currents (frequency of 113 or
119 Hz) between $10 - 100 \;\mu$A were applied. Comparing these
currents to the example above demonstrates that, even though $I_x$
should be adjusted as high as possible to improve the
signal-to-noise ratio, the sample temperature needs to be
monitored carefully.

A similar predicament holds for the magnitude of the magnetic
field. From eq.~(\ref{Hallvolt}), high fields appear preferable.
However, it is typically the regime of linear response, {\it i.e.}
the initial slope of $\rhoH$, one is interested in, see discussion
of eq.~(\ref{eq:Hall_coefficient_single_field}) below and
Fig.~\ref{fig:RhoH} for an example. This constrain often limits
the magnitude of $B_z$.

Any experimental setup to be used for Hall measurements on metals
has to be optimized for the measurement of small voltages. Besides
careful cabling we fund the usage of low-temperature transformers
\cite{llt} extremely helpful. These transformers allow for an
impedance change (we typically amplify the voltage by a factor of
100) very close to the sample and at low temperatures, {\it i.e.}
at low noise level. As a disadvantage, however, these transformers
require magnetic field-free conditions provided by their
superconducting housing and hence, the magnet used for Hall
measurements must have a compensated zone. If Hall effect and
magnetoresistance are to be measured simultaneously two
transformers need to be implemented making space in a dilution
refrigerator insert tight. Further amplification is provided by
low-noise voltage \cite{sr560} and, subsequently, lock-in
amplifiers. The setup retains phase control over the ac signals
which is of utmost importance for crosschecks on the measured Hall
voltage.

A careful look at the sample contacts in Fig.~\ref{fig:YRSsamp}
reveals an issue of far-reaching consequences: The voltage
contacts are usually not perfectly aligned perpendicularly with
respect to current $I_x$ (in contrast to patterned films). This
misalignment causes a magnetoresistive component as part of the
measured voltage $V_{\text{meas}}(B) = V_y(B) + V_x(B)$. To
separate the two components one can make use of the fact that the
Hall effect is antisymmetric, $V_y(B) = -V_y(-B)$, whereas the
magnetoresistive part is symmetric, $V_x(B) = V_x(-B)$.
Experimentally, this requires to measure $V_{\text{meas}}(B)$ at
positive \emph{and} negative magnetic fields, doubling the effort.
The Hall voltage is then obtained by $V_y(B) =
\frac{1}{2}[V_{\text{meas}}(B) - V_{\text{meas}}(-B)]$. As an
important way of crosschecking on our results we regularly compare
$V_x(B) = \frac{1}{2}[V_{\text{meas}}(B) + V_{\text{meas}}(-B)]$
to the directly measured ({\it e.g.\ }via the two lower contacts
in Fig.~\ref{fig:YRSsamp}) magnetoresistive voltage; these two
voltages are related by a simple factor which only depends on
sample geometry but not on magnetic field.

\begin{figure}%
  \begin{center}
    \includegraphics[width=5.0cm,clip]{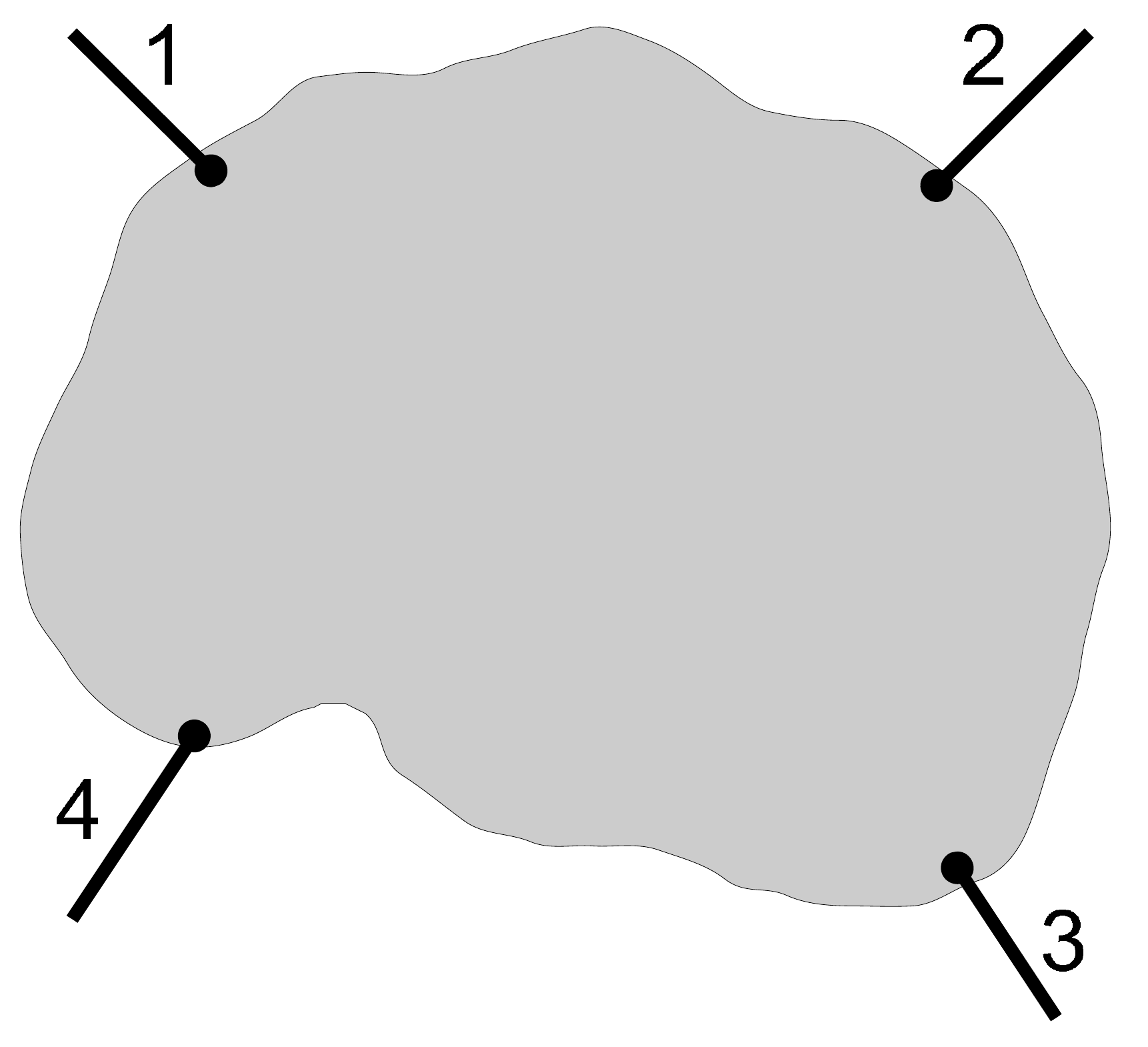}
    \caption{Sketch of a sample with contacts for van der Pauw
    measurements. An arbitrarily shaped sample of uniform
    thickness is connected via four contacts on its
    circumference. The resulting Hall voltage is obtained by
    swapping the contacts for current and voltage (see text).}
    \label{fig:vanderPauw}%
  \end{center}
\end{figure}

For Hall measurements on samples of irregular shape one can use
the so-called van der Pauw method \cite{vdp58} provided the sample
is of homogeneous thickness and of platelet-like form, {\it i.e.},
much thinner then wide. Four contacts are attached at arbitrary
positions on the circumference of the sample, as illustrated in
Fig.~\ref{fig:vanderPauw}. The van der Pauw method is handy for
tiny samples like those used in diamond anvil pressure cells. The
Hall voltage is extracted from two subsequent measurement
protocols: First, the current is passed through one set of
opposing contacts ({\it e.g.}, 1 and 3 in the figure) with the
voltage measured on the transverse contacts ($V_{24}$ in the
figure). In a second step, the former voltage contacts (2 and 4)
are used to inject the current and the voltage is measured on the
former current contacts, giving again the transverse voltage
($V_{13}$ in the figure). The Hall voltage is calculated as
$V_{\text H} = (V_{13}+V_{24})/2$. An often applied procedure to
separate magnetoresistivity contributions arising from the
arbitrary distribution of the contacts is via the determination of
the antisymmetric component of the Hall voltage under field
reversal.

The vectorial nature of the ingredients as described in section
\ref{sec:single} calls for a tensor notation in defining the
primary quantities of interest, specifically the resistivity
$\varrho$. Then, the Hall coefficient is given by $R_{\rm H} =
[\varrho_{yx}(B) - \varrho_{xy} (-B)]/2B$. Though it is the Hall
resistivity $\varrho_{xy}$ which is always measured in an
experiment, it is common practice to report experimental data in
the form of the Hall conductivity $\sigma_{xy}$ using a full
matrix inversion, {\it i.e.,} $\sigma_{xy} = \varrho_{xy} /
[\varrho_{xx}^2 + \varrho_{xy}^2]$. Considering the fact that
there is no preferred choice of notation in literature, this
review uses both these quantities interchangeably while describing
prior data. Note that this requires the knowledge of both,
$\varrho_{xx}$ and $\varrho_{xy}$ which is ideally measured
simultaneously. Along the same lines we will use the notations $H$
and $B$ interchangeably to denote the magnetic field.

\subsection{Advanced aspects of Hall effect measurements}
\label{sec:advexp}
\subsubsection{Single-field Hall experiments} \label{sec:single}
\begin{figure}%
\begin{center}
  \begin{tabular}{c @{\hspace*{2cm}} c}
  single field & crossed field \\[0.3cm]
    \includegraphics[width=4.0cm]{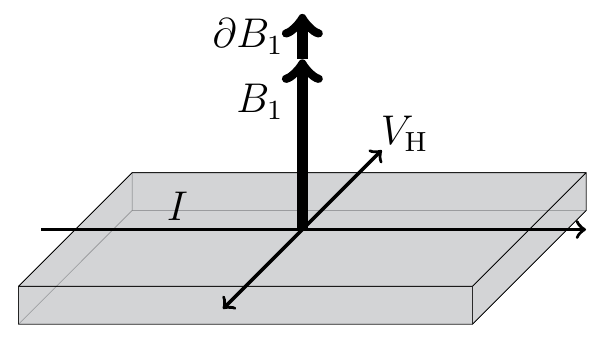} &
    \includegraphics[width=4.0cm]{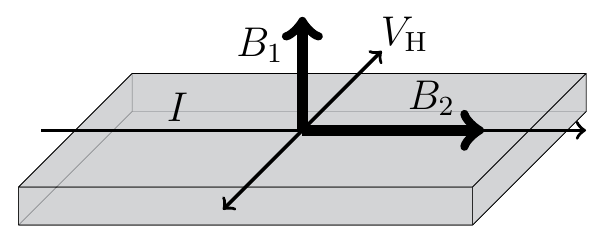} \\
    (a) & (b) \\
  \end{tabular}
\caption{Hall effect setups to study field-induced QCPs. (a)
Single-field setup with one magnetic field $B_1$. (b)
Crossed-field setup with the additional magnetic field $B_2$.}
\label{fig:Halleffectsetups}
\end{center}
\end{figure}

The traditional setup for Hall effect measurements was already
sketched in Fig.~\ref{fig:Hall}. In the following we will refer to
this as the single-field Hall setup with a single magnetic field
$B_1$ being applied, in an effort to distinguish it from the
crossed-field Hall setup discussed later, {\it cf.} section
\ref{sec:cross} and Fig.~\ref{fig:Halleffectsetups}.

For a more general derivation of the Hall voltage, eq.\
(\ref{Hallvolt}), we consider the geometry shown in Fig.\
\ref{fig:Hall} and specify the boundary conditions as
\begin{align}
\label{eq:cond_sf}
    {\pmb j}= \begin{pmatrix} j_x \\ 0 \\ 0 \end{pmatrix} &&
    {\pmb E}= \begin{pmatrix} E_x \\ E_y \\ 0 \end{pmatrix} &&
    {\pmb B}= \begin{pmatrix} 0 \\ 0 \\ B_1 \end{pmatrix}
\end{align}
with the current density ${\pmb j}$, the electrical field ${\pmb
E}$, and the magnetic field vector ${\pmb B}$. The conductivity
tensor defined via Ohm's law as
\begin{equation}
 {\pmb j} = \underline{\sigma} {\pmb E}
\label{eq:OhmsLaw}
\end{equation}
relates the current and the electrical field and incorporates
dependencies of its elements on the magnetic field ${\pmb B}$. In
the presence of one magnetic field in the $z$-direction only one
may reduce the conductivity tensor to a $2\times2$ form. By
further taking into account the Onsager relations
$\sigma_{xy}(B_1)= -\sigma_{yx}(B_1)$ \cite{Groot1954,Mazur1954}
and assuming an isotropic material with
$\sigma_{xx}=\sigma_{yy}=\sigma$, one derives a tensor with two
independent components
\begin{equation}
    \underline{\sigma} = \begin{pmatrix} \sigma  & \sigma_{xy} \\
    - \sigma_{xy} & \sigma \\ \end{pmatrix} \, .
\label{eq:sigma_sf}
\end{equation}
We note that moderate anisotropies may be incorporated without
altering the main result.  In order to obtain an expression for
the Hall coefficient we calculate the resistivity tensor
\begin{equation}
  \underline{\rho}= \begin{pmatrix} \rho & -\rho_{xy} \\
  \rho_{xy} & \rho \\ \end{pmatrix} = \frac{1}{\sigma^2+
  \sigma_{xy}^2} \begin{pmatrix} \sigma & -\sigma_{xy} \\
  \sigma_{xy} & \sigma\\ \end{pmatrix} \, . \label{eq:rho}
\end{equation}
The Hall resistivity $\rho_{\rm H}$ is identical to the element
$\rho_{xy}$ as this element reflects the relation of the
transverse electrical field to the longitudinal current
corresponding to the definition of the Hall coefficient. We are
now in the position to generalize eq.~(\ref{Hallvolt}) and obtain
the linear-response Hall coefficient as the corresponding element
of the tensor on the right hand side of eq.~(\ref{eq:rho})
\begin{equation}
    \RH (B_1)= \lim\limits_{B_1\to 0} \frac{\rhoH}{B_1} =
    \lim\limits_{B_1\to0}
    \frac{1}{B_1}\frac{\sigma_{xy}}{\sigma^2+\sigma_{xy}^2}
    \approx \lim \limits_{B_1\to0} \frac{1}{B_1}
    \frac{\sigma_{xy}}{\sigma^2} \; .
\label{eq:Hall_coefficient_single_field}
\end{equation}
The last approximation originates from the fact that the Hall
conductivity in metals is typically several orders of magnitude
smaller as the normal conductivity, {\it i.e.},
$\sigma_{xy}\ll\sigma$.

In accordance with eq.~(\ref{eq:Hall_coefficient_single_field}),
the Hall coefficient is normally measured isothermally as the
initial linear slope of the Hall resistivity with respect to the
magnetic field. For simple metals the Hall coefficient is a
constant which implies that the Hall resistivity is proportional
to the magnetic field, $\rhoH \propto B_1$, {\it cf.} eq.\
(\ref{Hallvolt}). From this equation it is also obvious that the
Hall coefficient \RH\ reflects the effective charge carrier
concentration which, for simple metals, does not depend on
magnetic field.

However, for the investigation of more complex materials and/or
phenomena such as quantum phase transitions the focus is often on
non-linear effects. For the study of quantum criticality we are
interested in metals which possibly change their Fermi surface
topology and hence, also the effective charge carrier
concentration. If this change is induced by a magnetic field, this
will lead to a non-linearity in the Hall resistivity. The overall
magnetic field $B_1$ applied will result in a tuning effect ({\it
cf.} section \ref{sec:HF}) whereas the probing of the Fermi
surface at this particular field strength shows up as the response
to infinitesimal fields $\partial B_1$ (see
Fig.~\ref{fig:Halleffectsetups}(a)). It is therefore
straightforward to define the differential Hall coefficient as
\begin{equation}
\TRH (B_1) = \frac{\partial \rho_{\text H}}{\partial B_1} \Big|_{B_1}
\label{eq:differential_Hall_coefficient}
\end{equation}
which allows to monitor the evolution of the Hall response across
a field-induced QCP. Hence, the disentanglement of the two effects
of the magnetic field might be done in the subsequent analysis by
numerical differentiation as we shall see below.

Based on linear response theory one may analyze the differential
Hall coefficient in more detail. The orbital deflection of the
charge carriers in an external magnetic field introduced by the
Lorentz force leads to a term which represents a generalized
definition of the Hall coefficient:
\begin{equation}
    \RH(B_1) = \left[ \frac{\partial \rhoH(B_1)}{\partial B_1}
    \right]_{\text{orb}}
\label{eq:orbital_Hall_contribution}
\end{equation}
In a simple metal the Lorentz force is just proportional to the
external magnetic field giving rise to a constant \RH\ as
expected. This orbital contribution corresponds to the probe in
our experiment, {\it i.e.}, to the Hall response. When the Fermi
surface topology changes, this is reflected in a change of the
orbital contribution.

For the investigation of materials in which the ground state is
field tuned, a second response to an external magnetic field
arises, which is termed Zeeman term.  Consequently, the
differential Hall coefficient reflects both the orbital and the
Zeeman contribution:
\begin{alignat}{2}
   \TRH &= \left[ \frac{\partial \rhoH(B_1)}{\partial B_1}
   \right]_{\text{orb}} &+ \left[ \frac{\partial \rhoH(B_1)}
   {\partial B_1}\right]_{\text{Zeemann}} \nonumber \\
   &= \RH(B_1) & + \left[ \frac{\partial \rhoH(B_1)}{\partial B_1}
   \right]_{\text{Zeemann}} \;.
\label{eq:Kubo}
\end{alignat}
We used eq.~(\ref{eq:orbital_Hall_contribution}) to relate the
first term to a linear-response of the system. The Zeeman term is
not known to be related to any measurable linear-response
quantity. Consequently, it is not straight forward to interpret
results of the single-field Hall experiment.

\subsubsection{Crossed-field Hall experiments} \label{sec:cross}
The recent interest in heavy-fermion systems stems mainly from
their model character in the investigation of QCPs. Although being
phase transitions at zero temperature quantum critical points lead
to unusual properties up to surprisingly high temperatures. This
issue becomes particularly apparent for the cuprate
superconductors for which quantum criticality is also discussed
(see, {\it e.g.} \cite{bro08} and section \ref{sec:cup}). We shall
see that the Hall effect turned out to be a sensitive tool to
explore the nature of a particular QCP.

Hall effect measurements provide currently the best probe to study
the Fermi surface evolution at quantum critical points. This is
mainly due to the experimental difficulties associated with other
probes: ARPES is limited to higher temperatures (compared to
magnetotransport experiments) whereas quantum oscillation
measurements can only be conducted in a range of high magnetic
fields and on very clean samples.

In order to distinguish whether the Fermi surface evolves
continuously or discontinuously one needs to have sufficient
resolution of the control parameter used to access the quantum
critical point. As the heavy-fermion systems involve magnetic
interactions, magnetic field tuning is the obvious choice to
achieve such high resolution. This, however, means that the
magnetic field has to cover two tasks, namely, tuning the material
across its QCP and probing its Fermi surface by generating a Hall
voltage. Two different approaches can be pursued to disentangle
these two effects.

Firstly, the conventional, single-field Hall effect setup as
described in section \ref{sec:single} might be employed. Here, the
tuning effect stems from the overall applied magnetic field $B_1$
whereas the probing of the Fermi surface at this particular field
strength shows up as the response to infinitesimal fields, see
Fig.~\ref{fig:Halleffectsetups}(a). The differential Hall
coefficient $\tilde{R}_{\text H}$ is given by eq.\
(\ref{eq:differential_Hall_coefficient}) and can be related to the
Fermi surface via the Kubo formalism \cite{pas-nat}. Hence, the
disentanglement of the two effects exerted by the magnetic field
$B_1$ is done in the subsequent analysis by numerical
differentiation.

Secondly, by using two magnetic fields one can disentangle the two
tasks already within the measurement. In order to do so, one
magnetic field $B_1$ is applied perpendicular to the current to
generate the Hall response whereas the second magnetic field $B_2$
is directed parallel to the current to tune the material, as
sketched in Fig.~\ref{fig:Halleffectsetups}(b). We refer to this
as crossed-field Hall effect setup. One precondition for this
disentanglement to successfully work out is, however, that the
tuning effect arising from $B_1$ has a negligible tuning effect to
the material under investigation compared to the tuning field
$B_2$. If the material exhibits a sufficiently high magnetic
anisotropy it can be utilized to ensure this condition. An
influence of $B_2$ on the Hall effect is generally small as it is
parallel to the current avoiding a Lorentz force acting on the
electrons. Consequently, tuning of the material is predominantly
achieved by $B_2$ whereas the Hall effect stems from $B_1$. The
power of the crossed-field setup lies in the fact that the Hall
effect represents the linear response of the system with respect
to the Hall field $B_1$: The Hall coefficient is extracted as the
initial slope of the Hall resistivity at a fixed tuning field
$B_2$,
\begin{equation}
R_{\text H}(B_2) = \lim_{B_1\to 0}\frac{\rho_{\text
H}(B_1,B_2)}{B_1}\;. \label{eq:Hallcross}
\end{equation}

If the geometry between Hall field $\partial B_1$, Hall voltage
and injected current is kept identical for both the single-field
and the crossed-field configurations, the orbital contribution of
the differential Hall coefficient $\RH(B_1)$ and the
linear-response Hall coefficient $\RH(B_2)$ respond to the
evolution of the Fermi surface. Differences may be ascribed to the
Zeeman term or magnetic anisotropies as discussed below and in
section \ref{sec:YRS}.

We shall scrutinize the influence of $B_2$ on the Hall response
arising from $B_1$ by analyzing the conductivity tensor for this
setup. For the given geometry the boundary conditions may be
framed as
\begin{align}
\label{eq:boundary_cond}
    {\pmb j}= \begin{pmatrix} j_x \\ 0 \\ 0 \end{pmatrix} &&
    {\pmb E}= \begin{pmatrix} E_x \\ E_y \\ 0 \end{pmatrix} &&
    {\pmb B}= \begin{pmatrix} B_2 \\ 0 \\ B_1 \end{pmatrix}
\end{align}
Clearly, the conductivity tensor $\underline{\sigma}$ relating the
current density and the electrical field via
eq.~(\ref{eq:OhmsLaw}) will be a $3\times3$ matrix. Only the
elements $\sigma_{xz} = \sigma_{zx} = 0$ can be omitted as no
magnetic field is applied in $y$-direction. Taking advantage of
the Onsager relations \cite{Groot1954,Mazur1954} one derives
\begin{equation}
\underline{\sigma} = \begin{pmatrix} \sigma_{xx} & \sigma_{xy} & 0\\
-\sigma_{xy} & \sigma_{yy} & \sigma_{yz} \\
0 & -\sigma_{yz}  & \sigma_{zz}\\ \end{pmatrix}
\label{eq:sigma_cf}
\end{equation}
The resistivity tensor $\underline{\rho}$ may again be derived via
inversion
\begin{equation}
\underline{\rho}=\frac{1}{\sigma_{xx}\sigma_{yy}\sigma_{zz}+\sigma_{yz}^2
\sigma_{xx}+ \sigma_{xy}^2\sigma_{zz}}
    \begin{pmatrix} \sigma_{yy}\sigma_{zz}+\sigma_{yz}^2
        & -\sigma_{xy}\sigma_{zz} & 0 \\
\sigma_{xy}\sigma_{zz} & \sigma_{zz}\sigma_{xx}
        & -\sigma_{yz}\sigma_{xx} \\
0 & \sigma_{yz}\sigma_{xx} & \sigma_{xx}\sigma_{yy}+\sigma_{xy}^2 \\
\end{pmatrix} \label{eq:rho2}
\end{equation}
from which the Hall coefficient can be read off as
\begin{equation}
    \RH = \lim\limits_{B_1 \to 0} \frac{\rho_{xy}}{B_1} =
    \lim\limits_{B_1 \to 0} \frac{1}{B_1} \frac{\sigma_{xy}}
    {(\sigma_{xx}\sigma_{yy}+\sigma_{yz}^2\sigma_{xx}
    \sigma_{zz}^{-1}+ \sigma_{xy}^2)} \;.
    \label{eq:HC2}
\end{equation}
Rearranging eq.~(\ref{eq:HC2}) allows a detailed comparison to the
single-field case ($B_2=0$):
\begin{equation}
    \RH = \lim\limits_{B_1 \to 0} \frac{1}{B_1}
    \frac{\sigma_{xy}}{\sigma_{xx}\sigma_{yy}+\sigma_{xy}^2}\left(1+
    \frac{\sigma_{yz}^2}{\sigma_{xx}\sigma_{zz}}\right)^{-1}\approx
    \frac{1}{B_1}\frac{\sigma_{xy}}{\sigma^2}\left(1+
    \frac{\sigma_{yz}^2}{\sigma^2}\right)^{-1} \;.
    \label{eq:HC_CrossedField}
\end{equation}
For the approximation in eq.~(\ref{eq:HC_CrossedField}) the same
symmetry simplification as for the case of the single-field Hall
experiment, $\sigma_{xx} \approx \sigma_{yy}$ and the assumption
of $\sigma_{xy} \ll \sigma_{xx}$ are used. Again, incorporating
moderate anisotropies does not alter the results. In fact, it will
become clear that one might utilize anisotropies to enable higher
resolution in the crossed-field Hall experiment.

The leading contribution to \RH\ in eq.\
(\ref{eq:HC_CrossedField}) stems from the unity within the sum and
matches the result of the single-field Hall experiment, eq.\
(\ref{eq:Hall_coefficient_single_field}). This contribution
corresponds to the Hall constant in a gedanken setup in which the
field $B_2$ does not generate any Lorentz force while fulfilling
the role of tuning the underlying state. The second part of the
sum, $\frac{\sigma_{yz}^2}{\sigma^2}$, represents a correction to
this case. In order to obtain an estimate for this correction we
ignore anisotropies in the material, {\it i.e.}, we assume that
the off-diagonal elements of the conductivity tensor are similar:
$\sigma_{yz}\approx\sigma_{xy}$. Thereby we derive that the
correction to the ideal case without any direct effect of $B_2$ on
the Hall coefficient is given by
\begin{equation}
 \frac{\sigma_{yz}^2}{\sigma_{xx}\sigma_{zz}} \approx
 \left(\frac{\rhoH}{\rho}\right)^2 \ll 1 \;.
\label{eq:correction}
\end{equation}
We estimate this correction to be much smaller than 1 using again
that for metals the off-diagonal elements of the conductivity
tensor are typically several orders of magnitude smaller than the
diagonal elements. Equation (\ref{eq:correction}) denotes the
square of the tangent of the Hall angle $(\tan \theta_{\text
H})^2$ which is typically of the order of $10^{-4}$ for metals
at low temperatures and small fields. Consequently,
this correction may be neglected.

These considerations of the transport theory emphasize that the
single-field Hall setup allows to study the Hall effect evolution
across a QCP by means of the differential Hall coefficient which
is composed of an orbital response and a Zeeman term. By contrast,
the crossed-field Hall experiment exclusively yields the
linear-response Hall coefficient associated with the orbital
effect. Consequently, the crossed-field Hall setup appears to be
superior. However, we shall see that the crossed-field setup is
experimentally more demanding. The consistency of the two setups
was tested by investigating a non-magnetic metal \cite{fri10}, see
section \ref{sec:comcross}.

\subsubsection{Realization of crossed-field Hall experiments}
\label{sec:crossexp} 
\begin{figure}[t]
\begin{center}
\includegraphics[width=5.6cm]{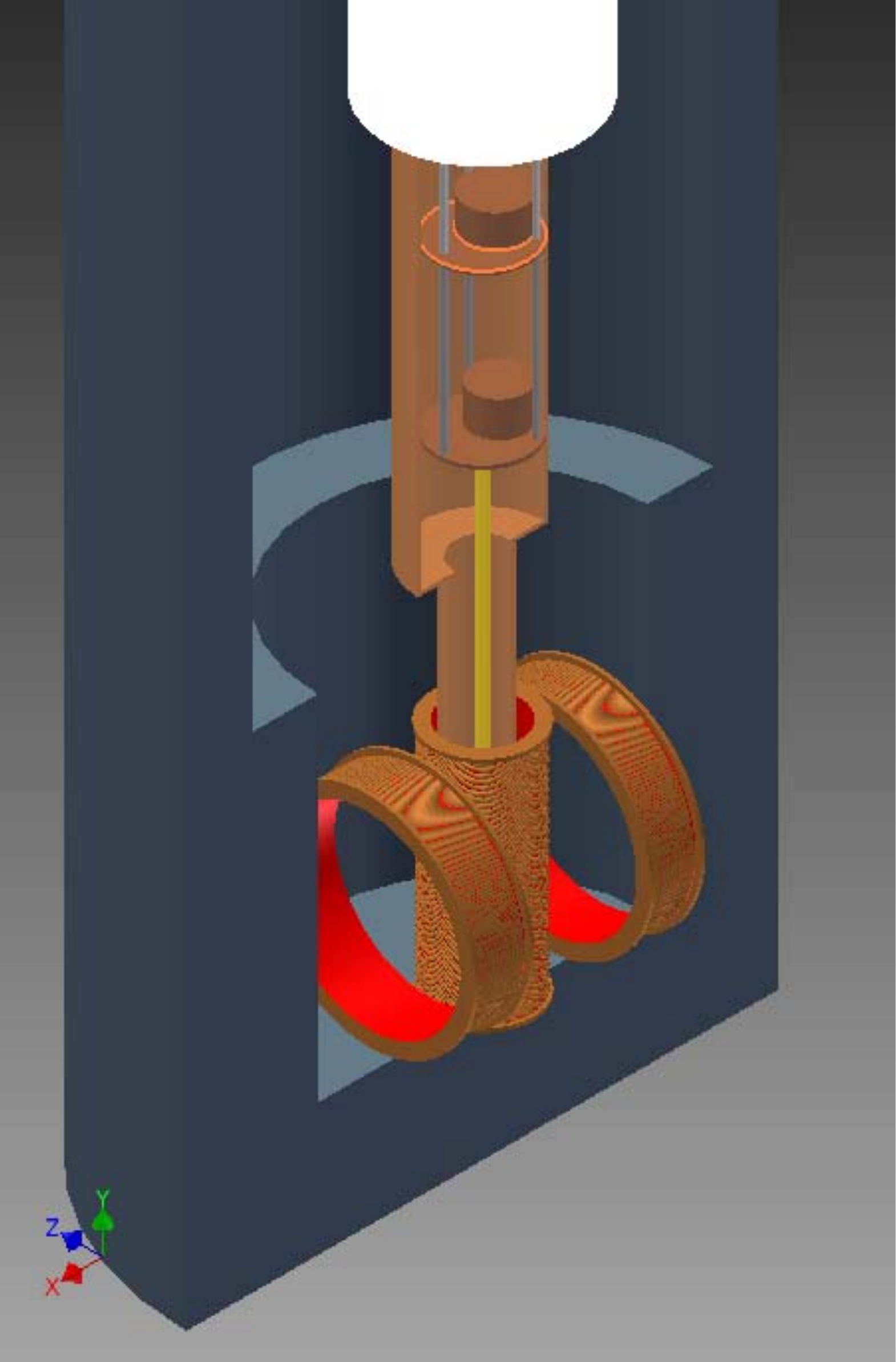}%
\caption{Experimental realization of the crossed-field Hall setup.
A $^3$He/$^4$He-dilution refrigerator is combined with a vector
magnet comprising a solenoid generating a field in the vertical
direction and a split pair generating a horizontal magnetic field,
parallel to the current injected into the sample (not shown).}
\label{fig:cfsetup}
\end{center}
\end{figure}

For many investigations access to lowest
possible temperatures is desired or even required. Hence, Hall
effect measurements are often performed in a
$^3$He/$^4$He-dilution refrigerator. For the case of the
crossed-field Hall setup a vector magnet comprising at least to
components is favorable. Such a setup is sketched in
Fig.~\ref{fig:cfsetup}. Here, the Hall field $B_1$ is generated by
a solenoid oriented vertically whereas the tuning field $B_2$ is
generated by a split pair oriented horizontally. Superconducting
magnets in persistent mode allow low noise measurements of the
Hall signal.

This setup is clearly superior to an earlier realization of
crossed fields \cite{pas-nat} in which a miniature superconducting
coil (providing $B_2$) was mounted inside a dilution refrigerator
equipped with a conventional solenoidal superconducting magnet
(used to apply $B_1$). In the latter case, problems arose due to
the limited space (limiting the magnitude of $B_2$), possible
quenching of the miniature coil and a difficult alignment of the
coils and the sample. We note that the tuning field $B_2$ should
be adjustable.

The Hall voltage is measured as a function of $B_1$ at several
field strengths of $B_2$. This allows to extract the Hall
coefficient $\RH(B_2)$ as the initial linear slope of the Hall
resistivity \rhoH\ with respect to $B_1$ at fixed $B_2$, {\it cf.}
eq.\ (\ref{eq:Hallcross}). Again, the Hall voltage is to be
measured at negative and positive fields $B_1$ to enable the
analysis of the antisymmetric part ({\it cf.} section
\ref{sec:techn}). As we shall see below it is favorable to measure
also at positive and negative $B_2$ to correct for misalignments.
Consequently, the Hall voltage has to be scanned on a close mesh
of $B_1$ and $B_2$ (and $T$ in order to allow an extrapolation to
$T\to 0$). The effort is partially reduced by the fact that it is
sufficient to restrict the range of $B_1$ to small fields as we
are only interested in the initial slope of $\left.
\rhoH(B_1)\right|_{B_2}$. The second magnetic field $B_2$,
however, gives rise to a large increase in measurement time.

\begin{figure}[t]
\begin{center}
\includegraphics[width=7.5cm,clip]{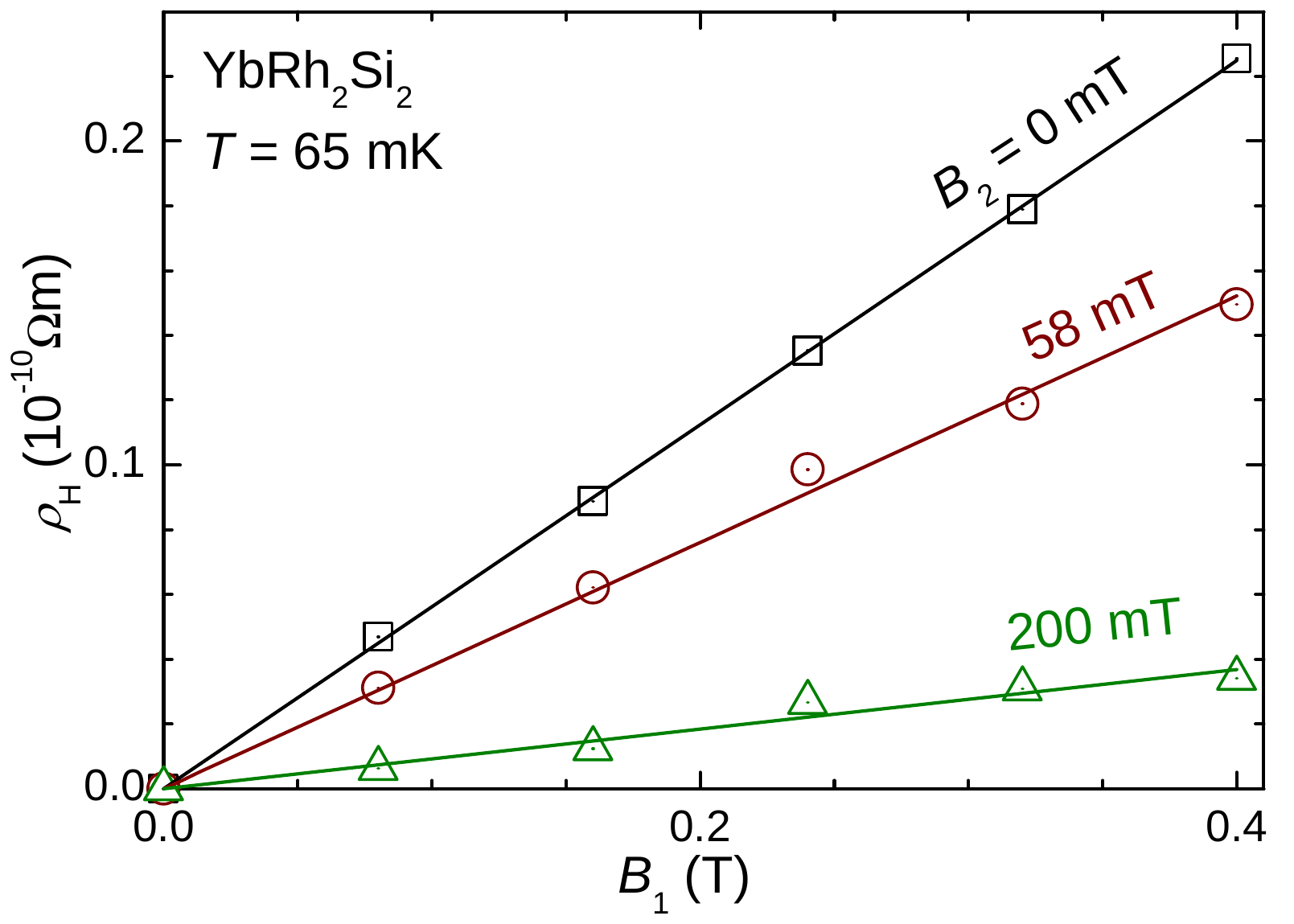}
\caption{Exemplary Hall resistivity isotherms for one \YRS\
sample. Solid lines mark linear fits to the data with the slope of
these lines corresponding to the linear-response Hall coefficient
\RH.} \label{fig:RhoH} \end{center}
\end{figure}

The experimental procedure is illustrated in Fig.~\ref{fig:RhoH}
for the measurements on \YRS\ at one particular temperature
($T=65$\,mK). Here, the Hall resistivity vs. $B_1$ clearly
exhibits a linear field dependence. Importantly, however, the
slope of these curves, {\it i.e.} \RH , is different for different
tuning fields $B_2$. The slope of $\rhoH(B_1)$ was extracted from
least square fits to determine the dependence of $\RH(B_2)$. This
is to be done for several temperatures in order to enable an
extrapolation to zero temperature.

For the conduction of the crossed-field Hall experiments it is
necessary to keep the tasks of the magnetic fields well separated.
As already pointed out, the tuning effect of the Hall field $B_1$
should be small compared to the tuning effect of the tuning field
itself. This may be assured by applying small field $B_1$ only,
but this also decreases the Hall signal and hence, the
sensitivity. Note that also for the single-field Hall setup one is
faced with this issue. For a material with a small critical field
it will be hard to resolve non-linear effects as the Hall signal
will be small around the critical field. Large critical fields on
the other hand will lead to an increase of the Zeeman response in
the single-field Hall setup and might easily be beyond the field
scale accessible with vector magnets needed for the crossed-field
Hall setup.

One may partially overcome the above described issues by taking
advantage of anisotropies in the materials response to external
fields. For the case of \YRS\ which is discussed in more detail in
section \ref{sec:YRS}, the magnetic anisotropy is such that a
field along the crystallographic $c$ direction has to be 11 times
larger compared to a field within the crystallographic $ab$ plane
to generate the same tuning effect on the ground state of the
material. Clearly, this favors an application of the Hall field
$B_1$ along the $c$ direction and, thus, in the case of the
crossed-field experiment, the tuning field $B_2$ within the easy
$ab$ plane. This allows to use higher fields $B_1$ inducing larger
Hall voltages without generating a substantial tuning effect. In
the case of \YRS\ it turned out that fields $B_1\lesssim 0.4$\,T
could be used without disturbing the tuning effect generated by
$B_2$. In fact, the Hall resistance remained linear also in
proximity to the critical value of the tuning field where the Hall
coefficient $\RH(B_2)$, and thus the slope of the Hall resistivity
considerably changes, as can be seen from Fig.~\ref{fig:RhoH}.

One of the major issues of the experimental setup as depicted in
Fig.~\ref{fig:cfsetup} is the alignment of the magnetic fields
with respect to the sample. This appears to be an issue for both
the single-field and crossed-field Hall setup as misalignments can
lead to unwanted effects. For the single-field Hall setup it is
important to have a precise alignment of $B_1$ perpendicular to
the current and to the Hall voltage contacts. Any component out of
this plane will generate a tuning effect without inducing a Hall
response. This effect will be amplified by a magnetic anisotropy.
However, one may take advantage of the magnetic anisotropy to
align the sample with respect to the magnetic field (or {\it vice
versa}). A proper orientation may be assured by monitoring
characteristic features related to the materials response to
magnetic fields as a function of angle between sample and magnetic
field. For the experiments on tetragonal \YRS\ a precision of
better than $0.5^{\circ}$ in alignment of $B_1$ parallel to the
magnetic hard $c$ direction was achieved by maximizing the
critical field while rotating the sample with respect to the field
direction \cite{Friedemann2009g}.

For the crossed-field Hall setup a misalignment might lead to a
mixing of the intended effects of the two magnetic fields. Any
component of $B_1$ within the plane of the current and Hall
voltage contacts acts as a tuning field, and a component of $B_2$
perpendicular to current and Hall voltage contacts contributes to
the Hall field $B_1$. Even small misalignments of the order of
$1^{\circ}$ may change the results dramatically in the presence of
a magnetic anisotropy (as in case of \YRS). In fact, a tilting of
the sample may lead to an antisymmetric part in the longitudinal
voltage as a function of $B_1$ if $B_2$ is finite. This is in
contrast to the fact that the longitudinal resistivity, {\it
i.e.}, the magnetoresistance $\rho_{xx}(B)$ is a symmetric
function of magnetic field. This effect arises solely from the
non-trivial combination of the two magnetic fields. For the
measurement of the Hall coefficient in the crossed-field setup
this is crucial because a longitudinal component will be present
on the Hall voltage contacts as well. This component is usually
separated by taking the antisymmetric part of the measured voltage
to derive the Hall resistivity. This, however, fails if the
longitudinal component already contains an antisymmetric part due
to sample misalignment. Such an antisymmetric part would
consequently add to the Hall coefficient. As a sample is manually
aligned to a precision of the order of several degree only one
might correct such a misalignment by tilting back the data. For
the experiments on \YRS, the uncertainty could be reduced to a
tilting in the plane spanned by $B_1$ and $B_2$. Consequently, it
was possible to analytically rotate back the Hall voltage
$V_{\text H}(B_1,B_2)$. Therefore, both the longitudinal voltage
$V_x$ and the Hall voltage $V_y$ were measured simultaneously as a
function of $B_1$ and $B_2$, with $B_2$ ranging from positive to
negative values. The analysis of the longitudinal voltage revealed
an antisymmetric component which could be omitted when rotating
the data by a small angle. The Hall voltage was then corrected by
the same angle prior to the calculation of the Hall resistivity
and the Hall coefficient \cite{Friedemann2009g}.

\subsubsection{Comparison of single and crossed-field Hall
experiments} \label{sec:comcross} 
\begin{figure}%
\begin{center}
\includegraphics[width=8.6cm,clip]{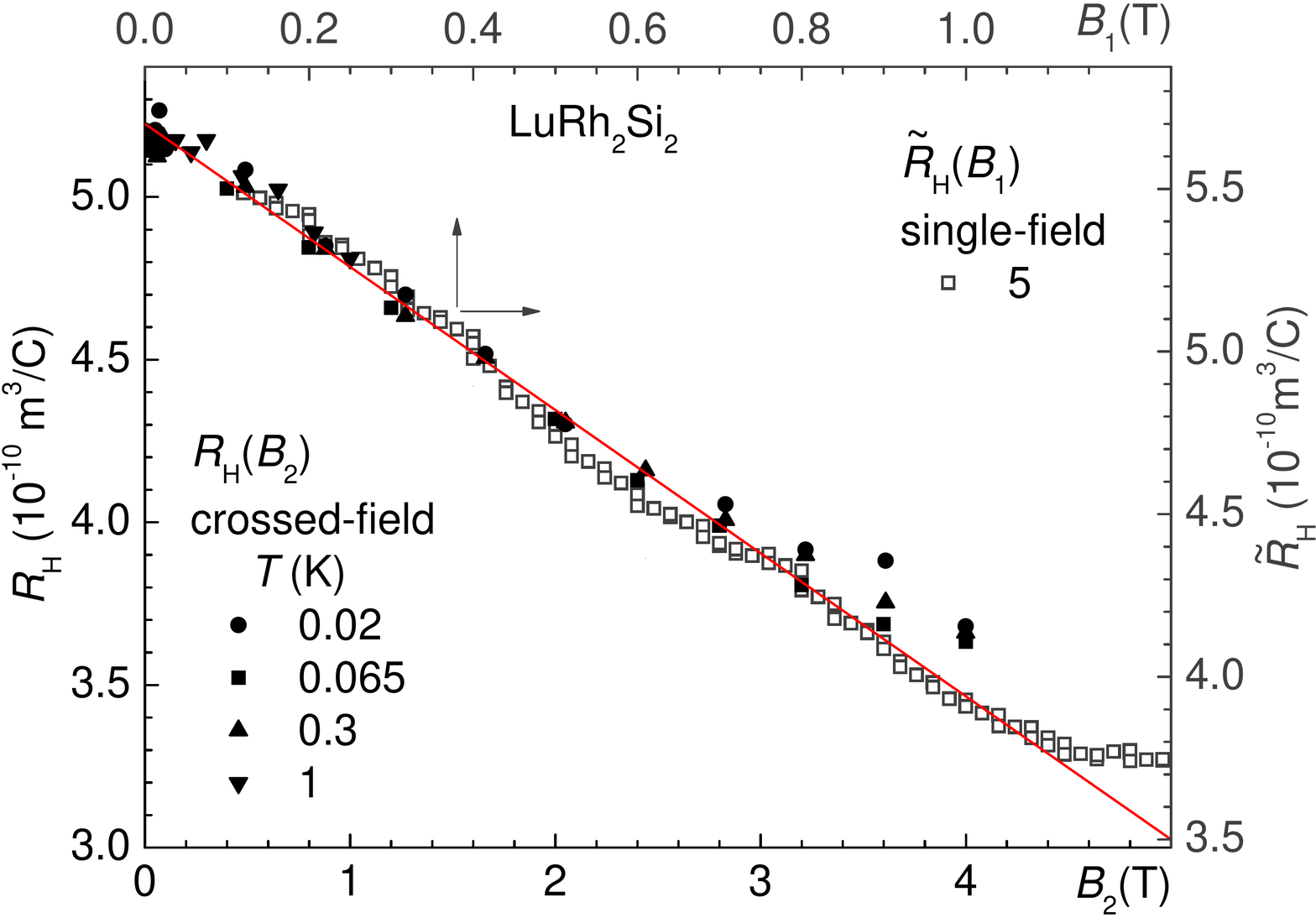}%
\caption{Comparison of crossed-field and single-field Hall
coefficient of \LRS\. Crossed-field results $\RH(B_2)$ are plotted
on the lower and left axis, whereas single-field results
$\TRH(B_1)$ are plotted on the upper and right axis. Solid line
emphasizes the linear evolution of both at small fields. The field
axis of the single-field data is scaled by a factor of 4 with
respect to the crossed-field data accounting for the anisotropy
observed in the field dependence of the Hall coefficient. In
addition, the ordinate of the crossed-field data is shifted by
${0.475\times10^{-10}}$ m$^3$/C compensating the reduction of \RH\
in the crossed-field experiment (see text). Reproduced with
permission from S. Friedemann {\it et al.}, Physica Status Solidi
B {\bf 247} (2010) 723 \cite{fri10}. Copyright \copyright\,2010
WILEY-VCH.} \label{fig:HE_LRS}
\end{center}
\end{figure}

The consistency of the
single-field and crossed-field Hall setup was checked for a
non-magnetic metal, \LRS\ \cite{fri10}. In these two measurements
a linear decrease of the differential and linear-response Hall
coefficient was revealed, respectively. The comparison of the
results is illustrated in Fig.~\ref{fig:HE_LRS}. First, the
single-field Hall results (open symbols) display a linear decrease
as the magnetic field is increased up to $B_1=1$\,T. Above this
field a saturation gradually sets in. Second, in the crossed-field Hall
experiment a linear decrease of $\RH(B_2)$ is observed over the
complete field range studied, {\it i.e.}, up to $B_2=4$\,T. The
fact that the change of the Hall coefficient is linear in both
experiments suggests a Zeeman splitting of the Fermi surfaces as
the underlying mechanism. The different field range over which the
linearity occurs might be related to the anisotropy of the
material: $B_1$ and $B_2$ are applied along different
crystallographic directions. In fact, the data collapse if the
crossed-field Hall data are scaled by $B_2/B_1=4$ (note that the
anisotropy in \LRS\ is not only smaller than in \YRS\ but also reversed).
The different
temperatures of the crossed-field and single-field Hall data do
not affect these considerations as both crossed-field and
single-field Hall results do not exhibit any temperature
dependence below 20\,K.

In the crossed-field experiment, the Hall field $B_1$ also
influences the tuning of the system due to the unfavourable
anisotropy. Therefore, a disentanglement of the two effects of the
magnetic field is incomplete in contrast to the case of \YRS, as
we shall see in section \ref{sec:qpt}. In fact, the results on
\LRS\ in the crossed-field configuration demonstrate that the
tuning effect can arise from both the dedicated tuning field $B_2$
as well as from the Hall field $B_1$. Here, the Hall coefficient
is deduced as the positive linear slope of the Hall resistivity
over a finite field range. In this field range, however, the Hall
resistivity is sublinear as best seen from the decrease of the
differential Hall coefficient (see Fig. \ref{fig:HE_LRS}).
Consequently, the slope of the linear fits and thus the
linear-slope Hall coefficient reflects an average of the
differential Hall coefficient over the field range considered for
the linear fits. This leads to a reduction of the linear-slope
Hall coefficient with respect to the differential Hall coefficient
which is reflected in Fig.~\ref{fig:HE_LRS} as a negative offset
of the crossed-field Hall results. This confirms that for the case
of \LRS\ in the crossed-field Hall experiment tuning stems from
both $B_2$ and $B_1$. The tuning contribution of $B_1$ can be
minimized by either going to smaller fields $B_1$ which makes the
Hall signal harder to detect, or by utilizing magnetic
anisotropies such as described for \YRS.

This section reveals the experimental difficulties associated with
the two setups. Although one should not underestimate the issues
of the single-field Hall setup, the crossed-field experiment is
much more involved. In case of the single-field setup (which is
very much established for the investigations of quantum critical
phenomena) one should consider the effort of low-temperature
measurements and precise orientation. For the crossed-field Hall
setup the effort is extended by the additional magnet needed and
the time consumption to scan the second magnetic field. However,
due to the difficulties with the interpretation of the
single-field Hall results it is necessary to combine them with
crossed-field Hall measurements. As we shall see in section
\ref{sec:YRS}, the crossed-field Hall effect measurements seem to
be the superior probe to reveal a Fermi surface reconstruction at
a QCP.

\section{Hall effect and Kondo-breakdown quantum criticality}
\label{sec:qpt}
\subsection{Hall effect evolution at the QCP in \YRS}
\label{sec:YRS} The heavy-fermion material which was very
extensively studied by Hall effect measurements across its QCP is
\YRS. We discuss the results on this material and subsequently
compare to other heavy-fermion materials. Hall effect measurements
on \YRS\ provide currently the best evidence for a Fermi surface
reconstruction at a QCP. In fact, these Hall effect measurements
allowed also to extract information on the quantum critical
scaling behavior as we shall see below.

\YRS\ is a prototypical material for the investigation of quantum
critical phenomena. The heavy-fermion character is evident from
various transport and thermodynamic properties. The resistivity
$\rho$ exhibits a logarithmic temperature dependence above 100\,K
associated with the single ion Kondo scattering. Below 100\,K,
$\rho(T)$ \cite{octav,Koehler2008} as well as the thermopower
\cite{Koehler2008} decrease as the temperature is reduced
reflecting the onset of coherent Kondo scattering and the
formation of composite quasiparticles. These effects, however,
involve all crystalline electric field (CEF) levels
\cite{Cornut72}. Upon cooling, the 4$f$ electrons increasingly
``condense'' into their CEF-derived Kramers doublet ground state
allowing the formation of a Kondo lattice below around 30 K
\cite{octav,Koehler2008}. These effects of combined CEF and Kondo
interactions have recently been demonstrated by scanning tunneling
spectroscopy  on \YRS\ \cite{ernst11}. Only at very low
temperatures of $\TN=70$\,mK orders \YRS\ antiferromagnetically
\cite{octav}. The N\'eel temperature \TN\ is continuously
suppressed by a small magnetic field leading to the QCP. This
field-induced QCP is accessed at a critical field of either
$B_{1\text c}= 660$\,mT if applied along the crystallographic $c$
direction, or $B_{2\text c}=60$\,mT if applied perpendicular to
the $c$ axis. This magnetic anisotropy is used to the advantage of
separating the two magnetic field tasks in the crossed-field
experiment ({\it cf.}~\ref{sec:crossexp}). The phase transition
appears to remain continuous down to at least 15\,mK
\cite{Kuechler2004} indicating gapless quantum fluctuations. Right
at the critical field pronounced non-Fermi liquid behavior is
observed over three decades in temperature down to the lowest
temperatures accessible \cite{Gegenwart2002}. This non-Fermi
liquid behavior is best explained to arise from the quantum
fluctuations within the QCP picture which assumes a continuous
phase transition at zero temperature. The non-Fermi liquid
behavior comprises a logarithmic divergence of the Sommerfeld
coefficient of the electronic specific heat, {\it i.e.} $ \gamma =
C_{\text{el}}/T \propto \log (T/T_0)$, and a linear temperature
dependence of the resistivity \cite{octav,Gegenwart2002}. At
fields larger than the critical field a Fermi liquid ground state
is recovered. Here, the resistivity follows a quadratic
temperature dependence and the Sommerfeld coefficient saturates.
The characteristics of this Fermi liquid phase are strongly
enhanced as typical for a heavy fermion material, the quadratic
coefficient of the resistivity and the Sommerfeld coefficient obey
values up to a factor 1000 larger than in non-correlated metals
\cite{Gegenwart2002}. This reflects an extremely large effective
mass of the composite quasiparticles.

Indications of an unconventional type of quantum criticality arose
from measurements of the Gr\"uneisen ratio \cite{geg03}. The
temperature dependence of the Gr\"uneisen ratio strongly underpins
the presence of a QCP as it obeys a power-law divergence, {\it
i.e.}, $\Gamma \propto T^{-x}$. The observed exponent of $x= 0.7$
is, however, in contrast to the predictions of the conventional
SDW QCP ($x=1$). This fact along with the outcome of $B/T$ scaling
of both the specific heat and the electrical resistivity
\cite{cus03} suggests that \YRS\ is an ideal candidate for
unconventional quantum criticality. This motivated extensive Hall
effect studies on \YRS\ \cite{pas-nat,Friedemann2010b}. As the
Hall coefficient is a rather complex quantity in its relation to
the Fermi surface ({\it cf.} section \ref{sec:HallFS}) we should
first note some peculiarities of the Hall effect in \YRS. First,
the anomalous Hall effect is shown to be dominant at high
temperatures but is negligible at lowest temperatures
\cite{Paschen2005}. In addition, we shall see that the anomalous
Hall contributions can not account for the signatures related to
quantum criticality. Second, we note that the low temperature Hall
coefficient of \YRS\ is found to exhibit sample dependencies. For
the almost compensated two-band material \YRS\ these sample
dependencies where found to arise from slight changes of the
relative scattering rates of the two dominating bands
\cite{Friedemann2010c}. These differences in the relative
scattering rates appear to originate from tiny variations in the
composition because they only affect samples from different
batches. A systematic study of different samples revealed the
sample dependencies to be associated with a background
contribution whereas the contributions related to the quantum
criticality are robust \cite{Friedemann2010b}.

The Hall effect on \YRS\ was studied in both the single-field and
the crossed-field geometry (explained in section
\ref{sec:advexp}). These studies are further corroborated by
longitudinal magnetoresistance measurements (see Ref.\
\cite{Friedemann2010e}). In the single-field Hall experiment, a
single magnetic field, $B_1$, applied along the $c$ axis was used
to both tune and probe the sample. For the crossed-field Hall
experiment, the tuning was carried out with an additional magnetic
field $B_2$ within the crystallographic $ab$ plane while a small
field $B_1$ was utilized to probe the system. The scales of the
tuning fields, {\it i.e.}, $B_1 \parallel c$ and $B_2 \perp c$, of
the two experiments differ by the magnetic anisotropy of the
material. This is quantified by the ratio of the critical fields
$B_{2\text c}/B_{1 \text c}=1/11$. We shall use this ratio to
convert the single-field Hall results to the equivalent
crossed-field scale. As detailed in section \ref{sec:cross} the
latter orientation allows to probe the system at larger fields
$B_1$ thus increasing the Hall signal. Fig.~\ref{fig:RhoH}
illustrates that in an extended range of low fields up to $B_1 =
0.4$ T, the Hall resistivity is linear with respect to $B_1$ even
close to the
\begin{figure}
\centering\includegraphics[width=10.8cm,clip]{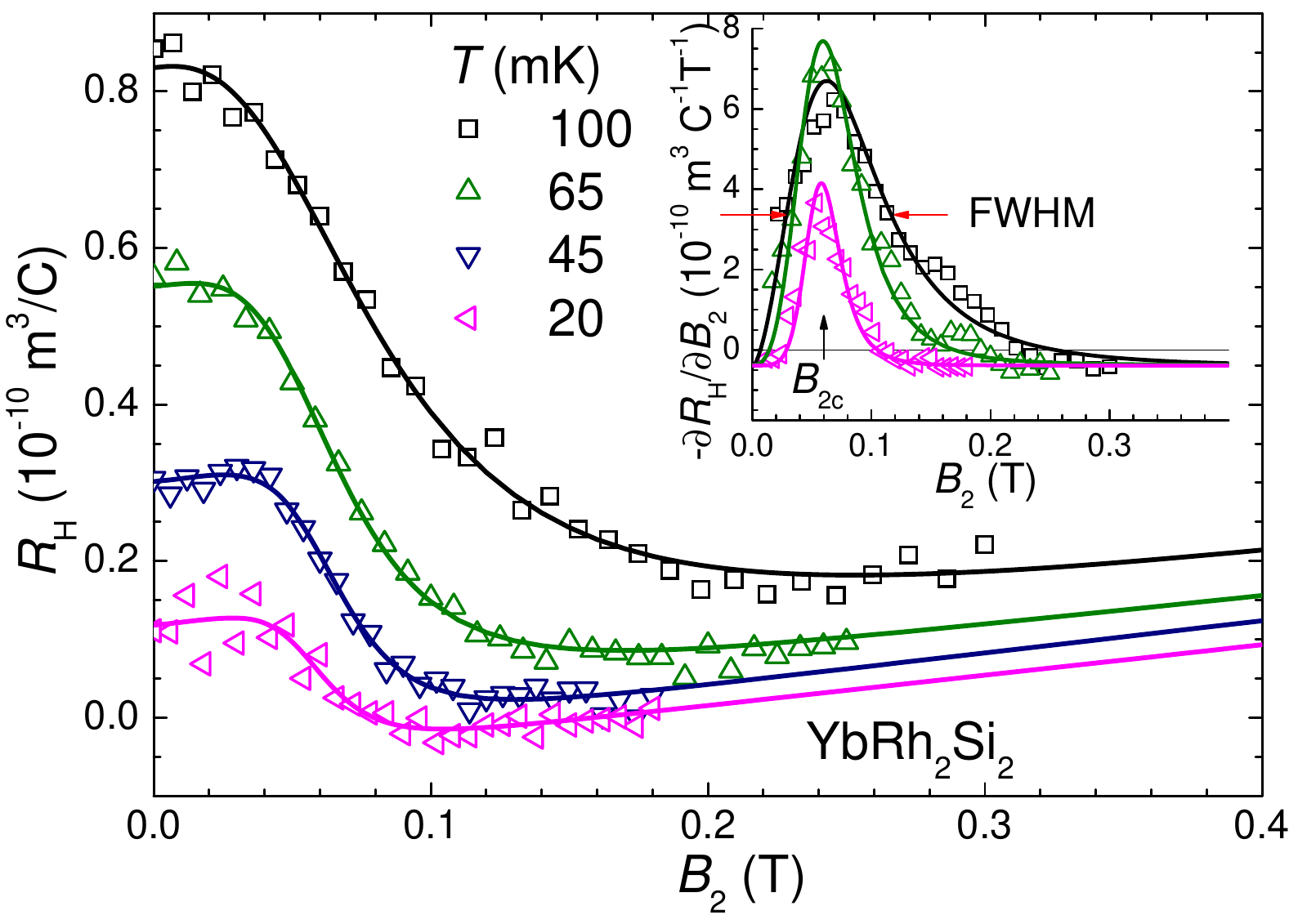}
\caption{Crossed-field Hall effect results: Linear-response Hall
coefficient measured in the crossed-field Hall experiment on
YbRh$_2$Si$_2$. The field dependence of the isothermal
linear-slope Hall coefficient is derived from linear fits to the
Hall resistivity $\left.\rhoH(B_1)\right|_{B_2,T}$ as detailed in
section \ref{sec:cross} (Fig.~\ref{fig:RhoH}). Solid lines are
best fits of eq.~(\ref{eq:crossover_function}) to the data
extending up to 2\,T. The inset shows the derivative $-\mathrm d
\RH /\mathrm d B_2$ of both the data (symbols) and the fits (solid
lines). Vertical arrow indicates the critical field $B_{2\text
c}$. Horizontal arrows specify the FWHM of the peak for one curve.
Reproduced from S. Friedemann {\it et al.}, Proceedings of the
National Academy of Science USA {\bf 107} (2010) 14547
\cite{Friedemann2010b}.} \label{fig:RHvsB_YRS}
\end{figure}
critical field $B_{2\text c}$ which, in consequence, allows to
reliably extract the linear-response Hall coefficient as a
function of $B_2$. The longitudinal magnetoresistance as shown in
detail in Ref.\ \cite{Friedemann2010e} was measured with field
perpendicular to the $c$-axis simultaneous to the crossed-field
Hall effect and is therefore associated with the equivalent field
scale $B_2$.

\begin{figure}
\centering\includegraphics[width=10.8cm,clip]{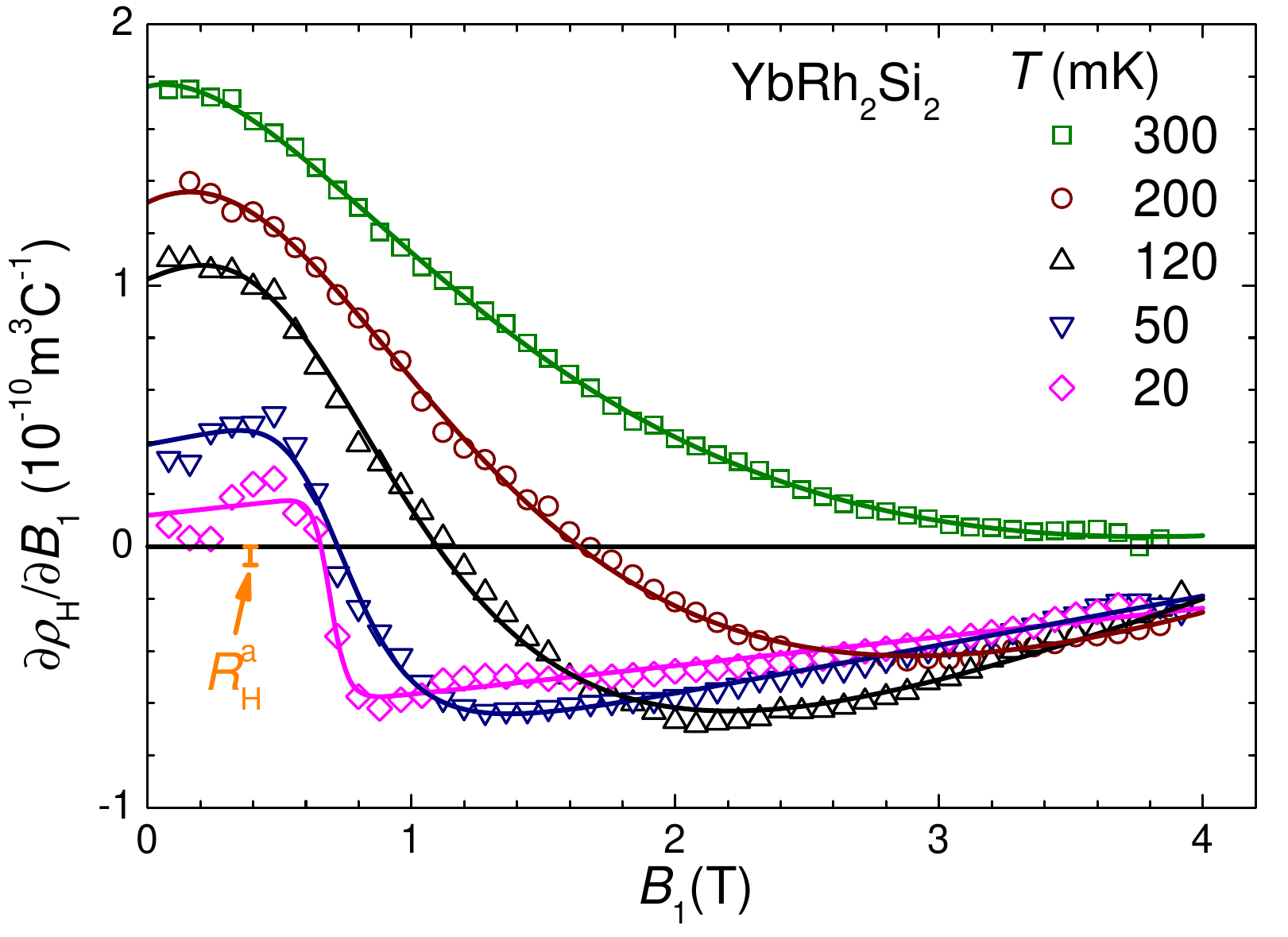}
\caption{Single-field Hall effect results: Isotherms of the
numerically derived differential Hall coefficient $\TRH(B_1)$ of
YbRh$_2$Si$_2$. Solid lines mark best fits of
eq.~(\ref{eq:crossover_function}) to the data. The vertical bar
denotes the size of the constant offset originating from an
anomalous contribution estimated from eq.~(\ref{eq:AHE}) via
$\RH^{\text a} = \partial \rhoH^{\text a} / \partial B_1$.
Reproduced with permission from S. Friedemann {\it et al.},
Journal of Physics: Condensed Matter {\bf 23} (2011) 094216
\cite{Friedemann2010e}. Copyright (2011) IOP Publishing Ltd.}
\label{fig:dRhoHdB1_YRS}%
\end{figure}

Figs.~\ref{fig:RHvsB_YRS} and \ref{fig:dRhoHdB1_YRS} display
selected isotherms of the  linear-response Hall coefficient
$\RH(B_2)$ and the differential Hall coefficient $\TRH(B_1)$
deduced in the crossed-field and single-field Hall experiment,
respectively. For plots of the magnetoresistance data we refer the
reader to Ref.\ \cite{Friedemann2010e}. All data sets exhibit
similar behavior comprising two main features. First, at small,
increasing fields the Hall coefficient and the magnetoresistance
decreases in a pronounced, crossover-like fashion. Second, at
elevated fields and low temperatures, the crossover of the Hall
coefficient is followed up by a linear increase. With decreasing
temperature the crossover becomes sharper in field, decreases in
height, and is shifted to lower fields in both experiments. In
fact, it approaches the critical field in all cases, {\it i.e.},
$B_{1\text c}$ and $B_{2\text c}$ for the single-field and
crossed-field Hall and magnetoresistance experiment, respectively.
Consequently, the crossover is related to the quantum critical
behavior. With the crossover being restricted to small fields the
linear behavior is revealed over an increasing field range as the
temperature is lowered. This suggests that the linear behavior
represents a background contribution to which the crossover is
superposed. The linearity of the background contribution suggests
that it originates from Zeeman splitting. Moreover, it is subject
to the above mentioned sample dependencies. The main features of
the curves, {\it i.e.} the crossover, by contrast, are not
affected by the sample dependencies. The decomposition of the Hall
coefficient into a critical and a background contribution is
further illustrated in the inset of Fig.~\ref{fig:RHvsB_YRS}.
Here, the derivative $-\partial \RH /\partial B_2$ is shown. The
crossover is reflected by a peak around the critical field
(indicated by a vertical arrow) whereas the background
contribution appears as a non-zero offset. Below, we shall examine
the characteristics of the crossover in the Hall coefficient in
more detail and relate it to the quantum critical contribution.

An anomalous contribution to the Hall effect in \YRS\ appears to
be negligible at lowest temperatures. The skew scattering ({\it
cf.} section \ref{sec:skew})
typically important in heavy fermion systems at high temperatures
was found to lead to pronounced signatures in $\RH(T)$ around
100\,K. At lowest temperatures, however, this contribution is
largely suppressed. The fact that the anomalous contribution to
the Hall resistivity
\begin{equation}
 \rhoH^{\text a} (B_1) = C \rho(B_1) \mu_0 M(B_1)
\label{eq:AHE}
\end{equation}
is essentially linear in field implies that it adds a small
constant offset to the Hall coefficient. Using the constant $C$
determined at higher temperatures \cite{Paschen2005} leads to a
value of $\rhoH^{\text a}= 0.07\times 10^{-10}$\,m$^3/$C which is
tiny compared to the large variation of the differential Hall
coefficient as illustrated by the vertical bar in
Fig.~\ref{fig:dRhoHdB1_YRS}. More importantly, the added offset to
the Hall coefficient does not affect the analysis of the critical
contribution. Particularly, the width, height and position of the
crossover are independent of any offset.

For the detailed analysis the empirical function
\begin{equation}
\RH(B_2) = \RH^{\infty}+mB_2-
\frac{\RH^{\infty}-\RH^0}{1+(B_2/B_0)^p}
\label{eq:crossover_function}
\end{equation}
was fitted to the data. This function simulates a crossover from
the zero-field value $\RH^0$ to the high-field value
$\RH^{\infty}$ superposed to a linear background $m B_2$. The
position of the crossover is given by $B_0$ and the sharpness is
determined by $p$ which is restricted to positive values. The
function describes the data very well as can be seen from
Figs.~\ref{fig:RHvsB_YRS} and \ref{fig:dRhoHdB1_YRS}. Analogous
fits to the differential Hall coefficient $\TRH(B_1)$ measured in
the single-field experiment yielded the zero-field and high-field
values $\TRH^{\infty}$ and $\TRH^0$, respectively.
Equation~(\ref{eq:crossover_function}) allows to deduce the
characteristics of the crossover, {\it i.e.}, the quantum critical
contribution. In the following we shall focus on a complete
characterization of the crossover with respect to its position,
width and height.

\begin{figure}
\centering\includegraphics[width=10.0cm,clip]{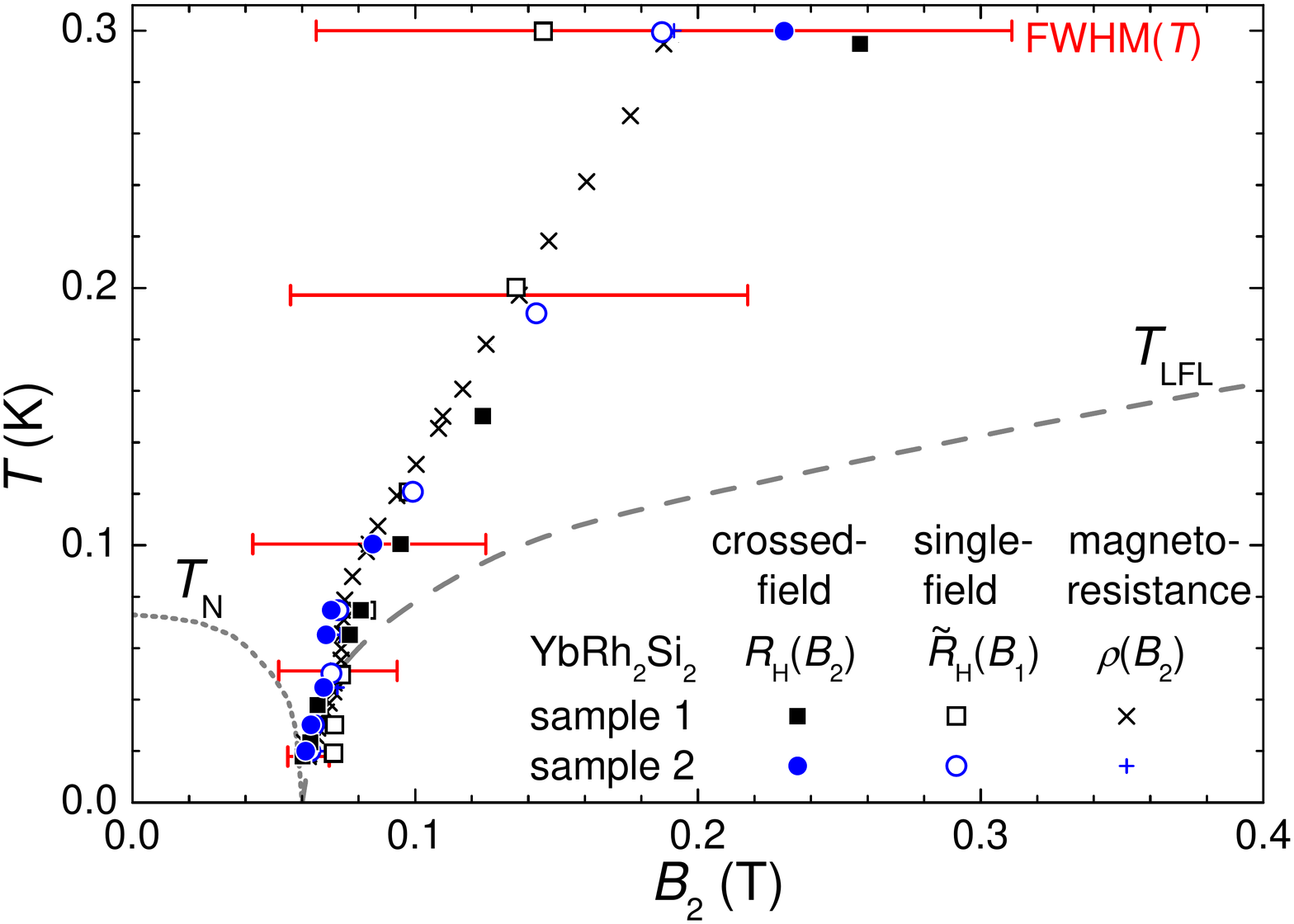}
\caption{Linkage of the Hall crossover to the QCP in
YbRh$_2$Si$_2$. The crossover field $B_0$ was extracted from fits
of eq.~(\ref{eq:crossover_function}) to the crossed-field
$\RH(B_2)$ (solid symbols) and single-field $\TRH(B_1)$ (open
symbols) Hall results as shown in Figs.~\ref{fig:RHvsB_YRS} and
\ref{fig:dRhoHdB1_YRS} as well as to magnetoresistance $\rho(B_2)$
(Ref.~\cite{Friedemann2010e}). Results on two samples with
different residual resistivities match within experimental
accuracy. The values deduced from the single-field Hall experiment
are scaled by $1/11$ to account for the magnetic anisotropy of the
material (see text). Red horizontal bars represent the FWHM at
selected temperatures. Error bars are omitted to avoid confusion
with the FWHM bars; with exception of the data at 0.3\,K and the
single-field result on sample 2 at 0.19\,K standard deviations are
smaller than the symbol size. Dotted and dashed lines mark the
boundary of the magnetically ordered phase and Fermi-liquid
regime, respectively, as taken from Ref. \cite{Gegenwart2002}.
Reproduced from S. Friedemann {\it et al.}, Proceedings of the
National Academy of Science USA {\bf 107} (2010) 14547
\cite{Friedemann2010b}.} \label{fig:PDRH_YRS}
\end{figure}

Firstly, the position of the crossover is tracked in the phase
diagram shown in Fig.~\ref{fig:PDRH_YRS} for two samples of very
different values of the residual resistivity which span the whole
range of sample dependencies observed for the low-$T$ Hall
coefficient ({\it cf.} Ref.~\cite{Friedemann2010c} with identical
nomenclature). The crossover field is found to match for both
samples and for three different experiments, {\it i.e.}, for
crossed-field and single-field Hall effect measurements and for
the analogous crossover in the magnetoresistivity. As the
temperature is lowered $B_0$ shifts to lower fields and converges
to 60\,mT, {\it i.e.}, it extrapolates to the QCP in the limit of
zero temperature. This reveals the linkage of the crossover in the
Hall coefficient (and magnetoresistance) to the quantum criticality.

\begin{figure}
\centering\includegraphics[width=13.6cm,clip]{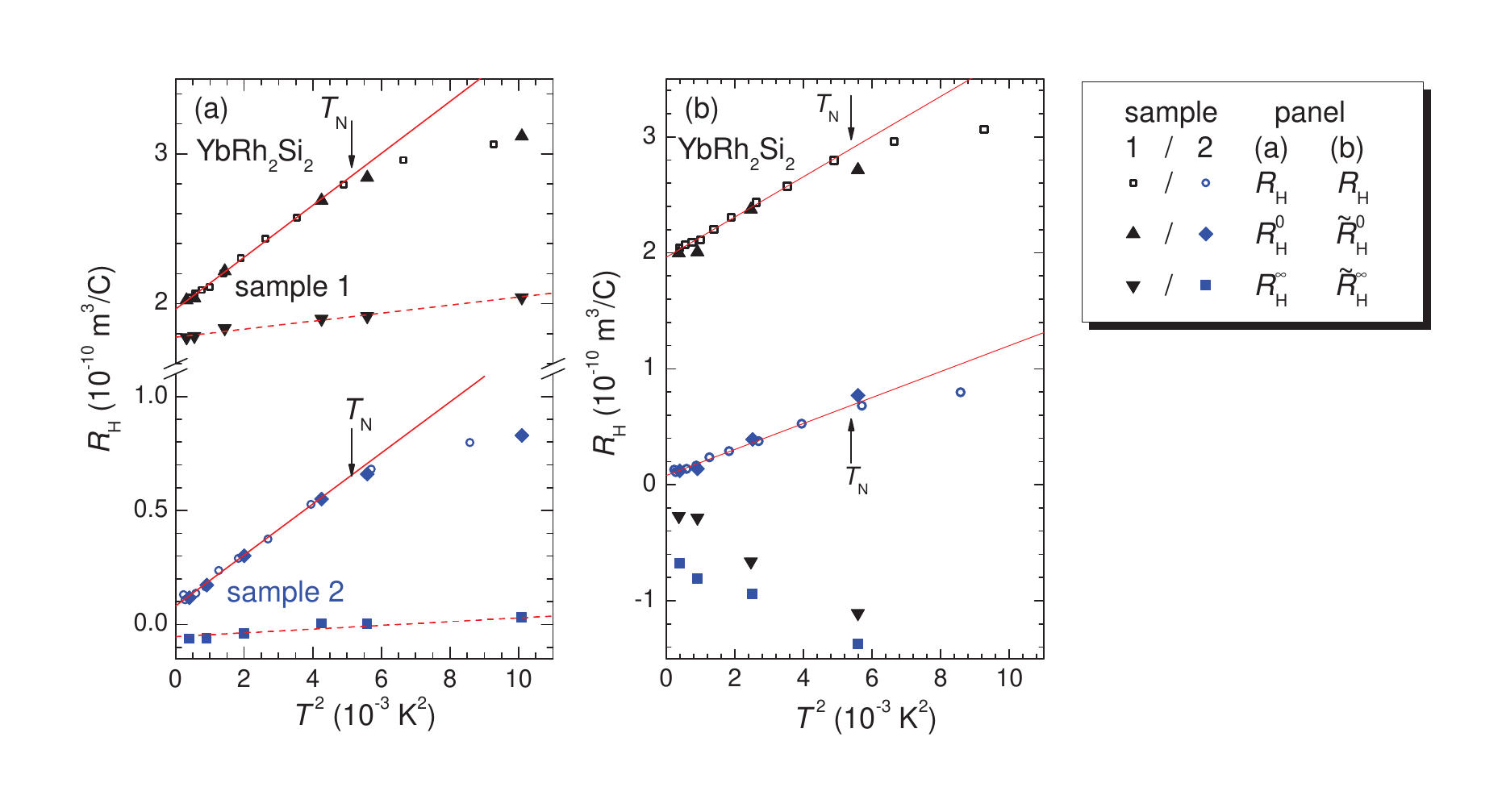}
\caption{Limiting parameters of the Hall crossover in
YbRh$_2$Si$_2$. The initial-slope Hall coefficient (open symbols)
is plotted against a quadratic temperature scale for two selected
samples. Solid lines mark best fits of a quadratic form
$\RH(T)=\RH(T=0) + A^{\prime} T^2$ to the initial-slope Hall
coefficient. The zero-field ($\RH^0$) and the high-field
($\RH^{\infty}$) values extracted from fits of
eq.~(\ref{eq:crossover_function}) to $\RH(B_2)$ are included in
(a). The corresponding values ($\TRH^0$ and $\TRH^{\infty}$)
extracted from the single-field Hall results
(Fig.~\ref{fig:dRhoHdB1_YRS}) are displayed in (b). Arrows mark
the N\'{e}el temperature. Reproduced with permission from S.
Friedemann {\it et al.}, Journal of Physics: Condensed Matter {\bf
23} (2011) 094216 \cite{Friedemann2010e}. Copyright (2011) IOP
Publishing Ltd.} \label{fig:RHiuRHo_YRS}
\end{figure}

Secondly, the height of the crossover is characterized by the
limiting parameters $\RH^0$ and $\RH^{\infty}$. Their temperature
dependencies are plotted in Fig.~\ref{fig:RHiuRHo_YRS} on a
quadratic temperature scale which allows to identify important
temperature dependencies as we shall see in the following. The
absolute values of $\RH^0$ and $\RH^{\infty}$ differ for the two
samples, in accordance with the sample dependencies that are
associated with the background contribution to \RH. The quantum
critical contribution is characterized by the difference
$\RH^0-\RH^{\infty}$. This difference decreases as the temperature
is lowered which can be seen from the $\RH(B_2)$ curves in
Fig.~\ref{fig:RHvsB_YRS}. In order to draw conclusions on the
evolution of the Fermi surface at the QCP we need to deduce this
difference in the limit $T\to 0$. Therefore, we utilize the
temperature dependencies found for the initial-slope Hall
coefficient. The zero-field value $\RH^0$ extracted from the fits
matches very well with the measured initial-slope Hall coefficient
in zero tuning field. We can therefore use the  quadratic
temperature dependence found for the initial-slope Hall
coefficient below \TN\ to reliably extrapolate to $T\to 0$ ({\it
cf.} solid lines in Fig.~\ref{fig:RHiuRHo_YRS})
\cite{Friedemann2010e}. The temperature dependence of the
high-field value $\RH^{\infty}$ is less pronounced and may well be
approximated with a quadratic form as well. Importantly, the
extrapolation of $\RH^0(T)$ and $\RH^{\infty}$ clearly indicates
that their difference persists in the extrapolation $T\to 0$, {\it
i.e.}, down to the QCP. This is not only true for
$\RH^0-\RH^{\infty}$ of both samples but also for the difference
$\TRH^0-\TRH^{\infty}$ of the limiting values extracted from the
single-field Hall experiment displayed in
Fig.~\ref{fig:RHiuRHo_YRS}(b) as well as for the corresponding
values of the crossover in magnetoresistance
\cite{Friedemann2010e}. In summary, we can conclude that a change
in the Hall coefficient persists at the QCP and is robust against
sample dependencies.

The absolute value of the crossover height, $\RH^0 -
\RH^{\infty}$, is difficult to evaluate because multiple bands
almost compensate each other at $E_F$ (despite the fact that the
valence of $\sim$2.9 indicates YbRh$_2$Si$_2$ to be a hole system
\cite{kum11}) and these bands are influenced by sample
dependencies \cite{Friedemann2010c}. Yet, the drop of $\RH$ at
$B_0$ is consistent with a change from two hole-like Fermi-surface
sheets dominating at $B < B_0$ to one hole/one electron sheet for
$B > B_0$ due to the hybridization between 4$f$ and conduction
electrons in the latter case.

\begin{figure}
\centering\includegraphics[width=8.8cm,clip]{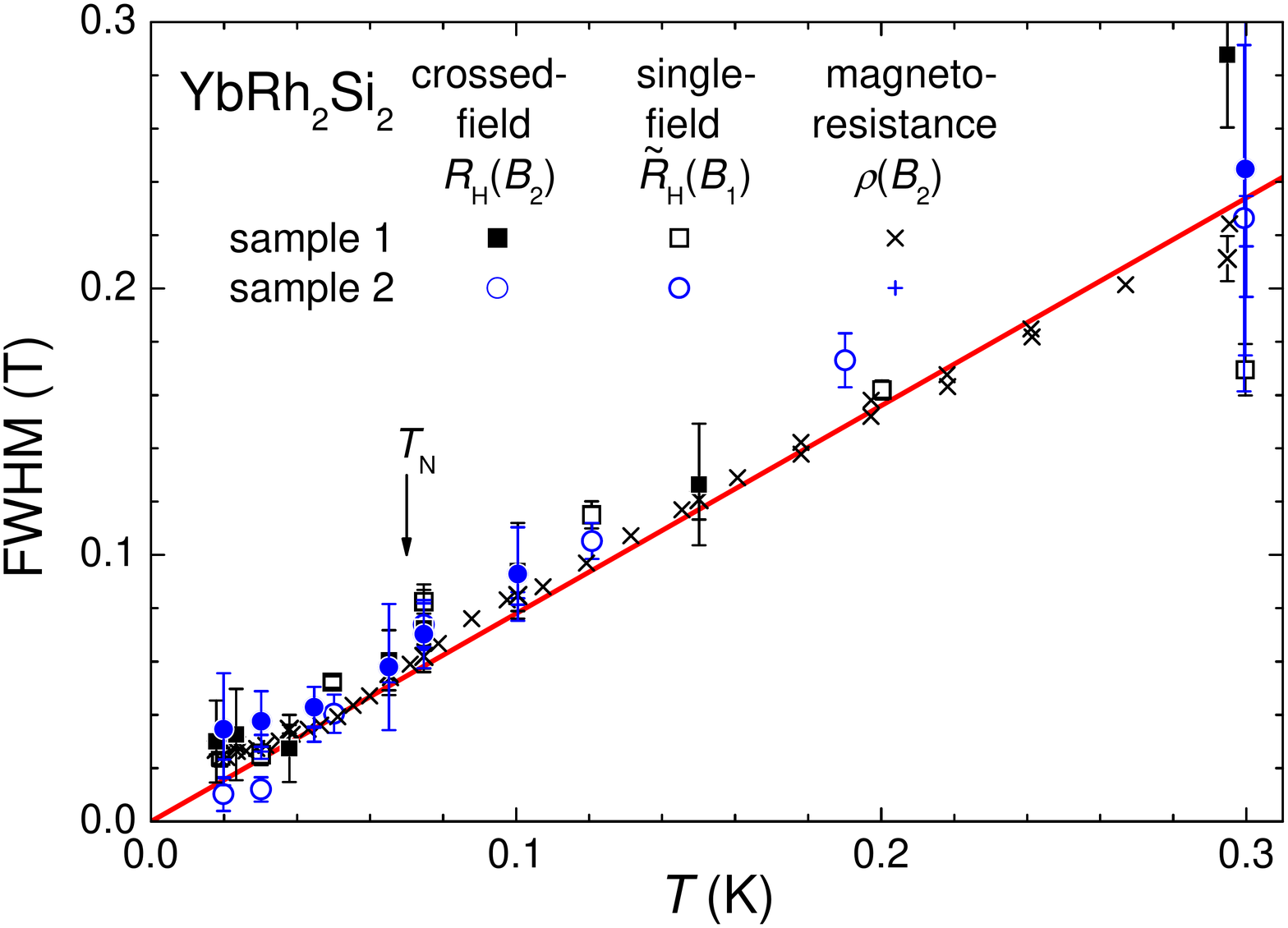}
\caption{Vanishing width of the Hall crossover in YbRh$_2$Si$_2$
after Ref.\ \cite{Friedemann2010b}. The full-width at half-maximum
(FWHM) was extracted from the derivative of the fits of
eq.~(\ref{eq:crossover_function}) to $\RH(B_2)$ in the
crossed-field experiment (closed symbols), to $\TRH(B_1)$ of the
single-field Hall data (open symbols) and magnetoresistance
(crosses). Inset of Fig.~\ref{fig:RHvsB_YRS} illustrates the
definition of FWHM. The values of the single-field experiment were
scaled by $1/11$ to account for the magnetic anisotropy. Solid
line marks a linear fit to all data sets up to 1\,K  which
intersects the ordinate at the origin within experimental accuracy
\cite{Friedemann2010e}. Vertical arrow marks the N\'{e}el
temperature. Reproduced from S. Friedemann {\it et al.},
Proceedings of the National Academy of Science USA {\bf 107}
(2010) 14547 \cite{Friedemann2010b}.} \label{fig:FWHM_YRS}
\end{figure}

Finally, we examine the width of the Hall crossover. This is
accomplished by determining the full-width at half-maximum (FWHM)
of the derivative of the curves fitted to $\RH(B_2)$, see inset of
Fig.~\ref{fig:RHvsB_YRS} for one exemplary temperature. The
temperature dependence of the FWHM is displayed in
Fig.~\ref{fig:FWHM_YRS}. Like for the crossover field, we find the
data sets of the different samples to be in excellent agreement.
This emphasizes that the Hall crossover is robust against sample
dependencies whereas the background contribution and the absolute
values of the Hall coefficient are indeed sample dependent. In
addition, the width extracted from the single-field and
crossed-field Hall effect measurements and from the analogous
crossover in the longitudinal magnetoresistivity match very well.
This indicates that we probe the same Fermi surface change with
both the Hall and the magnetoresistivity measurement.

As already deduced from the $\RH(B_2)$ curves, the width of the
crossover shrinks as the temperature is reduced. This width
reveals a fundamental property of the quantum critical Hall
crossover: Taking all these data sets together, the temperature
dependent width of the crossover is best described by a
proportionality to temperature, see the solid line in
Fig.~\ref{fig:FWHM_YRS}. No signature is observed in
$\text{FWHM}(T)$ at \TN. This is in contrast to $\RH^0$ and thus
also the height which changes to a quadratic temperature
dependence below \TN. A tendency towards saturation of
$\text{FWHM}(T)$ seems to set in for the crossed-field Hall and
longitudinal magnetoresistivity data below 30\,mK whereas the
single-field Hall-effect data continue the decrease linearly down
to the lowest temperature accessed. It is most likely that the
crossed-field Hall data are affected by the proximity to the
classical phase transition as at these lowest temperatures the
Hall crossover significantly interferes with the classical phase
transition at $\TN(B)$. This can be seen from the fact that at
lowest temperatures the FWHM extends substantially into the
ordered phase ({\it cf.} horizontal bars in
Fig.~\ref{fig:PDRH_YRS}). For the single-field Hall experiment the
classical fluctuations should be smaller according to this
reasoning. Such a difference may well originate from the different
orientation of the tuning field for the two experiments. For the
crossed-field Hall experiment with the tuning field in the
magnetically easy plane, the magnetization and thus the classical
magnetic fluctuations are almost one order of magnitude larger
compared to the single-field experiment with the tuning field
($B_1$) along the magnetic hard axis. Nevertheless, within the
experimental accuracy all data agree with the fit down to the
lowest temperatures. Consequently, the FWHM extrapolates to zero
in the limit $T\to0$. Together with the crossover field $B_0$
converging to the QCP and the step height remaining finite, the
vanishing width suggests that the crossover in the Hall
coefficient transforms into a discontinuity at the QCP. This
indicates a reconstruction of the Fermi surface.

The single-field and crossed-field Hall experiments reveal a
critical crossover in the Hall coefficient which matches in its
position and width. This is in accordance with the considerations
presented in section \ref{sec:advexp}. In particular it reflects
that according to eq.~(\ref{eq:Kubo}) the single-field Hall
results can be decomposed into an orbital contribution and a
Zeeman contribution; the former is expected to be identical to the
result of the crossed-field Hall experiment. Only the step height
appears to be different for the two experiments which should than
be associated with the Zeeman term. It remains to be understood in
detail how the step height is changed. One possibility is that it
may originate from the different field scales involved in the two
experiments.

A change of the scattering rates of the two dominating bands
crossing the Fermi level \cite{Friedemann2010c} can be ruled out
as the origin of a jump in the Hall coefficient. The sample
dependencies associated with the background contribution are
explained incorporating different relative scattering rates of the
two bands which arise from minute differences in the composition.
The Hall crossover however originates from the quantum
criticality. Furthermore, the tuning parameter is varied
continuously and tunes the material through a continuous phase
transition. A change in the scattering rates at the latter can
thus not lead to a discontinuous evolution of the Hall
coefficient.

The abrupt change of the Hall coefficient is associated with a
Fermi surface collapse which is incompatible with the expectations
for an SDW QCP. For an SDW QCP a smooth evolution is expected
theoretically \cite{col01,Norman2003,Bazaliy2004}. The Fermi
surface of a SDW state is reconstructed from that of the
paramagnetic state through a band folding. When the SDW order
parameter is adiabatically switched off, the folded Fermi surface
is smoothly connected to the paramagnetic one. As a result, the
Hall coefficient does not show a jump provided the nesting is not
perfect \cite{Bazaliy2004}.  In fact, for the prototypical
material of itinerant antiferromagnetism, elemental Chromium, the
Hall coefficient seems to evolve smoothly through the QCP. This is
seen from studies in which the magnetism is suppressed by
hydrostatic pressure or through a combination of Vanadium
substitution with pressure \cite{Lee2004,Jaramillo2010}.

We can rule out that the discontinuity in the Hall effect is
related to a breakdown of the low-field limit as it might occur at
a field induced SDW QCP. It was pointed out theoretically that for
a magnetic-field driven QCP the non-zero critical field can give
rise to a small discontinuity in the magnetotransport coefficients
even in the case of an SDW QCP \cite{Fenton2005} (see section
\ref{sec:spinfluct}). If the energy scale of the vanishing
magnetic order becomes smaller than the energy scale associated
with the Lorentz force the weak field limit breaks down and this
may lead to a non-linearity in magnetotransport. In the Hall
coefficient the non-linearities of the conductivities can add to a
jump which would be smeared by disorder. Up to date, however, such
a jump in \RH\ at an SDW QCP has not been seen in any other
quantum critical system. For the case of \YRS\ we exclude this
effect by using the crossed-field geometry. This setup allows to
apply a vanishingly small Hall field $B_1$ while the material is
tuned to its QCP via a critical value of the field $B_2$. Thus,
the energy scale associated with the Lorentz force (arising from
$B_1$) is negligibly small avoiding the above described breakdown
of low-field magnetotransport. In fact, the Hall resistivity
remains linear even at the critical value of the tuning field
$B_2$ as can be seen from Fig.~\ref{fig:RhoH}. In addition, the
signatures due to low-field breakdown are expected to change with
residual resistivity. Consequently, the fact that the Hall
crossover is observed for two samples with residual resistivities
differing by a factor two discards this scenario. Furthermore,
with the width and position of the crossover of the single-field
experiment being identical to the crossed-field Hall results even
the single-field results seem to be unrelated with a breakdown of
low-field magnetotransport.

The interpretation of the results from Hall effect measurements in
terms of a Kondo breakdown are corroborated by thermal transport
\cite{pfau12} and thermodynamic measurements \cite{geg07}. In
particular, the isothermal, linear magnetostriction,
$\lambda_{[110]}(B,T) = \partial \ln L /
\partial B$ (where $L$ is the length along the [110] direction
and $B\!\perp\! c$), has been measured down to temperatures of 20
mK. Interestingly, $\lambda_{[110]}(B,T)$ can be fit by a
crossover function analogous to eq.~(\ref{eq:crossover_function})
and yields crossover field values $B_0^{\lambda}$ in good
agreement with those obtained from Hall effect measurements
\cite{geg07}. We note that all the low-temperature measurements on
YbRh$_2$Si$_2$ so far (including, {\it e.g.}, magnetostriction,
heat capacity and susceptibility) indicated a second order phase
transition at $B_0$.

Very recently, Hall effect measurements at {\em high} magnetic
fields have been conducted on YbRh$_2$Si$_2$ \cite{pfau12un}.
Beyond 12~T, the Hall resistivity $\rho_{xy}$ is independent of
temperature within the measured range 0.05~K $\le T \le$ 2.75~K.
This is consistent with a suppression of the Kondo state at these
high fields. The Hall coefficient $R_{\rm H}$ exhibits several
small features which are in good agreement with changes in the
topology of the calculated Fermi surface \cite{zwick12}. The
overall increase of $R_{\rm H}$ in the field range between about
$6-10$ T may signal the de-renormalization of the Kondo
quasiparticles. Most importantly, renormalized band structure
calculations \cite{zwick12} indicate the crossing of the Fermi
level of one of the spin-split bands at above 10~T which is in
line with our Hall results, and a Lifshitz transition at these
high fields. We emphasize that---in contrast to these results at
{\em high} magnetic fields---a Lifshitz transition at {\em small}
fields $B_0$, {\it i.e.} close to the QCP, is unlikely,
specifically in view of the observed linear increase of the width
of the Hall with temperature, FWHM $\propto T$, {\it cf.}
Fig.~\ref{fig:FWHM_YRS}.

Clearly, it is desirable to support the results of the Hall
measurements by other, possibly more direct, probes of the Fermi
surface in YbRh$_2$Si$_2$, like ARPES or dHvA effect measurements.
However, given the low temperatures and small magnetic fields at
which consequences of the QCP are expected to be observed a
confirmation of the Kondo breakdown scenario by these methods is
not suitable at present. Nonetheless, ARPES measurements have been
conducted down to about 10 K in zero magnetic field and clearly
demonstrated the (dynamical) hybridization between Yb 4$f$ and
conduction electrons \cite{danz2011}. Moreover, dHvA effect
measurements \cite{rourke08} indicated the presence of a large
Fermi surface at about 8 T, {\it i.e.}, in the Fermi liquid regime
of Fig.~\ref{sdwlocal}.

\subsection{Comparison to other candidates for Kondo breakdown}
\label{sec:HFcomp} A continuous evolution of the Fermi surface was
identified in CeRu$_2$Si$_2$ with the help of Hall effect
measurements \cite{Daou2006}. In fact, dHvA measurements suggested
a localization of $f$ electrons at the metamagnetic quantum phase
transition in resemblance to the Kondo breakdown scenario. Hall
effect measurements down to below 10 mK, however, revealed a
Lifshitz transition as the underlying mechanism: One of the
spin-split branches of a Fermi surface sheet shrinks to a point at
the metamagnetic critical field giving rise to signatures in both
the resistivity and the Hall resistivity. At high temperatures a
pronounced maximum is observed in the Hall resistivity which
shrinks as the temperature is lowered and transforms to a kink at
lowest temperatures. Similar behavior might be present in UPt$_3$
where only a maximum at elevated temperatures was observed
\cite{Kambe1999}. The small kink as observed in CeRu$_2$Si$_2$
would not be resolved in the high-field measurements on UPt$_3$.
Importantly, the slope of the Hall resistivity, equivalent to the
differential Hall coefficient, approaches identical values in
CeRu$_2$Si$_2$ on either side of the transition which basically
rules out a Fermi surface reconstruction. This is to be contrasted
with the data for \YRS\ which show different values of the
differential Hall coefficient on either sides of the QCP. In
addition, such a Lifshitz transition would lead to drastic changes
in the signatures in the Hall effect upon variation of the
residual resistivity which is in disagreement with the fact that
identical signatures are observed for \YRS\ in two samples with
residual resistivities differing by a factor of almost 2.

A more complicated Lifshitz transition was recently suggested to
explain the measured Hall effect of \YRS\ \cite{Hackl2010}. Here,
it is assumed that there is a hitherto unknown ultra-narrow band
in the band structure and that one of its Zeeman-split components
sinks below (or above) the Fermi energy by the critical field.
However, this proposal invokes an unphysically small bandwidth of
about 5~$\mu$eV and is unlikely to be relevant to YbRh$_2$Si$_2$.

Field induced quantum phase transitions were also studied in two
other heavy fermion compounds by means of Hall effect
measurements, namely URu$_2$Si$_2$ \cite{oh07} and YbAgGe
\cite{Budko2005a}. In URu$_2$Si$_2$ the Hall resistivity nicely
tracks the different phases traversed when increasing the magnetic
field to beyond 35~T. For each of the phases a proportionality of
the Hall resistivity to the magnetic field corresponding to a
constant Hall coefficient reflects the different Fermi surface
configurations. The transition between these phases occurs
abruptly indicating discontinuous phase transitions. In fact, the
Hall effect shows no signature at the putative QCP underlying the
field-induced phase which surrounds it as a dome. This emphasizes
the unique role of \YRS\ in which the QCP is not masked by an
emergent phase providing the chance to study the Hall effect
across the QCP.

\begin{figure}
\includegraphics[height=9.6cm,clip]{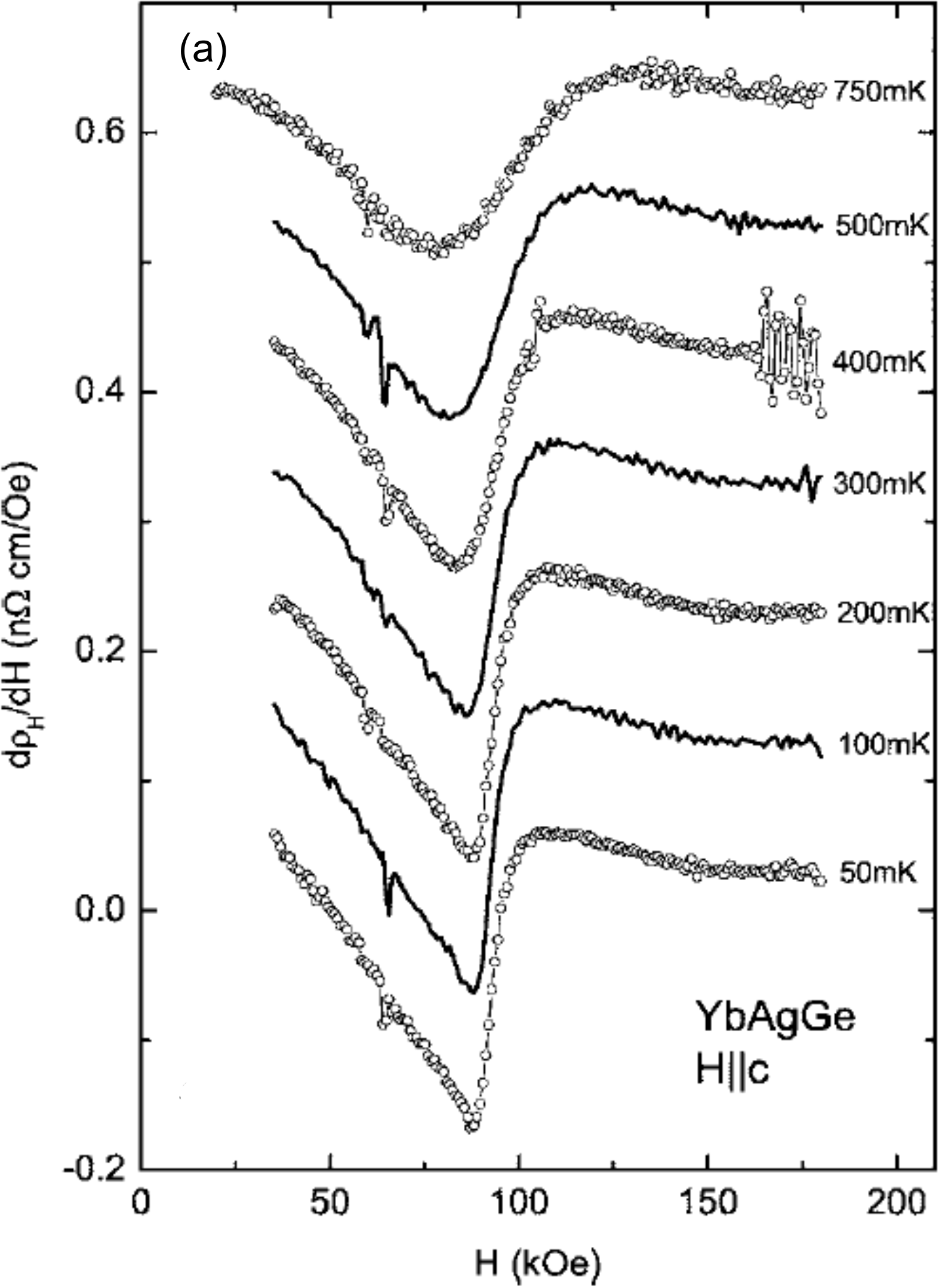}
\hfill
\includegraphics[height=9.6cm]{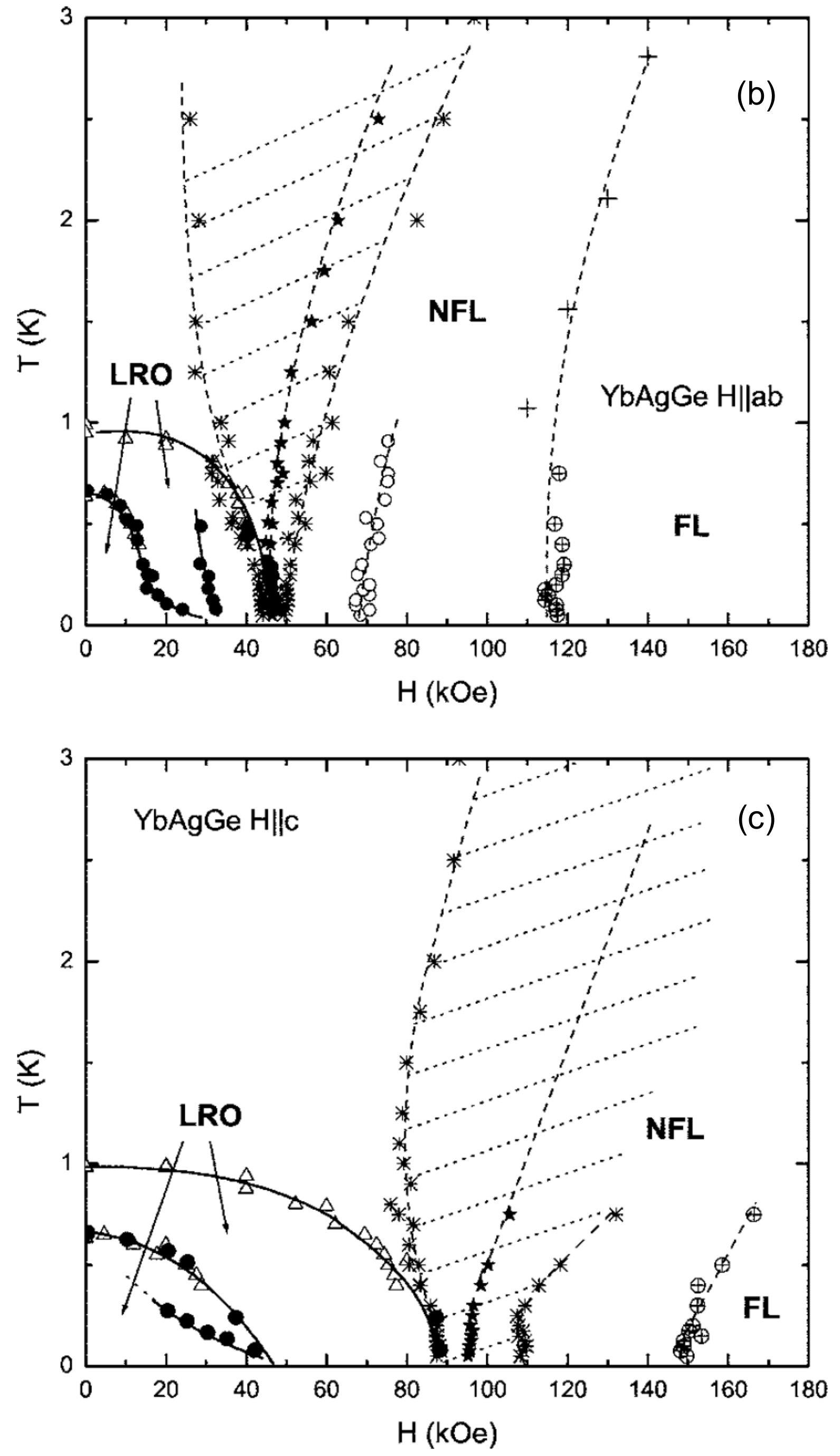}
\caption{(a) Selected isotherms of the differential Hall
coefficient as a function of magnetic field for YbAgGe after
Ref.~\cite{Budko2005a}. Magnetic field is oriented parallel to the
crystallographic $c$ axis. Except for $T=50$\,mK curves are
shifted by 0.1\,n$\Omega$cm/Oe increments for clarity. Phase
diagram for fields perpendicular (b) and parallel (c) to the $c$
axis. LRO denotes the long range magnetically ordered phases.
Stars represent the position of the crossover in the differential
Hall coefficient and crosses depict the onset and end of the Hall
crossover, respectively. They define the width of the Hall
crossover which is emphasized by the shaded region. NFL and FL
correspond to the non-Fermi-liquid and Fermi-liquid regime as
deduced from further features in the differential Hall
coefficient. Reprinted figure with permission from S.L. Bud'ko
{\it et al.}, Physical Review B {\bf 72}, 172413 (2005)
\cite{Budko2005a}. Copyright \copyright\,(2005) by the American
Physical Society.} \label{fig:YbAgGe}%
\end{figure}

In YbAgGe, the differential Hall coefficient shows a crossover
like structure similar to that observed in \YRS. Exemplary curves
are reproduced in Fig.~\ref{fig:YbAgGe}(a). The position of the
crossover is tracked in the phase diagram shown in
Figs.~\ref{fig:YbAgGe}(b) and (c) for two different field
orientations. It is apparent that the crossover position is linked
to the suppression of long range magnetic order, similar to what
was observed in \YRS. However, this seems to be different for the
two different orientations. For the field perpendicular to the
crystallographic $c$ direction the position of the Hall crossover
seems to converge directly to the magnetic instability whereas for
fields along the $c$ axis a finite distance appears to persist in
the extrapolation to zero temperature. This is in contrast to the
results on \YRS\ which show identical behavior for both field
orientations ({\it cf.} Fig.~\ref{fig:PDRH_YRS}). The main
difference to \YRS\ is, however, that for YbAgGe a finite width of
the crossover persists down to zero temperature. This is most
prominent for fields along the $c$ axis as can be seen from the
finite range of the shaded area at lowest temperatures. Like the
position of the crossover also the width is different for
different orientations of the magnetic field and can not be scaled
by the magnetic anisotropy, in contrast to the findings for \YRS.
In summary, these findings represent more complex behavior and, in
particular, the finite width favors an SDW scenario for YbAgGe.

Quite recently, the Kondo breakdown scenario has been suggested to
be applicable to the cubic heavy fermion compound
Ce$_3$Pd$_{20}$Si$_6$ \cite{paschen12}. Although the material
crystallizes in a rather complex structure the magnetic Ce ions
are arranged on two cubic lattices and in cage-like environments
typical for the clathrates \cite{deen10}. The results of the
single-field Hall effect measurements can nicely be analyzed by
eq.~(\ref{eq:crossover_function}) indicating similarities to the
Kondo breakdown in YbRh$_2$Si$_2$, {\it cf.}
Fig.~\ref{fig:PDRH_YRS}. In particular, the crossover in the Hall
resistivity also appears to sharpen to a discontinuity in the
extrapolation to zero temperature indicating a Fermi surface
reconstruction at the critical field of the antiferromagnetic
state. In contrast to \YRS, however, the Fermi surface
reconstruction does \textit{not} coincide with the transition into
a paramagnetic ground state. Rather, a second ordered phase with
higher transition temperature and extending to high magnetic field
encloses both the antiferromagnetic phase and the Fermi surface
reconstruction. The finite-temperature crossover line associated
with the change of the Fermi surface crosses the phase boundary of
this second phase at finite temperature in the $B$--$T$ phase
diagram of Ce$_3$Pd$_{20}$Si$_6$.

The different behaviors of YbRh$_2$Si$_2$ and
Ce$_3$Pd$_{20}$Si$_6$ have been discussed in the framework of a
global phase diagram in which the degree of quantum fluctuations
of the local-moment magnetism frustration is plotted against the
normalized Kondo coupling \cite{si06,coleman10,si10}. Both the
quantum fluctuations and the Kondo effect weaken magnetic order
whereas solely the Kondo coupling is predicted to induce a
Fermi-surface reconstruction.

\begin{figure}
\centering\includegraphics[width=6.0cm,clip]{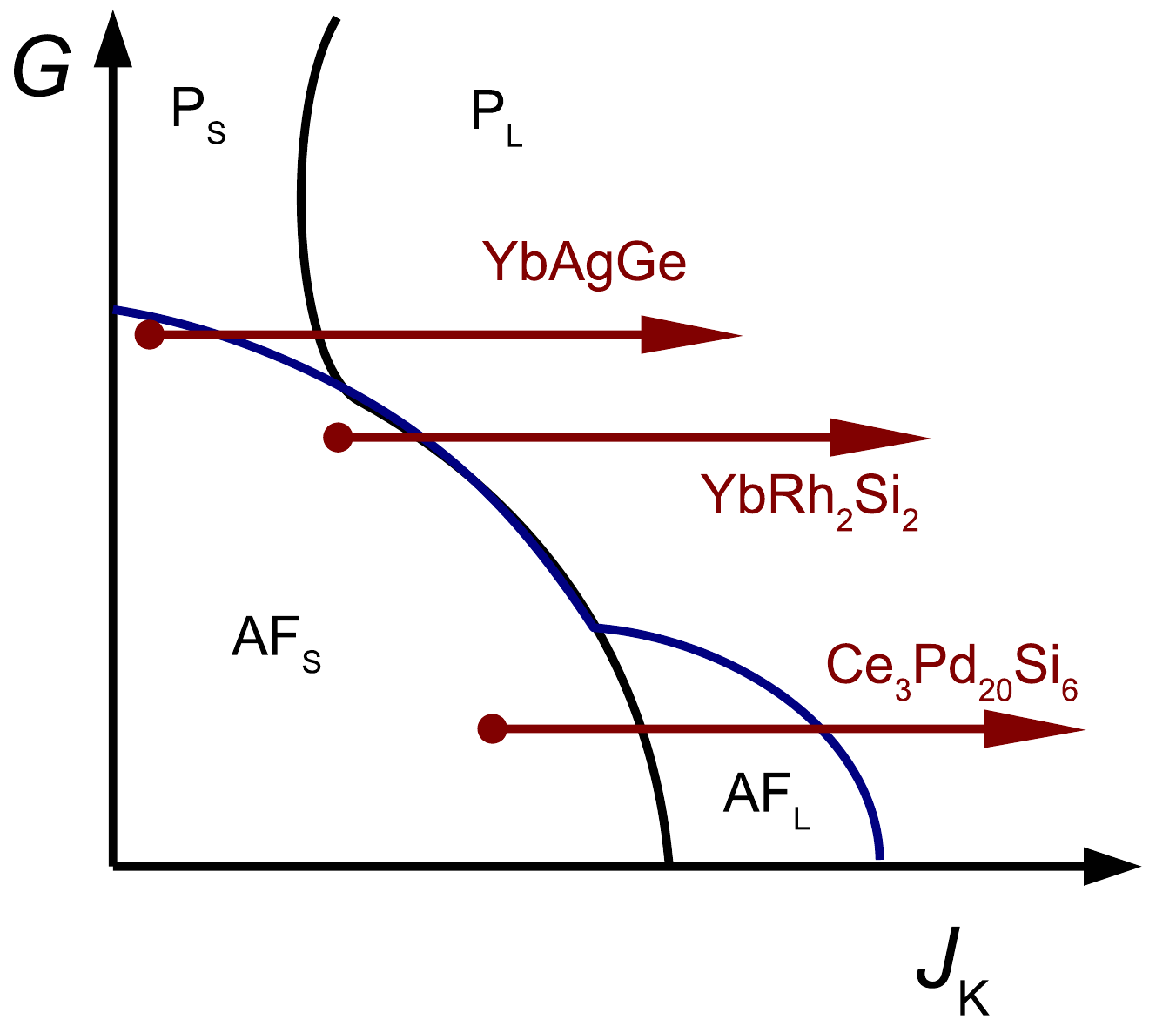}
\caption{Global phase diagram for heavy-fermion systems. Both
magnetic frustration ($G$) and Kondo coupling ($J_{\mathrm K}$)
suppress the antiferromagnetic (AF) ground state giving way to a
paramagnetic phase (P). The Fermi surface reconstruction
associated with the breakdown of the Kondo effect marks the
crossover from the large Fermi surface (subscript L) which
includes the $f$ electrons to the small Fermi surface (subscript
S) excluding $f$ electrons. The heavy-fermion compounds YbAgGe,
\YRS, and Ce$_3$Pd$_{20}$Si$_{6}$ are placed according to the
observed relation between magnetic order and signatures in the
Hall effect. This is justified by structural arguments (see text)
giving rise to different levels of quantum frustration. Figure
based on Q. Si, Phys. Stat. Solidi B {\bf 247}, p. 476, 2010
\cite{si10}. Copyright (2010) by the WILEY-VCH.}
\label{fig:global}
\end{figure}

Due to its cubic structure Ce$_3$Pd$_{20}$Si$_6$ is expected to
exhibit rather three-dimensional, {\it i.e.} small, quantum
fluctuations. Therefore, it is naturally settled in the
low-frustration regime of the global phase diagram where the Kondo
breakdown occurs within the ordered phase, as sketched in
Fig.~\ref{fig:global}. However, it is important to scrutinize this
picture further by investigating the NFL behavior associated with
the Kondo breakdown and clarifying the nature of the
high-temperature ordered phase in Ce$_3$Pd$_{20}$Si$_6$.

\YRS\ has a tetragonal structure and may, therefore, exhibit more
two-dimensional fluctuations. This increase in the influence of
fluctuation upon lowering the dimensionality would place \YRS\ at
a somewhat higher frustration level than Ce$_3$Pd$_{20}$Si$_6$. In
this regime of the global phase diagram a coincidence of the Kondo
breakdown and the magnetic-to-paramagnetic transition is predicted
just as observed in \YRS. The separation of the Kondo breakdown
and the magnetic transition observed in \YRS\ under chemical
\cite{Friedemann2009} and hydrostatic \cite{Tokiwa2009a} pressure
indicate that in \YRS\ the frustration can be fine-tuned thereby
changing its level of frustration and consequently its vertical
position in the global phase diagram.

The hexagonal structure of YbAgGe makes this material a promising
candidate for further increase of frustration. This would settle
YbAgGe in the regime of the global phase diagram where the
magnetism is suppressed before the Kondo breakdown is reached. The
fact that the signature in the Hall effect for fields parallel to
the $c$-direction of YbAgGe is separated from the magnetically
ordered phase could be in accordance with this prediction.
However, a more detailed analysis of the temperature evolution of
the signatures in the Hall effect may possibly allow to scrutinize
whether this system undergoes a Fermi surface reconstruction.

\subsection{Hall effect and scaling behavior} \label{sec:scal}
The proportionality to temperature of the FWHM of the Hall
crossover in \YRS, {\it cf.} Fig.~\ref{fig:FWHM_YRS}, allows
further insight into the nature of the quantum criticality. In
order to understand the relevance of this proportionality we have
to recall what the Hall coefficient is sensitive to. For the given
Fermi liquid ground states on either side of the QCP, the Hall
coefficient probes the Fermi surface. $\RH^0$ is assigned to the
small Fermi surface on the low-field side, and $\RH^{\infty}$ is
assigned to the large Fermi surface on the high-field side in
agreement with renormalized band structure calculations
\cite{Friedemann2010c}. At zero temperature, the transition from
one Fermi surface to the other is sharp leading to a step in the
Hall coefficient. This is in accordance with the experimental
observation of a vanishing FWHM. At finite temperature---in the
absence of a phase transition---a crossover from one Fermi surface
to the other is expected. This is reflected by the crossover in
the Hall coefficient. The width of this crossover is determined by
the single-electron excitations across the Fermi surface. In the
quantum critical regime they obey a linear-in-temperature
relaxation rate as a result of the underlying $E/T$ scaling
expected for the Kondo-breakdown QCP \cite{Friedemann2010b}. These
relaxation processes allow a continuous connection of the
different Fermi surfaces at finite temperatures. As the Hall
crossover appears to mirror the Fermi surface crossover, the
linear-in-temperature FWHM hints towards an linear-in-temperature
relaxation rate of the single electron states and thus towards an
energy over temperature scaling of the electronic excitations
\cite{Friedemann2010b}. A conventional (3D SDW) QCP, by contrast,
is predicted to obey $E/T^x$ scaling with $x > 1$ which would lead
to a super-linear $T$ dependence of both the relaxation rate and
the Hall crossover width.

The systematic study of the Hall effect and magnetoresistivity in
\YRS\ provides currently the best evidence for a Fermi surface
reconstruction at a QCP. The discontinuity of the Hall coefficient
in the limit of zero temperature reflects this change of the
electronic structure. Moreover, the scaling analysis of the Hall
crossover width at finite temperatures sheds further light on the
unconventional nature of the quantum criticality in \YRS. The
proportionality of the FWHM to temperature observed over almost
two decades hints towards a linear-in-temperature relaxation rate
of the single electron states which itself is linked to $E/T$
scaling of the electronic excitations. Consequently, \YRS\ seems
to be the first material where both the fundamental signatures of
the breakdown of the Kondo effect, a collapse of the Fermi surface
and an $E/T$ scaling of the critically fluctuating Fermi surface
have been observed. This indicates that the macroscopic
scale-invariant fluctuations arise from the microscopic many-body
excitations associated with the collapsing Fermi surface. The
linkage between microscopics and macroscopics provided by the
fundamental $E/T$ scaling is expected to be a keystone to
understand the physics of correlated materials. This might apply
to a wide class of materials with abnormal finite temperature
behavior for which the evolution of the Fermi surface as a
function of a control parameter plays a central role like, {\it
e.g.}, in the case of the high-temperature cuprate
superconductors.

The crossover in the Hall effect defines a new energy scale called
$T^{\star}(B)$ which is confirmed by other transport measurements
as well as by thermodynamic measurements \cite{geg07}. This energy
scale vanishing at the QCP is reminiscent of the additional local
energy scale proposed in the Kondo breakdown which relates to the
break up of the quasiparticles. Interestingly, sufficient changes
in the unit cell volume lead to a detachment of the
antiferromagnetic QCP from the $T^{\star}(B)$ line. This was
realized by isoelectronic substitution on either Rh
\cite{Friedemann2009} or Si \cite{Custers2010} sites, hydrostatic
pressure \cite{Tokiwa2009a} or a combination of hydrostatic and
chemical pressure \cite{Nicklas2010}. For lattice expansion, the
$T^{\star}(B)$ line is separated from the magnetic QCP with the
intermediate regime featuring non-Fermi-liquid behavior in a
finite range of the tuning parameter. It still has to be tested
whether the scaling behavior of the Hall crossover persists under
lattice changes which might help to understand the nature of the
emergent non-Fermi-liquid phase. While the existing experiments
have indicated a finite range of chemical pressure in which the
two concur \cite{Friedemann2010d}, this issue deserves further
investigation, especially under external pressure. It is also
important to examine how this relates to the unconventional
critical scaling seen in both the specific heat and Hall effect
\cite{Friedemann2010f}.

\section{Hall effect in systems with 115 type of structure}
\label{sec:fluc}
\subsection{Influence of magnetic fluctuations on superconductivity}
\label{sec:flucsup} As stated in section \ref{sec:HF}, a
remarkable manifestation of quantum many body physics has been the
gradual realization that the apparently inimical phenomena of
superconductivity and magnetism could be intimately related. At
face value, the coexistence of these two diverse electronic ground
states seems unlikely, since the prerequisites for these states
(itinerant and localized electrons, respectively) appear to be
contradictory. Moreover, it was also shown \cite{gin57} that even
small amounts of magnetic impurities were detrimental for the
stability of the superconducting condensate in the conventional
(BCS) superconductors known at the time. Experimentally, this was
supported by early investigations using binary rare earth alloys
of the form La$_{1-x}R_x$, ($R$ denotes other members of the
Lanthanide series), where the suppression of superconductivity in
La appeared to be directly correlated with the 4$f$ spin
configuration of the dopant \cite{mat59}. The effect of impurities
has been treated within the theory by Abrikosov and Gor'kov
\cite{abrik}. At a microscopic level, the drastic suppression of
superconductivity due to the presence of a magnetic entity can be
understood to occur as a result of a ``pair-breaking'' interaction
that arises from the interaction of the magnetic moment ($\mu$)
with the spin of the conduction electron (via the Zeeman
interaction) or that of its vector potential with the momenta
($\pmb p$) of the electrons. The influence of an internal magnetic
field generated by a system with (or close to) magnetic order on a
BCS superconductor was examined by Berk and Schrieffer
\cite{ber66}, who showed that strong (ferro)magnetic exchange
would suppress the superconducting ground state. Using the example
of Pd, which was known to have an exchange-enhanced spin
susceptibility, it was suggested that only a non-superconducting
state was feasible for reasonable values of phonon interaction
strengths \cite{ber66}.

A relatively simple manner in which magnetism and
superconductivity can coexist is by spatial separation, wherein
the magnetic sublattice is relatively decoupled from the sea of
conduction electrons. This is the case in a variety of systems
like the Chevrel phases (of the form $M_x$Mo$X_8$; where $M$ is a
rare earth metal, and $X$ is a chalcogenide) \cite{fis78}, the
rhodium borides \cite{mat77,pri88}, the ruthenocuprates
\cite{kla08} and the borocarbides \cite{gup08}. However, the heavy
fermion superconductors are distinct in the sense that here
magnetism and superconductivity both involve the same set of $f$
electrons, and it was the discovery of heavy fermion
superconductivity in CeCu$_2$Si$_2$ which unambiguously
demonstrated that superconductivity was clearly sustainable in an
inherently magnetic environment \cite{ste79}. Moreover, it was
also obvious that these strongly renormalized heavy electrons were
a necessary condition for superconductivity, since the homologous
LaCu$_2$Si$_2$ where the 4$f$ electrons, and consequently, the
heavy fermion state were absent did not exhibit a superconducting
ground state. Long-range magnetic order is in close vicinity to
the superconductivity in CeCu$_2$Si$_2$, as is evidenced by the
fact that a magnetically ordered state emerges in homogeneous
CeCu$_2$Si$_2$ samples with very small Cu deficit \cite{ste05j}.
This relation between unconventional superconductivity and
magnetism is now well recognized, and the heavy fermion systems
have turned out to be an interesting testing ground where such
emergent behavior is observed. Within the same ThCr$_2$Si$_2$
structure, a number of other Ce based systems have now exhibited
unconventional superconductivity, thought to be mediated by
antiferromagnetic fluctuations. For instance, CeNi$_2$Ge$_2$
exhibits ambient pressure superconductivity with a $T_{\rm c}
\approx 0.3$ K \cite{gro00}, whereas ambient-pressure magnetic
order gives way to pressure-induced superconductivity in
CePd$_2$Si$_2$ ($T_{\rm N} \approx 10$ K)and CeRh$_2$Si$_2$
($T_{\rm N} \approx 36$ K) \cite{mat98,mov96}. The closely related
CeCu$_2$Ge$_2$ was also shown to be superconducting at external
pressures of the order of 7 GPa \cite{jac92}, though the physics
of this system, like CeCu$_2$Si$_2$, is complicated by the
presence of two superconducting domes \cite{yuan} that possibly
arise from different underlying mechanisms: (i) from a magnetic
instability as in the above mentioned systems, and (ii) from
charge density fluctuations due to a valence transition in
Ce$^{3+}$ \cite{onod02,hol04}. A number of U-based systems also
exhibit unconventional superconductivity, notable amongst them
being UPt$_3$ \cite{ste84} and URu$_2$Si$_2$ \cite{pal85,schl86}.
In UPt$_3$ a superconducting condensate was seen to emerge at
$T_{\rm c} \approx 0.5$ K from within a magnetically ordered
ground state which sets in at around 17.5 K. Though initial
analysis of the specific heat was used to speculate on the
possibility of ferromagnetic spin fluctuations in this system
\cite{vis84,bro86}, the antiferromagnetic nature of these
correlations was established by the use of neutron scattering
measurements \cite{aep85,aep87}. A low temperature superconducting
ground state is also observed to emerge from within an ordered
magnetic state in the hexagonal UPd$_2$Al$_3$ and UNi$_2$Al$_3$,
with $T_{\rm c}$-values of 2 K and 1.1 K, respectively
\cite{gei91,gei91b}. A valuable addition to this class of
materials has been the Ce$M$In$_5$ ($M$ = Rh, Ir or Co) family of
compounds, discussed in detail in the subsequent sections.

In all the systems mentioned above, superconductivity is observed
in the presence (or in the vicinity) of long-range
antiferromagnetic order. However, in a small number of heavy
fermion systems, the superconducting condensate is known to form
in the presence (and through the mediation) of ferromagnetic
fluctuations. Needless to say, this places more stringent
conditions on the symmetry of the superconducting order parameter,
since the spin-singlet state (in which the pairing quasiparticles
have opposite spins) is ruled out. In unconventional
superconductors in the presence of strong antiferromagnetic
fluctuations, stabilization of these spin-singlet Cooper pairs
with a non-zero angular momentum is most likely in a $d$-wave
state. It is this aspect of heavy fermion superconductivity that
binds it with the superconducting cuprates. However, when Cooper
pair formation is mediated by ferromagnetic fluctuations, the
spin-triplet state is most feasible \cite{mac00} in which the
Cooper pairs have an effective odd angular momentum (analogous to
that seen in the superfluid $^3$He phase). The existence of
superconductivity in the vicinity of ferromagnetic order was first
demonstrated in UGe$_2$, where application of external pressures
greater than 1 GPa was seen to result in the stabilization of a
low-temperature superconducting ground state from within the
ferromagnetically ordered state \cite{sax00}. The orthorhombic
URhGe and UCoGe also belong to the same class, with the
superconducting condensate being stabilized from within the
ferromagnetic ordered state at ambient pressures
\cite{aok01,huy07}.

\subsection{Interplay of magnetism and superconductivity in
Ce$M$In$_5$} \label{sec:fluc115} The Ce$M$In$_5$ (where $M$ = Co,
Rh or Ir) family of heavy fermion systems have emerged as a focus
of intense investigations since the turn of the century.
Structurally, these systems can be looked upon as layered variants
of the CeIn$_3$ system, and can be generally described as
Ce$_n$$M_m$In$_{3n+2m}$, where $n$ layers of CeIn$_3$ alternate
with $m$ layers of $M$In$_2$ \cite{gri79}. CeIn$_3$ is an ambient
pressure antiferromagnet, with a transition temperature of 10 K.
On the application of pressure, non-Fermi liquid behavior emerges
near 2 GPa, where long-range antiferromagnetic order is suppressed
to zero temperatures, and superconductivity with a maximum $T_{\rm
c} \sim 0.2$ K emerges \cite{mat98,wal97}. Current interest in the
Ce$M$In$_5$ family of systems (equivalent to the $n = m = 1$
variants of the general form described above) was sparked off with
the observation of pressure-induced superconductivity in
CeRhIn$_5$, with a superconducting transition temperature $T_{\rm
c} \approx 2.1$ K at 2.1 GPa \cite{heg00}. Subsequently,
ambient-pressure superconductivity was discovered in the Ir- and
Co-based systems \cite{pet01a,pet01b}. Superconductivity has also
been discovered in the $n$ = 2, $m$ = 1 variants, with
Ce$_2$RhIn$_8$ being superconducting at $T_{\rm c} \sim 1.1$ K at
1.6 GPa \cite{nic03} and the systems Ce$_2$CoIn$_8$ and
Ce$_2$PdIn$_8$ becoming superconducting at $p = 0$ and below
$T_{\rm c} \approx 0.4$ K \cite{che03} and $T_{\rm c} = 0.68$ K
\cite{kacz09}, respectively. Recently, the first Ce-based
(pressure induced) heavy-fermion superconductor of the type $n$ =
1 and $m$ = 2 was discovered: CePt$_2$In$_7$ ($T_{\rm c} = 2.1$ at
$p \approx 3.1$ GPa) \cite{bau10}.

The flurry of attention is primarily due to the intricate
interplay between magnetism and superconductivity which this
family of heavy fermion metals have demonstrated. For instance,
CeCoIn$_5$ is an ambient-pressure superconductor with a $T_{\rm c}
\approx 2.3$ K which is the highest reported value among all Ce
based heavy fermion systems to date. Magnetism lies in close
proximity to superconductivity in this system, and a magnetic
field-induced quantum critical point can be approached by the
application of magnetic fields of the order of the superconducting
upper critical field ($\mu_0 H_{\rm c2} \approx$ 5 T)
\cite{pag03}. Several measurements have indicated that the
superconducting gap function has line nodes and is most likely to
have a $d$-wave symmetry \cite{mat06}. Closely related is the
system CeRhIn$_5$, which has an ambient pressure antiferromagnetic
transition at about 3.8 K. Though the application of pressure
suppresses long-range antiferromagnetism, access to the quantum
critical point is hindered by the onset of superconductivity
\cite{heg00}. The presence of this magnetic instability has been
deduced by specific heat measurements which also point towards the
existence of a quantum tetracritical point where four distinct
phases (namely, a non-magnetic phase, a superconducting phase, a
magnetically ordered phase, and a region within which
superconductivity coexists with magnetic order) meet \cite{par06}.
The existence of this magnetic instability at 2.3 GPa is also
reinforced by de Haas-van Alphen measurements indicating an abrupt
change in the band structure \cite{shi05}. The other
ambient-pressure superconductor in this series, namely CeIrIn$_5$,
has remained more enigmatic. For instance, unlike its Co
counterpart, superconductivity in this system appears to be well
separated from the magnetic instability which in the case of
CeIrIn$_5$ is suggested to be metamagnetic in origin \cite{cap04}.
The symmetry of the superconducting gap in CeIrIn$_5$ has also
been a matter of dispute. For instance, measurements of thermal
conductivity indicated that the superconducting gap in CeIrIn$_5$
has a $d_{x^2 - y^2}$ symmetry \cite{kas08} which is similar to
that seen in CeCoIn$_5$ \cite{izawa}. This was in contrast to
earlier reports which had ascribed the superconducting gap of
CeIrIn$_5$ to be a hybrid one with an $E_g$ symmetry \cite{sha07}.
A $d_{x^2 - y^2}$ symmetry of the gap in both these systems would
imply that superconductivity in both the superconducting members
of the Ce-115 family are likely to be mediated by
antiferromagnetic fluctuations in spite of the apparent
differences in the position of the superconducting regime with
respect to the onset of magnetic order.

\begin{figure}
\centering\includegraphics[width=9.6cm,clip]{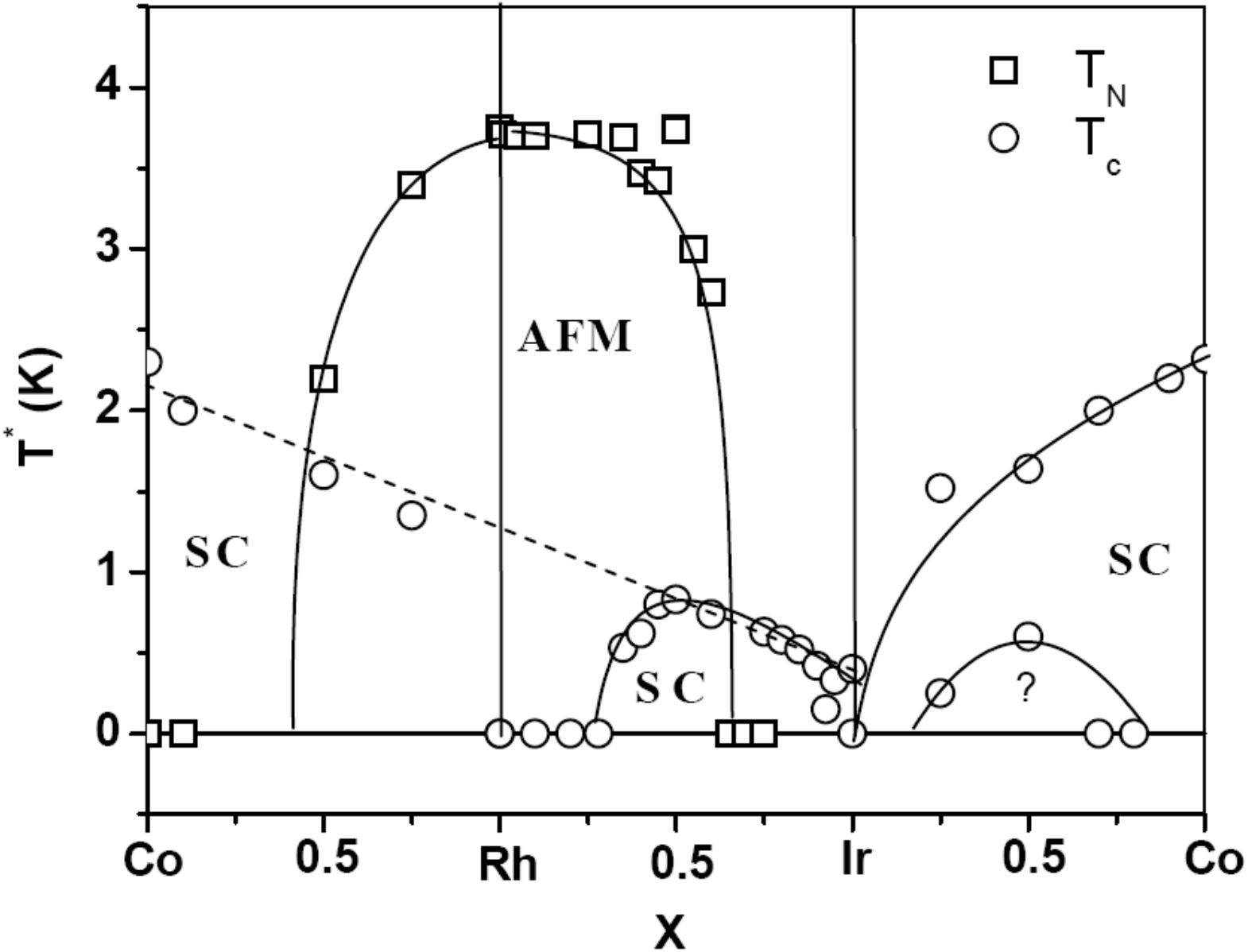}
\caption{\label{fig:pag02} Phase diagram of the system
Ce$M_{1-x}M'_x$In$_5$ ($M, M'$ = Co, Rh, Ir). Reprinted from
Physica B {\bf 312-313}, P.G. Pagliuso {\it et al.}, p.129
\cite{pag02}. Copyright \copyright\, (2002) with permission from
Elsevier.}
\end{figure}

The possible difference between superconductivity in the Co and Ir
systems is clearly reflected in the phase diagram of solid
solutions of Ce$M_{1-x}M'_x$In$_5$($M, M'$ = Co, Rh, Ir) as shown
in Fig.~\ref{fig:pag02}, with a possible
superconductor-superconductor critical point at
CeRh$_{0.1}$Ir$_{0.9}$In$_5$ \cite{pag02}. This difference is also
obvious in doping studies which show that antiferromagnetic order
emerges cleanly from within the superconducting ground state in
CeCo(In$_{1-x}$Cd$_x$)$_5$, whereas these two ground states appear
to be separated from each other in the CeIr(In$_{1-x}$Cd$_x$)$_5$
series of systems \cite{pah06}. An overview of some of the
physical properties of the members of the Ce$M$In$_5$ family
appears in Ref.~\cite{sar07}.

\subsection{Hall effect measurements on Ce$M$In$_5$ systems}
\label{sec:Hall115}
\subsubsection{Scaling relations in Hall effect} \label{sec:comp}
Considering the fact that the Ce$M$In$_5$ systems have been
extensively investigated in the recent past, it is not surprising
that a substantial amount of Hall data exists in these systems.
\begin{figure}
\centering\includegraphics[width=13.8cm,clip]{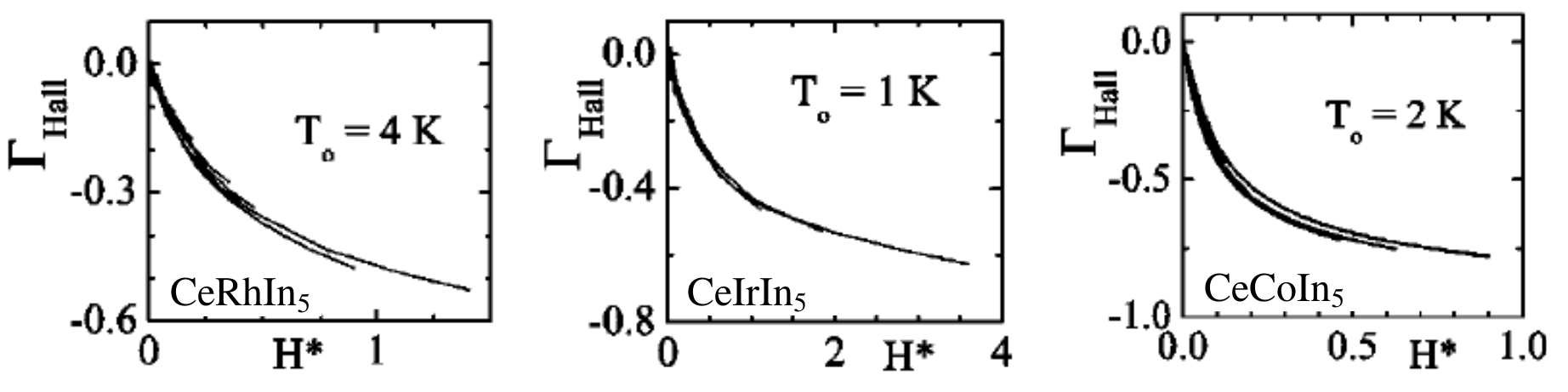}
\caption{\label{fig:GammaH} Scaling of the relative change in Hall
coefficient ($\Gamma_{\rm Hall}$) as a function of the transformed
field parameter $H^*$ for the Ce$M$In$_5$ family. Reprinted figure
with permission from M.F. Hundley {\it et al.}, Physical Review B
{\bf 70}, 035113 (2004) \cite{hun04}. Copyright \copyright\,(2004)
by the American Physical Society.}
\end{figure}
These systems have also provided overwhelming evidence for the
presence of antiferromagnetic fluctuations, as deduced from Hall
effect measurements. Work by Hundley and co-workers \cite{hun04}
in the high-temperature regime (2~K~$\le T \le$ 325 K)
demonstrated that the field-dependent Hall data in the Ce$M$In$_5$
systems satisfied a scaling relationship that is similar to the
one used in analyzing the single-impurity magnetoresistance data.
Defining the relative change in the Hall coefficient as
\begin{equation}
\Gamma_{\rm Hall}(H) = \left[ R_{\rm H}(H) - R_{\rm
H}(H\!\rightarrow\! 0) \right] / R_{\rm H}(H\! \rightarrow\! 0)
\label{eqGammaH} \,\, ,
\end{equation}
it was demonstrated that isotherms of $\Gamma_{\rm Hall}$ plotted
as a function of a transformed field parameter $H^*$ (defined as
$H^* = H /(T + T_0)^{\beta}$, where $H$ is the applied magnetic
field and $T_0$ and $\beta$ are scaling parameters) could be made
to collapse on top of each other. These scaled data plots, which
were obtained by fixing $\beta =$ 2 and varying $T_0$ are shown in
Fig.~\ref{fig:GammaH}. This scaling analysis yielded $T_0 =$ 4 K,
1 K and 2 K for the Rh, Ir and Co systems, respectively. These
$T_0$ values are in reasonable agreement with those estimated for
the Kondo temperature from the analysis of specific heat data.

The measured Hall coefficient of the afore-mentioned systems were
also expressed as a sum of a skew scattering term and a second one
proportional to the Hall coefficient in the non-magnetic
La$M$In$_5$ systems, using a relation:
\begin{equation}
R_{\rm H}^{\rm (Ce)}(H,T) = R_{\rm H}^{skew}(T) + \alpha_f(H,T)
R_{\rm H}^{\rm (La)}(T) \,\, .
 \label{eq:skew} \end{equation}
Here, $\alpha_f$ was defined as the $f$-electron Hall weighting
function. It was shown that $\alpha_f$ grows anomalously with the
onset of Kondo coherence---the magnitude of which could be
suppressed by the application of a large magnetic
field---demonstrating the influence of antiferromagnetic
fluctuations on the measured Hall response. This line of analysis
bears some similarity to the two-fluid Kondo lattice model
proposed by Nakatsuji and coworkers \cite{nak04}, who modeled the
Kondo lattice system into a component analogous to the Kondo
single impurity phase, and one analogous to a coherent heavy
fermion phase, with the temperature evolution of the system being
determined by a mixing parameter $f(T)$. A key aspect is the fact
that whereas the characteristic energy scale in the Kondo
single-impurity regime is the single ion Kondo temperature $T_K$,
in the heavy fermion fluid it is the intersite coupling
interaction $T^*$. Using La dilution on the Ce site of CeCoIn$_5$,
it was also demonstrated that the condensation of the coherent
heavy fermion component is incomplete at the onset of
superconductivity. The need for incorporating these effects in
combination was of course realized much earlier, for instance in
the analysis of neutron scattering data on the system CeCu$_6$
\cite{aep86} where a Hamiltonian which accounted for the influence
of both the single-impurity Kondo interaction and the RKKY
interactions was used. The momentum-dependent magnetic scattering
observed in this system was used to infer on the presence of
antiferromagnetic correlations which could be quenched by the
application of a magnetic field. The presence of these two
contributions and a similar behaviour was also established for the
material CeRu$_2$Si$_2$ \cite{ros88}.

\subsubsection{CeCoIn$_5$} \label{sec:Co115}
\begin{figure}
\centering\includegraphics[width=12.8cm,clip]{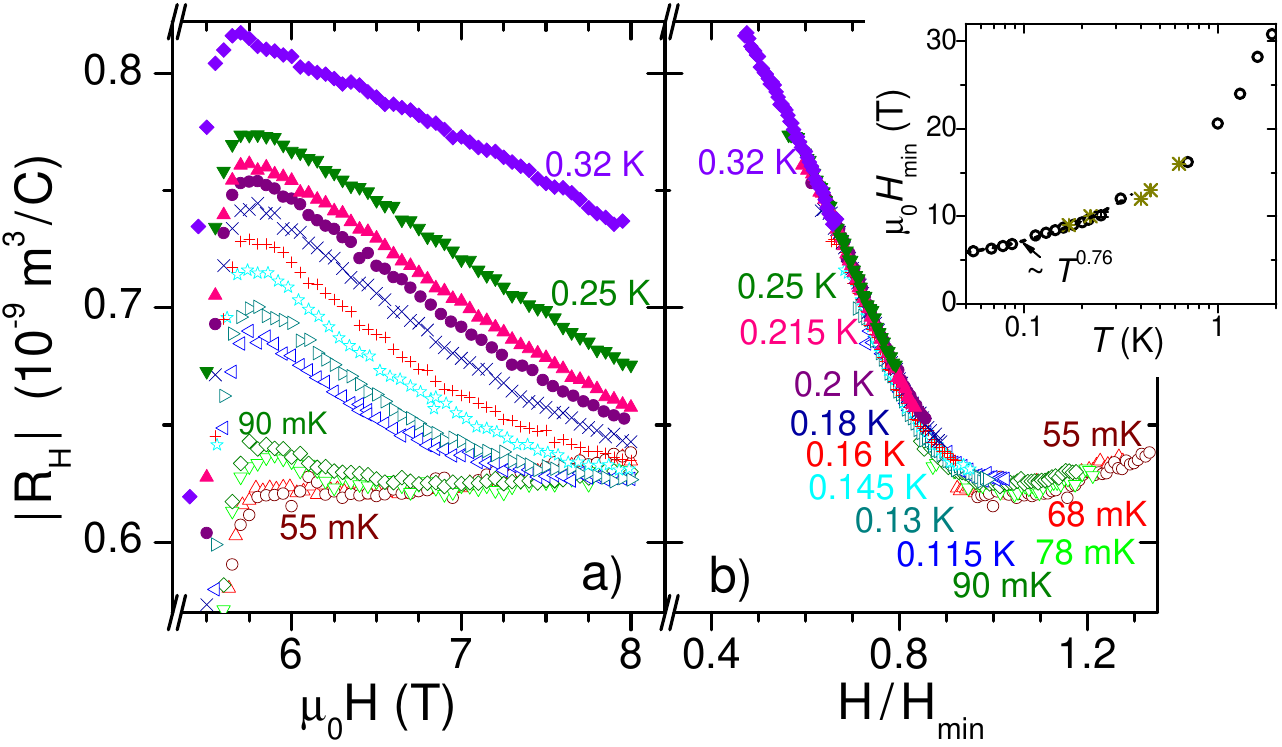}
\caption{\label{fig:scaleCo} (a) $R_{\rm H}$ as measured in
CeCoIn$_5$. (b) Same data scaled by a normalized field $H/H_{\rm
min}$ to obtained maximum overlap. The temperature dependence of
$H_{\rm min}$ is depicted in the inset, with the solid line
representing a power law fit. Reprinted figure with permission
from S. Singh {\it et al.} Physical Review Letters {\bf 98},
p.057001, 2007 \cite{sin07}.}
\end{figure}

\begin{figure}
\centering\includegraphics[width=8.8cm,clip]{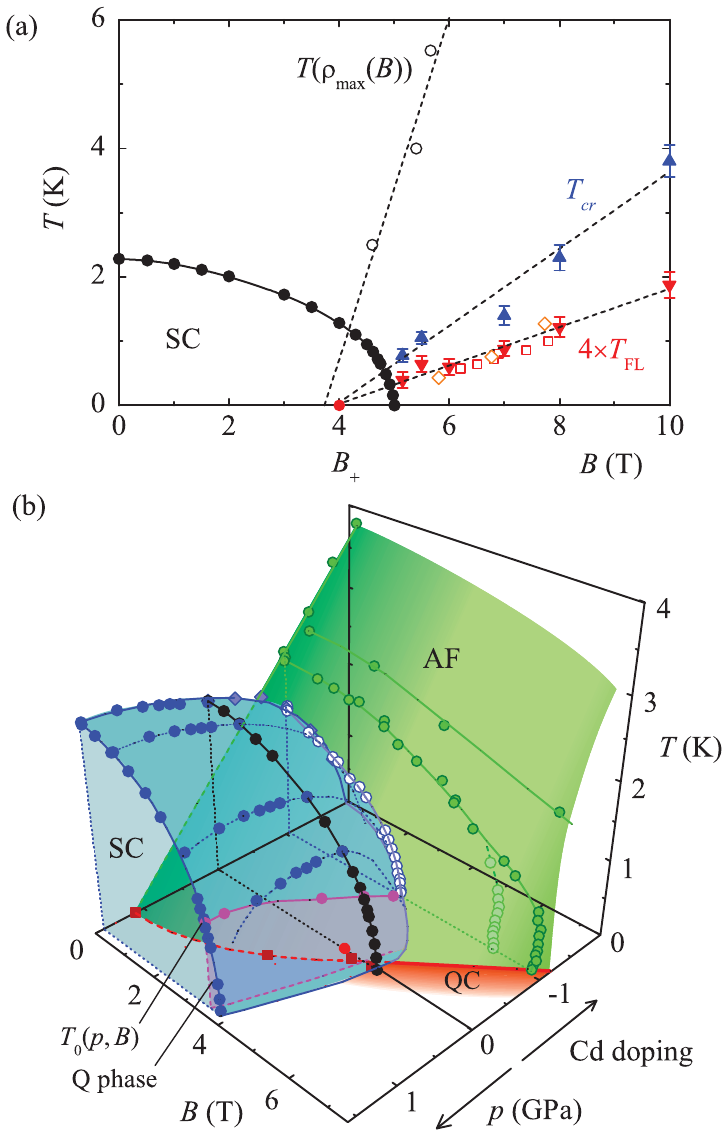}
\caption{\label{fig:BTphase} (a) $B$-$T$ phase diagram of
CeCoIn$_5$ at ambient pressure. $T_{cr}$ indicates the change in
critical behavior \cite{zau11} and $T(\rho_{\rm max}(B))$ refers
to the position of a maximum in the field-dependent resistance
\cite{pag03}. Fermi liquid behavior is inferred below $T_{\rm FL}$
from thermal-expansion measurements in magnetic field \cite{zau11}
(filled triangle), Hall effect \cite{sin07} (open squares) and
resistivity measurements \cite{ron05} (open diamonds). A QCP
inside the superconducting (SC) phase (located at the field $B_+$)
is inferred. (b) $p$-$B$-$T$ phase diagram of CeCoIn$_5$. Shown
are results for superconducting transition (black and blue) and
N\'{e}el temperatures (green) for hydrostatic and negative
chemical pressure in CeCoIn$_{5-x}$Cd$_x$ to demonstrate the
relation of antiferromagnetism (AF) and superconductivity.
Reprinted figure with permission from S. Zaum {\it et al.},
Physical Review Letters {\bf 106}, 087003 (2011) \cite{zau11}.
Copyright \copyright\,(2011) by the American Physical Society.}
\end{figure}

A scaling related to one just described in the previous section
was also observed in the low-temperature regime (0.05 K $\le T
\le$ 5 K) of CeCoIn$_5$ \cite{sin07} where it was shown that the
Hall coefficient could be scaled into a single generic curve by
normalizing their field dependencies by a factor $H_{\rm min}$, as
is shown in Fig.~\ref{fig:scaleCo}. Moreover, the $H_{\rm
min}$-values which were determined experimentally at low
temperatures from a dip in the measured $R_{\rm H}$ (and was
thought to represent the inverse of the effective mobility
$\mu_{\rm eff}$), could be fit reasonably to a power law, {\it
i.e.}, $H_{\rm min} = a + bT^c$. This power law fit as shown in
the inset of Fig.~\ref{fig:scaleCo} yielded $a = 4.0 \pm 0.7$ T, a
result that was taken to be a signature of the fact that the
magnetic instability in this system and the superconducting upper
critical field lie at separate field values. This conjecture has
now been confirmed using measurements of magnetization
\cite{yos-pc}, of the volume thermal expansion \cite{zau11}, of
vortex core dissipation \cite{thu12}, of longitudinal
magnetoresistance \cite{how11} and of field-dependent entropy
\cite{tok12} indicating that the putative QCP lies within the
superconducting regime. The critical field for this
zero-temperature transition was estimated from the former three
types of measurement to be about $(4.1 \pm 0.2)$ T which is in
excellent agreement to the value deduced from the analysis of the
Hall data. This can be seen in the $B$-$T$ phase diagram Fig.\
\ref{fig:BTphase}(a) which was taken from Ref.\ \cite{zau11}.
Here, data from our Hall measurements, from resistance
\cite{ron05} and magnetoresistance $T(\rho_{\rm max}(B))$
\cite{pag03} as well as from thermal expansion measurements
\cite{zau11} are compiled.

\begin{figure}
\centering\includegraphics[width=10.8cm,clip]{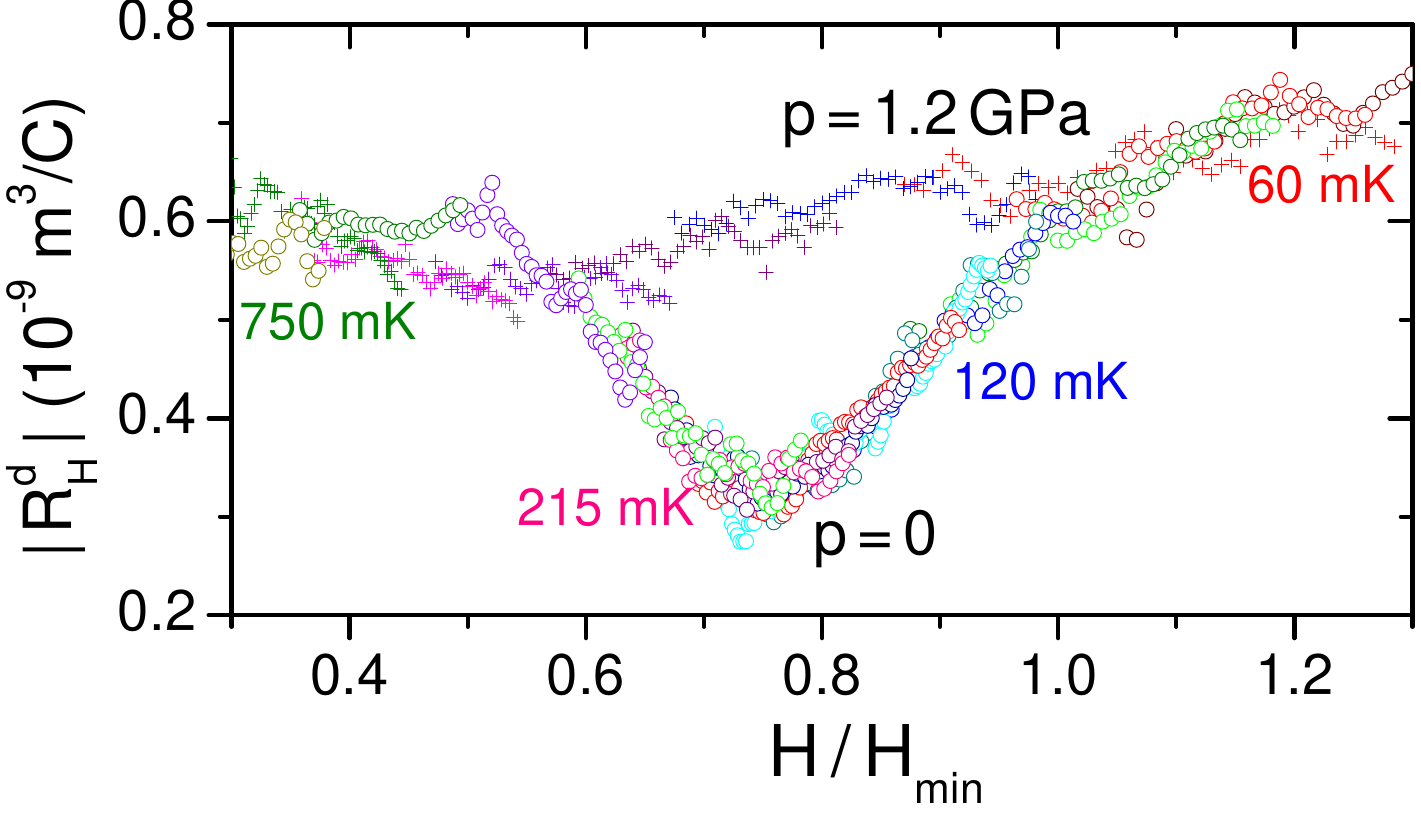}
\caption{\label{fig:scalepr} Observation of a pressure dependent
feature in the differential Hall conductivity in CeCoIn$_5$. The
baseline represents Landau Fermi liquid behavior, whereas
deviations from it arise due to antiferromagnetic fluctuations.
Reprinted from \cite{sin07}.}
\end{figure}

An interesting manifestation of the possible influence of
antiferromagnetic fluctuations on the Hall response was the
observation of a pressure dependent feature in the differential
Hall coefficient ($R_{\rm H}^d = \partial \rho_{xy}(H,T) /
\partial H$) in CeCoIn$_5$ \cite{sin07}. Using the same $H_{\rm
min}$-values as described in the preceding paragraph, it was shown
that $|R_{\rm H}^d|$ could be scaled on a curve as shown in
Fig.~\ref{fig:scalepr}. It was suggested that this dip in $|R_{\rm
H}^d|$ was related to the influence of antiferromagnetic spin
fluctuations or a spin-density-wave-driven gap on the Fermi
surface. This suggestion was based on the fact that this feature
was seen to vanish at applied pressures of the order of 1.2 GPa,
which forms the baseline for this curve along with the data at
high ($H > 1.1 H_{\rm min}$) and low ($H < 0.5 H_{\rm min}$)
values of the normalized field. This baseline was thus interpreted
to be a reflection of the Fermi liquid regime, and deviations from
it were suggested to represent the onset of non Fermi liquid
characteristics arising as a consequence of antiferromagnetic
fluctuations. The suppression of these magnetic fluctuations by
use of pressure in CeCoIn$_5$ has been documented earlier, and is
in reasonable agreement with this scenario \cite{ron06}. An
alternate means of suppressing the incipient antiferromagnetic
fluctuations in these systems is by the application of magnetic
fields. This was demonstrated by the combined investigation of the
electrical and thermal Hall conductivities ($\kappa_{xy}$ and
$\sigma_{xy}$, respectively) in the CeCoIn$_5$ system by Onose and
coworkers \cite{ono07}. Separating out the influence of the
electronic contribution ($\kappa_e$) and that due to
charge-neutral (spin) excitations ($\kappa_b$) to the thermal
conductivity it was demonstrated that the latter charge neutral
term displays an anomalous field dependence. It was suggested that
these spin excitations are primarily responsible for the
scattering of charge carriers in the non Fermi liquid regime of
the phase diagram. Applied magnetic fields rapidly quench these
excitations, resulting in a rapid enhancement in the mean free
path $\ell$, which in turn is manifested in the form of novel
scaling relationships between the magnetoresistance and the
thermal and Hall conductivities.

Besides the above mentioned studies which consider the influence
of magnetic fluctuations on the scattering of charge carriers,
Hall measurements have also been used to infer a possible
modification of the Fermi surface, which then manifests itself in
the form of anomalous transport properties. This aspect will be
dealt with in section \ref{sec:sces}, in combination with, and in
relation to, Hall effect measurements on the high temperature
superconducting cuprates.

A number of attempts were made to tailor the balance between
magnetism and superconductivity in the CeCoIn$_5$ system by the
use of site-specific dopants. For instance, Sn substitution on the
Indium site was seen to suppress the superconducting transition in
this system without observable increment in the extent of magnetic
fluctuations \cite{bau05}. A novel perspective in this family of
systems was gained by the doping of Cd in the series
CeCo(In$_{1-x}$Cd$_x$)$_5$ \cite{pah06}. Here, magnetism is seen
to evolve from within the superconducting ground state as a
function of increasing Cd substitution. Preliminary Hall effect
measurements on a member of this series
CeCo(In$_{0.925}$Cd$_{0.075}$)$_5$ revealed an electron dominated
Hall response similar to that seen in the parent system
\cite{nai09c}. Additional features corresponding to the possible
field-induced destabilization of the antiferromagnetic ground
state was also observed. Similar observations in the
magnetoresistance were then used, in conjunction with neutron
diffraction measurements, to chart out the temperature-magnetic
field phase diagram in this system \cite{nai10}. The intertwinned
relation of antiferromagnetism and superconductivity is also
suggested by the $p$-$B$-$T$ phase diagram, reproduced from
\cite{zau11} in Fig.\ \ref{fig:BTphase}(b). Here, superconducting
transition and N\'{e}el temperatures are compiled for hydrostatic
pressure on CeCoIn$_5$ and negative chemical pressure in
CeCoIn$_{5-x}$Cd$_x$.

\subsubsection{CeIrIn$_5$} \label{sec:Ir115}
As mentioned in section \ref{sec:fluc115}, the magnetic
instability is known to be well separated from superconductivity
in CeIrIn$_5$. Moreover, the magnetic instability appears to be
metamagnetic in origin \cite{tak01,kim02,cap04}; a factor which
modulates the temperature-magnetic field phase diagram in this
material. For instance, unlike the case of CeCoIn$_5$, the
application of an external magnetic field is seen to suppress,
rather than stabilize the Fermi liquid ground state. The
metamagnetic nature of the putative quantum phase transition in
CeIrIn$_5$ was also reflected in the scaling analysis of the Hall
effect along the lines mentioned in the preceding section where
the experimental $R_H$ curves within the Kondo coherent regime
could be scaled to a generic curve \cite{nai09b}. However, in this
case the coefficient yielded a negative value, which was ascribed
to the fact that there is no magnetic instability in the immediate
vicinity of superconductivity in CeIrIn$_5$. Analysis of the Hall
effect was also used to deduce the presence of a precursor state
to superconductivity in this material \cite{nai08}. This was
accomplished by monitoring the field dependence of the Hall angle
$\theta_H$ in relation to the inverse of the applied field.
Deviations from linearity were then used to mark out the existence
of a hitherto unidentified state which precedes the formation of a
superconducting condensate in this system. The boundary of this
precursor state in the temperature-field phase diagram could be
scaled onto the boundary of the superconducting regime implying
that they could originate from the same underlying phenomena. The
existence of this state was in line with prior observations in the
related compound CeCoIn$_5$, where a precursor state to
superconductivity was inferred from measurements of the Nernst
effect \cite{bel04l} and resistivity under pressure \cite{sid}.

The anomalous transport properties specifically of CeIrIn$_5$ are
discussed in relation to those of the cuprates in section
\ref{sec:cupHF}.

\subsubsection{Comparative remarks} \label{sec:comp115}
\begin{figure}
\centering\includegraphics[width=12.8cm,clip]{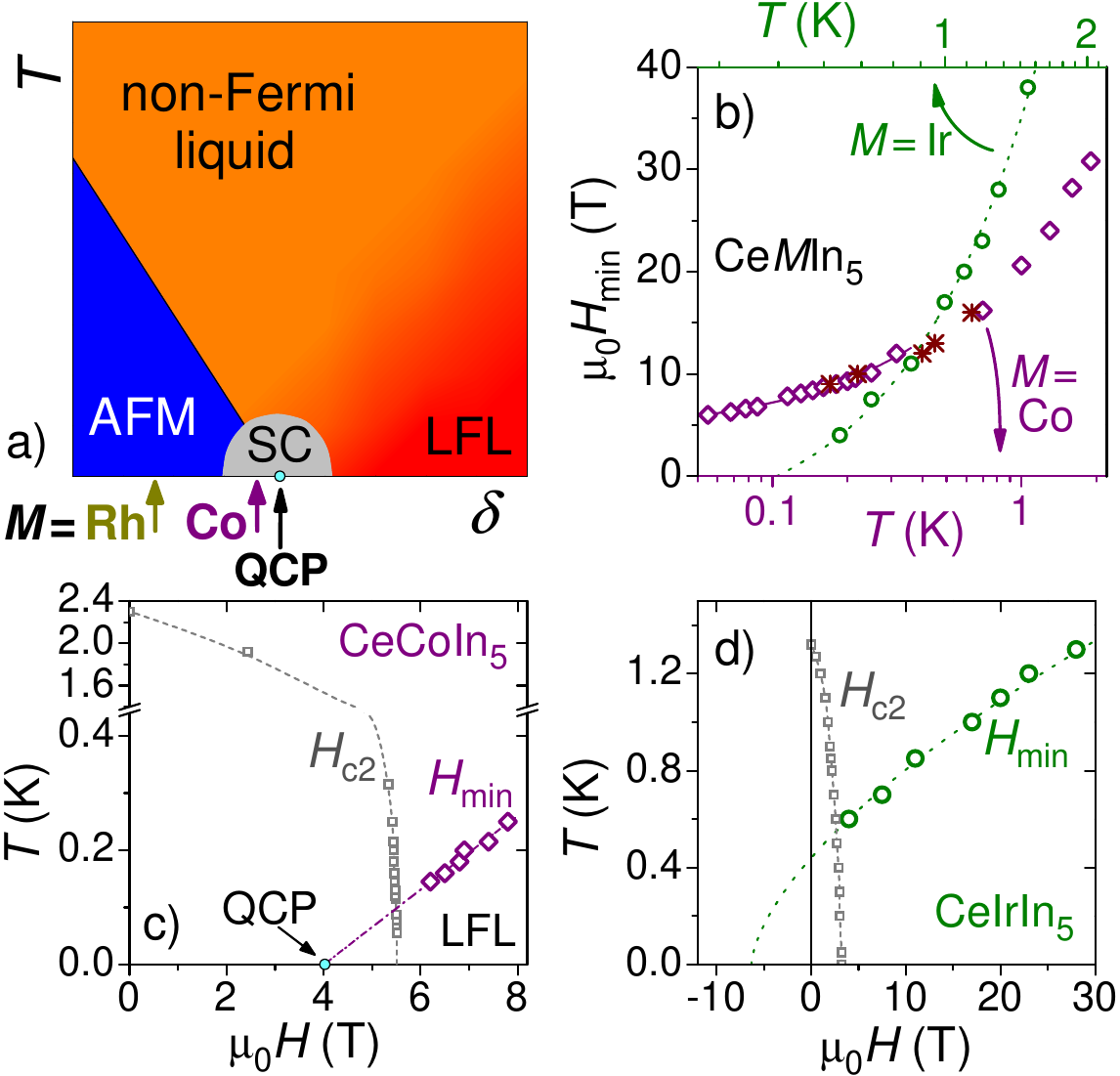}
\caption{\label{fig:115CoIr} (a) Schematic phase diagram of an
archetypical heavy fermion superconductor (SC). The
antiferromagnetic (AFM) and Landau Fermi liquid phases as well as
the quantum critical point (QCP) are indicated. The positions of
Ce$M$In$_5$ ($M =$ Co, Rh) in their ground states are sketched
out. CeIrIn$_5$ cannot be placed within this schematic due to the
first-order nature of its magnetic instability. $\delta$ denotes
the experimental control parameter which can be magnetic field or
pressure. (b) Comparison of the scaling fields $H_{\rm min}$ for
CeCoIn$_5$ and CeIrIn$_5$, see Figs.~\ref{fig:scaleCo},
\ref{fig:scalepr}. Note the break in the temperature scale. (c)
$H\! -\!T$ phase diagram of CeCoIn$_5$ with the QCP at around 4 T.
(d) Phase diagram of CeIrIn$_5$ without indication of a QCP.}
\end{figure}

As discussed in section \ref{sec:HF} the competition of RKKY and
Kondo interaction may result in quantum critical behavior. Here, a
QCP is expected at that value of the experimental control
parameter at which the antiferromagnetic order vanishes. However,
the excessive entropy related to the QCP is often avoided by the
formation of superconductivity [{\it cf.}
Fig.~\ref{fig:115CoIr}(a)]. Nonetheless, the QCP still lurks
around as manifested by non Fermi liquid behavior near the value
of the experimental parameter at which the QCP is expected but at
temperatures high enough to suppress superconductivity
\cite{mat98,bro08}.

At temperatures above $\sim$50 K, the Hall coefficient in
Ce$M$In$_5$ with $M$ = Co, Ir, Rh is dominated by skew scattering
\cite{hun04}. Below this temperature, {\it i.e.}, in a range 2 K
$\lesssim T \lesssim$ 50 K, a $T^2$ dependence of the cotangent of
the Hall angle, $\cot \theta_H$, was interpreted to be caused
mainly by the complex band structure of these compounds. This was
due to the fact that the non-magnetic analogs La$M$In$_5$ (with $M
=$ Co, Ir and Rh) also exhibited a $T^2$-dependence of the Hall
angle in a similar temperature regime. In systems with complex
band structures as has been deduced for the La-115 systems
\cite{hal09}, a temperature-dependent Hall response could arise
due to different temperature dependencies in the contributing
electron and hole bands, see section \ref{sec:spinfluct}. In the
magnetic Ce$M$In$_5$ systems, these anomalies would be accentuated
due to the addition of strong Kondo and $f$ electron interactions.

At even lower temperatures the results of Hall effect measurements
CeCoIn$_5$ could be used to gauge the crossover from non Fermi
liquid to Landau Fermi liquid behavior \cite{sin07}, see section
\ref{sec:Co115} and Figs.~\ref{fig:115CoIr}(b) and (c). It could
be shown that the latter behavior is recovered once the
antiferromagnetic spin fluctuations are suppressed either by
pressure or magnetic field. As outlined in section
\ref{sec:Co115}, an extrapolation of these results indicated a QCP
at around 4 T. Consequently, the quantum critical field is {\em
smaller} than $H_{\rm c2}$. This important result renders the
resulting phase diagram Fig.~\ref{fig:115CoIr}(c) to be in good
qualitative agreement with the generic one,
Fig.~\ref{fig:115CoIr}(a), if the zero field state of CeCoIn$_5$
is assumed to be located at a position marked by the arrow in the
latter.

Applying the same scaling procedure of the Hall coefficient as
outlined for CeCoIn$_5$ results in a phase diagram for CeIrIn$_5$
as presented in Fig.~\ref{fig:115CoIr}(d), {\it cf.} also section
\ref{sec:Ir115}. Clearly, there is no physical meaning of an
extrapolation to a negative quantum critical field. But such an
extrapolation reveals the close affinity between these two members
of the Ce$M$In$_5$ family: even though there is no indication for
a QCP in CeIrIn$_5$ the zero-field position within the generic
phase diagram as marked by the arrow in Fig.~\ref{fig:115CoIr}(a)
suggests that the Ir compound nicely fits into this picture. Such
a comparison is corroborated by the presence of strong
antiferromagnetic fluctuations \cite{koh} in both compounds and
similarities in superconductivity \cite{izawa,kas08} and Fermi
surfaces \cite{shi02,sar07}. This implies that the balance between
Kondo and RKKY interaction (section \ref{sec:HF}) is shifted
towards the former in case of the Ir compound. This notion is
supported by the ratio $T_{\rm RKKY} / T_{\rm K}$ being smaller
for Ir ($\approx 7$) if compared to the Co case ($\approx 25$).
Within these considerations even CeRhIn$_5$---which exhibits an
antiferromagnetic ground state and superconducts upon application
of pressure \cite{par06}---fits into the picture ($T_{\rm RKKY} /
T_{\rm K} \approx 130$). It should be noted, however, that
Fig.~\ref{fig:115CoIr}(a) does {\em not} imply that changing the
element $M$ in Ce$M$In$_5$, application of pressure and magnetic
field are interchangeable parameters; rather the arrows only
indicate the zero-field ground state properties of the respective
material.

\section{Comparison to Hall effect of other correlated materials}
\label{sec:sces}
\subsection{Copper oxide superconductors and related systems}
\label{sec:copper}
\subsubsection{Cuprates} \label{sec:cup}
A discussion of the Hall effect in the heavy fermion metals would
not be complete without taking note of how measurements of the
Hall effect has influenced the understanding of other strongly
correlated electron systems, especially of the high temperature
superconducting cuprates. A priori, there appears to be very
little in common between these two classes of materials: the heavy
fermion metals are inherently metallic systems with a dense array
of magnetic ions, whereas the cuprates are doped Mott insulators.
However, the emergence of a superconducting condensate in both
sets of systems has revealed a number of striking parallels. For
instance, the inherently quasi--2D nature of electronic
correlations, the proximity of long range antiferromagnetism (or a
spin density wave) to superconductivity, a scaling of the observed
transition temperatures with the Fermi energy, and the
unconventional (possibly $d$ wave) nature of superconductivity in
these materials all suggest an underlying similarity in the
physics of these materials. A welcome addition to this narrative
has been the recent discovery of superconductivity in the
oxy-pnictides and related systems.

The discovery of superconductivity in La$_{2-x}$Ba$_x$CuO$_4$ by
Bednorz and M{\"u}ller \cite{bed86} fuelled an unprecedented
interest into many families of the layered cuprates in an attempt
to understand the mechanism of superconductivity, as well as to
synthesize materials with increasing high superconducting
transition temperatures $T_c$. Initial Hall effect measurements in
these systems were aimed at discerning how these systems differed
from conventional metals, and to monitor the sign and evolution of
the charge carriers as a function of temperature and doping.
Subsequent availability of high-quality single crystalline
specimens stirred up the utility of the Hall effect as a probe of
the Fermi Surface topology.

\begin{figure}
\centering\includegraphics[width=6.8cm,clip]{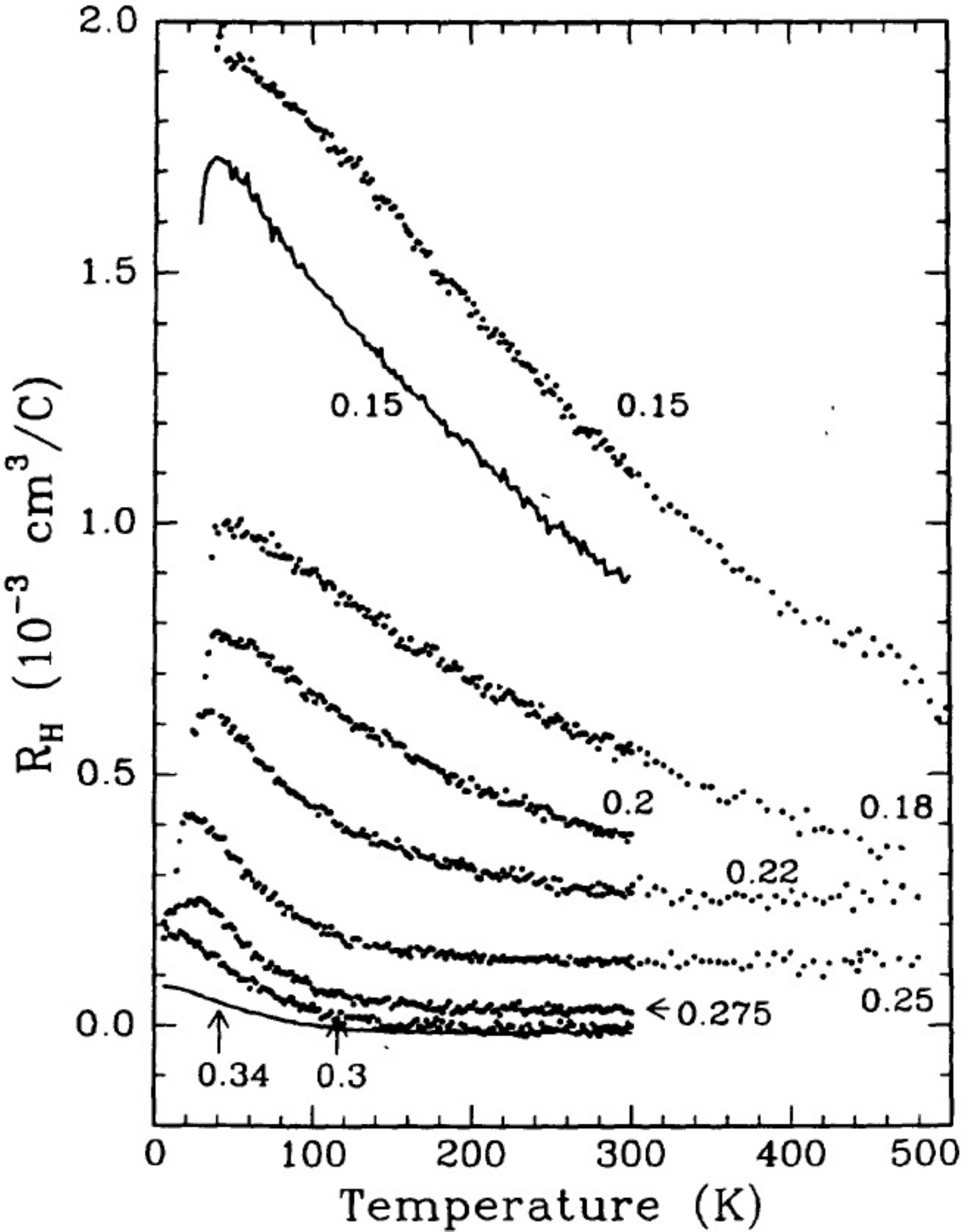}
\caption{\label{hwang} Temperature dependence of the Hall
coefficient $R_H$ as measured in La$_{2-x}$Sr$_x$CuO$_4$
exemplifying the temperature dependence of the Hall effect in the
hole doped cuprates. Reprinted figure with permission from H.Y.
Hwang {\it et al.}, Physical Review Letters {\bf 72}, 2636 (1994)
\cite{hwa94}. Copyright \copyright\,(1994) by the American
Physical Society.}
\end{figure}

In a simple (uncorrelated) metal, where the electron relaxation
times are isotropic on all points of the Fermi surface, the Hall
coefficient $R_H$ would be expected to be independent of
temperature. This scenario was thought to be valid for reasonably
high temperatures ($T \geq \theta_D$, with $\theta_D$ being the
Debye temperature) where phonons were the dominant scattering
mechanism which ensured a reasonably isotropic electron scattering
\cite{zim61}. In practice, $R_H$ was seen to be weakly dependent
on temperature above a characteristic temperature $T = s
\theta_D$, where $s$ varies from 0.2 to 0.4 in different metals
like Cu, Ag, Cd, and Mg \cite{dug69,hur76,mac77}. Below this
characteristic temperature, the observed temperature dependence
was thought to arise from an increasing anisotropy in an otherwise
isotropic Fermi Surface due to residual impurity scattering. Not
surprisingly, early Hall measurements on the hole doped cuprates
indicated that the scenario in this class of materials was
dramatically different. Measurements on La$_{2-x}$Sr$_x$CuO$_4$
\cite{ong87}, YBa$_2$Cu$_3$O$_{7-\delta}$
\cite{che87,wan87,fri88,sha89} and Tl$_2$Ca$_2$Ba$_2$Cu$_3$O$_x$
\cite{cla88} family of systems exhibited a pronounced dependence
of the Hall effect on both the temperature as well as the extent
of hole doping. Typically, the Hall voltage was seen to be
positive, Fig.~\ref{hwang}, which at least in the case of
La$_{2-x}$Sr$_x$CuO$_4$ was thought to arise from the formation of
hole pockets at the corners of the first Brillioun Zone. It was
suggested that the area of these pockets increased with increasing
hole concentration, which also accounted for the doping dependence
of $R_H$.

However, experimental signatures like monotonically increasing
$R_H$ with decreasing temperature in all the measured cuprates
could not be understood within the simple picture based on
electron-phonon scattering. For instance, in YBa$_2$Cu$_3$O$_7$
the temperature dependence of $R_H$ was seen to extend up to 360 K
\cite{chi91b}. A simple electron-phonon scattering mechanism as
described for simple metals would then warrant an effective
$\theta_D > 900$ K, whereas estimates of $\theta_D$ from heat
capacity measurements indicated a value which was of the order of
400 K. Measurements on thin films of YBa$_2$Cu$_3$O$_{6+\delta}$
indicated that this temperature dependence of the Hall effect
continued up to about 700 K \cite{dav88}. A logical extension was
the use of a two-band picture in which the Hall effect from the
electrons and the holes partially cancel each other. Thus, the two
band Hall effect could be given as ({\it cf.} eq.~(\ref{twoband})
on p.~\pageref{twoband})
\begin{equation}
R_H = \frac{\mu_h - \mu_e}{e n (\mu_h + \mu_e)}
\end{equation}
where $\mu_h$ and $\mu_e$ refer to the hole and electron
mobilities, respectively. In this case, the observed net $R_H$
could have a temperature dependence if $\mu_h$ and $\mu_e$ have
different temperature dependencies, and there existed two nearly
symmetric bands which fulfil the above equation in a wide range of
temperature \cite{pen87,sha87}. However, the fact that this
temperature dependence of $R_H$ was seen in structurally diverse
systems (with different associated Fermi surfaces) like
YBa$_2$Cu$_3$O$_7$ and Tl$_2$Ca$_2$Ba$_2$Cu$_3$O$_8$, would mean
that this rather delicate balance had to exist in all these
cuprates in a large part of their phase diagram. The relatively
weak pressure dependence of $R_H$ as measured in
YBa$_2$Cu$_3$O$_7$ \cite{par88} would also warrant that $\mu_h$
and $\mu_e$ have similar pressure dependencies as well. Thus, it
was evident that the two band scenario was implausible.

Another potential mechanism for the observed temperature
dependence in $R_H$ could have been the magnetic skew scattering,
described earlier in section \ref{sec:skew}. If scattering from
the magnetic impurities were relatively independent from each
other, the anomalous part of the Hall effect could scale with the
(temperature dependent) paramagnetic magnetic susceptibility
$\chi$. The validity of this scenario was tested by extending Hall
measurements to higher fields, where the Zeeman energy $g \mu_B B$
was clearly higher than the thermal energy $k_B T$. Here, $g$
refers to the g-factor, $\mu_B$ is the Bohr Magneton, $B$ is the
applied magnetic field, $k_B$ is the Boltzmann constant. At large
values of $B$, {\it i.e.} $B > k_B T / g \mu_B$, the paramagnetic
susceptibility (and thus the anomalous Hall effect) should
saturate, and should be clearly discernible in field sweep
experiments. However, it was observed that in
La$_{2-x}$Sr$_x$CuO$_4$, the Hall voltage as measured at 52 K
remained linear up to fields of the order of 12 T. Measurements on
the electron doped system Nd$_{2-x}$Ce$_x$CuO$_4$ indicated that
this linear behaviour in the Hall voltage persisted up to 20 T
even at temperatures as low as 1.2 K, whereas a typical g-factor
value of 2, at $T$ = 2 K should have resulted in a saturation in
$\chi$ at moderate fields of the order of 1.5 T.

These anomalous transport properties where the resistivity and the
Hall effect had different temperature dependencies could not be
accounted for using either the conventional Boltzmann-Bloch models
or by incorporating magnetic skew scattering. An additional
mystery in the cuprates was the pronounced sensitivity of the Hall
effect to in-plane disorder. For instance, the addition of Zn or
Co impurities in the CuO$_2$ planes in YBa$_2$Cu$_3$O$_7$ was seen
to result in the suppression of the ``Hall slope'' defined as
$d(1/eR_H)/dT$ \cite{cla89}. An important advance in explaining
this anomalous behaviour in the cuprates was Anderson's conjecture
that there exist two transport relaxation times in the cuprates
which \emph{independently} influence the Hall effect and the
resistivity in these systems \cite{and91}. This was in response to
the observation that the Hall angle $\theta_H$, as defined in
eq.~(\ref{HallAngle}), has an unambiguous $T^2$-dependence in an
extended temperature regime, and appeared to be an intrinsic
parameter in this class of materials. In normal metals, there is
one transport relaxation time $\tau_{\rm tr}$ for electrons on all
parts of the Fermi surface, which governs \emph{both} the
resistivity and the Hall effect. Thus the conductivity
$\sigma_{xx}$ is proportional to $\tau_{\rm tr}$ and has a
$T^{-1}$-dependence (since $\hbar / \tau_{\rm tr} \propto k_B T$),
whereas the Hall conductivity $\sigma_{xy}$ is proportional to
$\tau_{\rm tr}^2$, and thus has a $T^{-2}$-dependence. However,
using a Luttinger liquid formalism, Anderson suggested that
transport in these materials were governed by quasiparticles
(spinons and holons) which carry the spin and charge degrees of
freedom, and which \emph{independently} influence the Hall effect
and resistivity. According to this picture, the Hall effect is
governed by the transverse relaxation rate $\tau_H$ which is
determined by scattering between spin excitations and has a
$T^{-2}$-dependence, whereas $\sigma_{xy}$ is proportional to
$\tau_{\rm tr}\cdot t_H$, and thus has a $T^{-3}$-dependence.

\begin{figure}
\centering\includegraphics[width=7.2cm,clip]{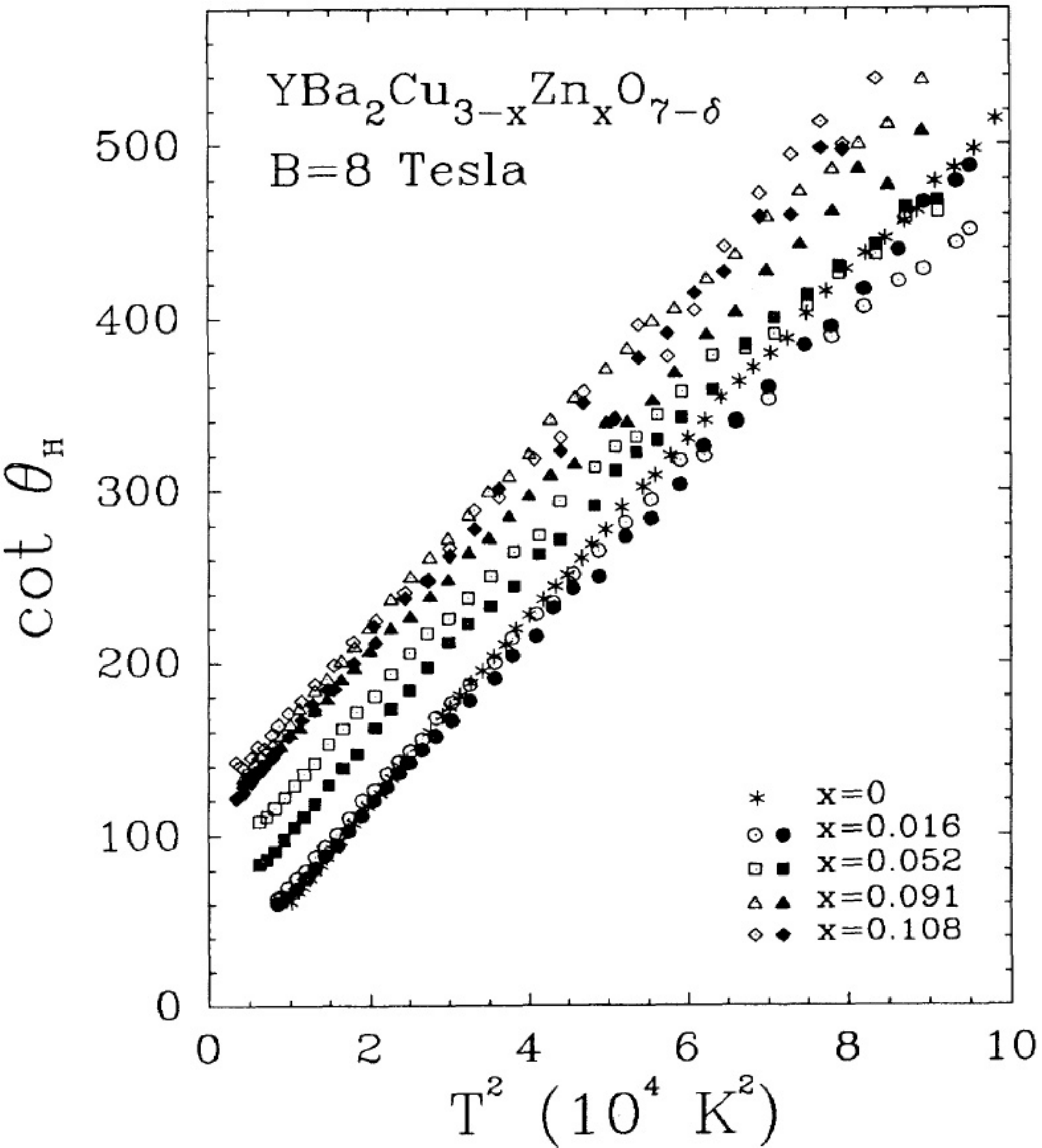}
\caption{\label{chien} Temperature dependence of the Hall angle as
measured in a series of Zn doped YBa$_2$Cu$_3$O$_7$ samples
demonstrating the $T^2$ dependence of $\theta_H$ independent of
the impurity concentration. Reprinted figure with permission from
T.R. Chien, Z.Z. Wang, and N.P. Ong, Physical Review Letters {\bf
67}, 2088 (1991) \cite{chi91}. Copyright \copyright\,(1991) by the
American Physical Society.}
\end{figure}

More importantly, the Hall angle $\theta_H$ of
eq.~(\ref{HallAngle}) relies on $\tau_H$ alone and is thus a
quantity of fundamental interest. Moreover, it has a $T^2$
dependence, and is independent of the impurity scattering which
only adds a temperature independent term to both the scattering
rates. This, being an easily verifiable prediction, was confirmed
on Zn substituted YBa$_2$Cu$_3$O$_7$ \cite{chi91}, as is shown in
Fig.~\ref{chien}. This $T^2$ dependence of the Hall angle was
shown to persist up to temperatures as large as 500 K in oxygen
reduced samples of the same series by Harris and co-workers
\cite{har92}. Measurements of the optical conductivity confirmed
the non-Drude nature of charge dynamics in thin films of
YBa$_2$Cu$_3$O$_7$ and could be qualitatively explained on the
basis of the two distinct scattering rates \cite{kap95}. The $T^2$
dependence of the Hall angle is now known to be endemic to many
cuprate superconductors and was used extensively to emphasize on
the anomalous transport properties of many cuprate families.
Prominent examples include the case of Zn, Co, Fe and Pr
substituted YBa$_2$Cu$_3$O$_7$ \cite{chi91,jia92,lan94,car92}, Y,
Ni and Zn substituted Bi$_2$Sr$_2$CaCu$_2$O$_8$ \cite{ken92}, and
Fe, Co, Ni, Zn and Ga substituted La$_{1.85}$Sr$_{0.15}$CuO$_4$
\cite{xia92}. An interesting manifestation of the existence of two
time scales was the reformulation of the Kohler's scaling rule in
terms of the Hall angle. In conventional metals, the Kohler's rule
mandates that the orbital contribution to magnetoresistance
$[\rho_{xx}(H) - \rho_{xx}(0)] / \rho_{xx}(0)$ can be scaled as a
function of the term $H / \rho_{xx}(0)$. Harris and co-workers
demonstrated that this scaling clearly breaks down in the case of
YBa$_2$Cu$_3$O$_7$ and La$_{2-x}$Sr$_x$CuO$_4$ \cite{har95}. They
also showed that the temperature dependence of the
magnetoresistance varies with the square of the Hall angle, which
is in line with the scenario of two scattering relaxation rates. A
phenomenological transport equation was
developed~\cite{coleman_1996a,coleman_1996b} for the cuprates
which allowed distinct scattering rates for $\tau_{tr}$ and
$\tau_H$ as envisaged in the spin-charge separation scenario. This
reproduced the violation of Kohler's rule and made predictions for
the thermopower~\cite{clayhold_1998a} but required the two
scattering rates to differ significantly in magnitude which was
later shown not to be consistent with optical
transport~\cite{kap95}.

Besides the spin charge separation scenario, a prime candidate in
explaining the anomalous transport properties in the cuprates is
the nearly antiferromagnetic Fermi liquid (NAFFL) theory as
described in a series of papers by Pines and others
\cite{mon91,mon92,mon93,mon94a,mon94b,sto97}. This model relies on
an anisotropic reconstruction of the (otherwise isotropic) Fermi
surface in the presence of (antiferro-)magnetic spin fluctuations.
This is achieved by the formation of ``hot'' spots and ``cold''
regions on the Fermi surface, with the hot spots being located at
positions in the momentum space where the antiferromagnetic
Brillouin Zone intersects the Fermi Surface. All the transport
properties would thus be normalized with respect to these
(modified) scattering rates. It was suggested \cite{sto97} that
the Hall effect in these NAFFL systems could be broadly classified
into three temperature regimes, with $\sigma_{xy}$ varying as
$T^{-4}$, $T^{-3}$ and $T^{-2}$ with increasing temperatures.
Here, the crossover points between these regimes depend on details
of the band structure. In addition to these theories, which rely
on the Hall angle being linearly dependent on the scattering rate
$\tau_H$, the marginal Fermi liquid theory by Varma and co-workers
relies on a square scattering response ($\tau^2 H)$
\cite{var89a,var89b,var01,aba03}. Another avenue of interpreting
the anomalous magnetotransport measurements within the Fermi
liquid formalism is the use of current vertex corrections (CVC)
championed by Kontani and co-workers \cite{kon08}. This refers to
the current driven by quasiparticles which propagate though
collision events in the Fermi liquid. In a liquid with incipient
antiferromagnetic fluctuations, these currents can have a
pronounced momentum distribution which then correspondingly
modifies the observed magnetotransport properties (see, however,
section \ref{sec:spinfluct}).

Measurements of the Hall effect, in particular the Hall angle, was
also used to speculate on the existence of the pseudogap in some
of these cuprates. For instance Abe and co-workers used deviations
from the $T^2$-behavior in Zn doped YBa$_2$Cu$_3$O$_7$ as a
measure of the pseudogap temperature \cite{abe99}. The direction
of change in $\theta_H$ at the onset of the pseudogap was used to
suggest that the opening of the gap was associated with an
enhancement of $\tau_H^{-1}$ and a reduction in $\tau_{\rm
tr}^{-1}$. A similar deviation of $\theta_H$ at the onset of the
pseudogap was also reported in the underdoped
YBa$_2$Cu$_3$O$_{6.63}$ system \cite{xu99}. The temperature scale
of the pseudogap was also used to scale the Hall number $n_H$ and
the Hall angle in films of YBa$_2$Cu$_3$O$_7$ indicating that the
pseudogap is an intrinsic energy scale of these systems
\cite{wuy96}. Interestingly, a similar line of analysis was used
earlier in crystals of La$_{2-x}$Sr$_x$CuO$_4$ where the
temperature dependent $R_H$ was scaled using an arbitrarily
defined temperature scale $T^*$, where $T^*$ was thought to
approximately represent a crossover between a temperature
independent to a temperature dependent $R_H$ \cite{hwa94,nis93}.

\subsubsection*{Fermiology in the Cuprates}
The availability of extremely high-quality single-crystal
specimens, coupled with sensitive new Hall effect measurements
have thrown new light on the nature of the Fermi surface in
cuprates. Whereas systems in the overdoped regime of the cuprate
phase diagram behave like metals to a reasonable approximation,
the underdoped cuprates are characterized by Fermi surfaces made
up of discontinuous ``Fermi Arcs'' as deduced from spectroscopic
measurements \cite{nor98,shen05}. An important advance was the
observation of quantum oscillations as measured in the Hall
resistivity in the underdoped  YBa$_2$Cu$_3$O$_{6.5}$,
establishing the existence of a continuous Fermi surface in this
class of systems \cite{doi07}. The Hall response was observed to
be negative (unlike the typical positive $R_H$ observed in other
hole-doped cuprates), and this was attributed at that time to the
influence of the vortex liquid phase which persists in this region
of the phase diagram. Moreover, the low frequency of the observed
Shubnikov--de Haas oscillations indicated the presence of a Fermi
surface made up of small pockets. Observation of similar
oscillations in the magnetoresistance of YBa$_2$Cu$_4$O$_8$
indicated that the existence of these small pockets were
independent of the details of the band structure and could
probably be a generic feature of many underdoped cuprate
superconductors \cite{ban08}. Monitoring the sign of the Hall
response in the magnetic field regime where these oscillations
were observed, LeBoeuf and co-workers concluded that these small
pockets were electron like, rather than being hole like
\cite{leb07}. This is surprising considering the hole doped nature
of these materials and also because of the fact that the existence
of a cylindrical, hole-like Fermi surface in the overdoped
cuprates has been documented \cite{mac96}. This could imply that
there exists a Fermi surface reconstruction as a function of hole
doping, with a possible zero-temperature instability demarcating
these two metallic states. Recently, LeBoeuf and coworkers have
reported on a Hall effect investigation on the YBa$_2$Cu$_3$O$_y$
system with varying hole doping per planar Cu atom ($p$) from
0.078 to 0.152 \cite{leb11}. In the $T \! \rightarrow \! 0$ limit,
the Hall coefficient $R_H$ was seen to change from positive to
negative, as the doping is reduced below 0.08. This was suggested
to arise due to the presence of a Lifshitz transition which
modifies the Fermi surface topology. The electron pocket which is
observed to persist till the highest investigated doping level ($p
= 0.152$), was conjectured to arise from an ordered state (like
some form of stripe order \cite{cha10}) which breaks the
translational symmetry of the system.

\subsubsection{Comparison of cuprates and heavy-fermion systems}
\label{sec:cupHF} The insight gained in investigating the
anomalous transport properties of the underdoped
\begin{figure}
\centering\includegraphics[width=6.4cm,clip]{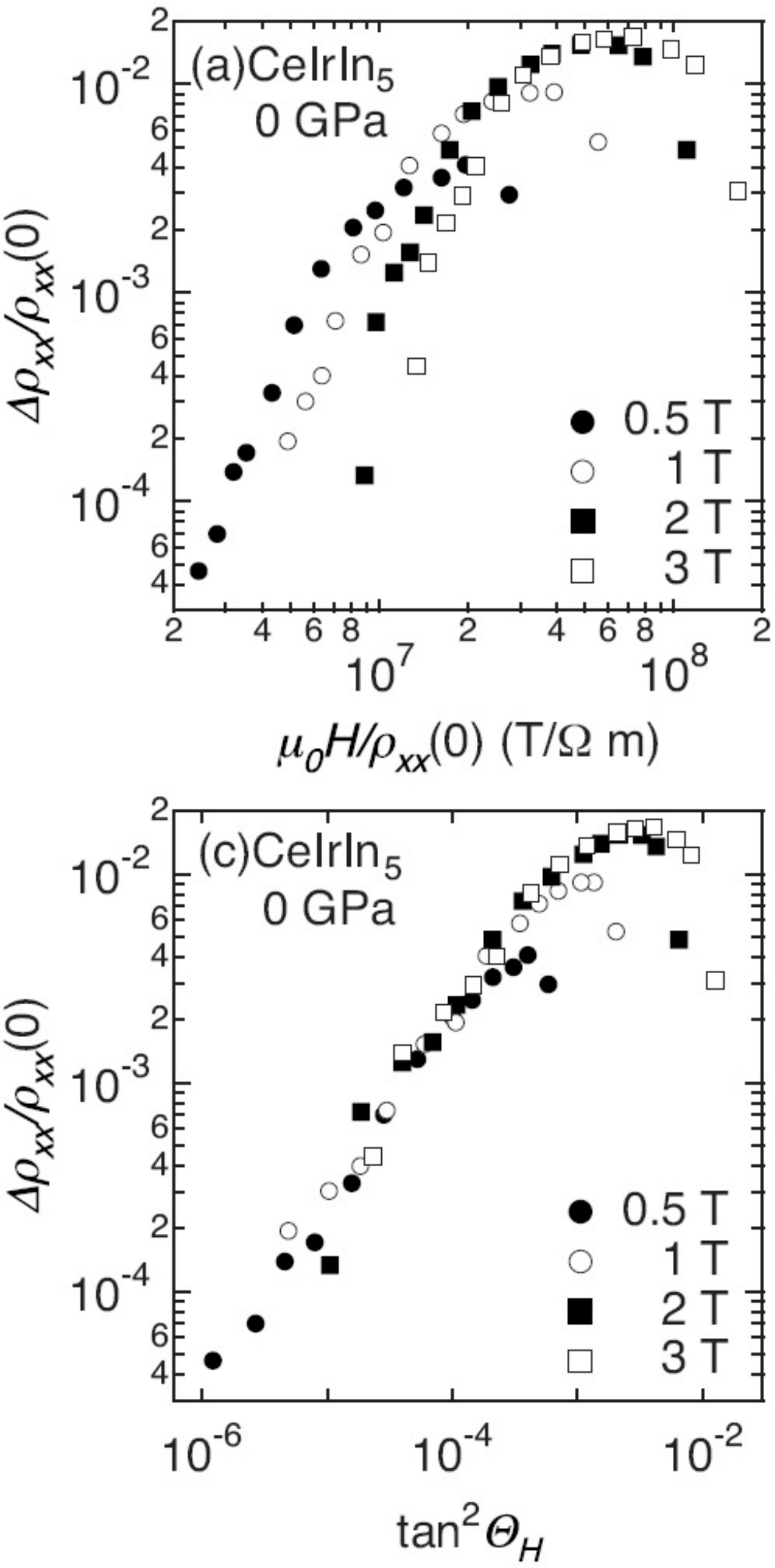}
\caption{\label{naka3} The violation of Kohler's scaling (upper
panel) and its reformulation in terms of the Hall angle (lower
panel) as deduced in CeIrIn$_5$. Reprinted figure with permission
from Y. Nakajima {\it et al.}, Physical Review B {\bf 77}, 214504
(2008) \cite{nak08a}. Copyright \copyright\,(2008) by the American
Physical Society.}
\end{figure}
cuprates have found resonance in attempting to understand the
nature of pronounced non Fermi liquid characteristics observed in
some heavy fermion metals. This is especially so in the case of
the magnetotransport in Ce-115 systems, section \ref{sec:Hall115},
where an acute interplay between superconductivity and quantum
magnetism has revealed some striking parallels with that observed
in the cuprates. For instance, in both the CeCoIn$_5$ and
CeIrIn$_5$ systems, a quasi-linear temperature dependence of the
resistivity is observed, with this non-Fermi liquid characteristic
being driven due to the influence of incipient antiferromagnetic
fluctuations \cite{pet01a,pet01b}. The Hall effect as measured in
these systems also shows a pronounced temperature dependence. More
importantly, the Hall angle $\theta_H$ clearly varies as $T^2$
indicating the disparate nature of scattering processes that drive
the resistivity and Hall effect in these systems
\cite{hun04,nak07,nak04a,nak06,nai08,nak08a}. This is also
reinforced by the analysis of the magnetoresistance where the
Kohler's scaling is violated, and the scaling procedure is only
seen to work when equated in terms of the Hall angle as is shown
in Fig.~\ref{naka3}.

In CeIrIn$_5$, an additional anomaly in the form of a breakdown of
the modified Kohler's scaling is observed very close to the onset
of superconductivity. However, this is due to the influence of a
precursor state to superconductivity in this system, which
influences the resistive and Hall scattering rates in disparate
fashions \cite{nai08}. The existence of this precursor state was
inferred from the field dependence of the Hall angle, and was
thought to represent an anisotropic reconstruction of the Fermi
surface prior to the formation of the superconducting condensate.
In line with prior investigations in the cuprates where the energy
scale associated with the pseudogap was used to scale the
resistivity and the Hall effect \cite{wuy96}, a model-independent
single parameter scaling of the Hall angle was observed in
CeIrIn$_5$ implying that the precursor state represented an
intrinsic energy scale of the system \cite{nai09a}. Interestingly,
neither the resistivity nor the Hall effect could be individually
scaled using the precursor state, implying that the onset of this
precursor state selectively influences the Hall channel alone.
This is in consonance with that observed in the underdoped
YBa$_2$Cu$_3$O$_{6.63}$ where the opening of the pseudogap was
seen to significantly reduce the Hall response leaving the
resistivity relatively unaffected \cite{xu99}. Unlike the systems
CeCoIn$_5$ and CeIrIn$_5$, which lie on the paramagnetic side of
the putative Quantum critical point, CeRhIn$_5$ orders
antiferromagnetically at ambient pressures. Application of
external pressure shifts this system towards a superconducting
ground state, and in this regime the pronounced non-Fermi liquid
\begin{figure}
\centering\includegraphics[width=12.6cm,clip]{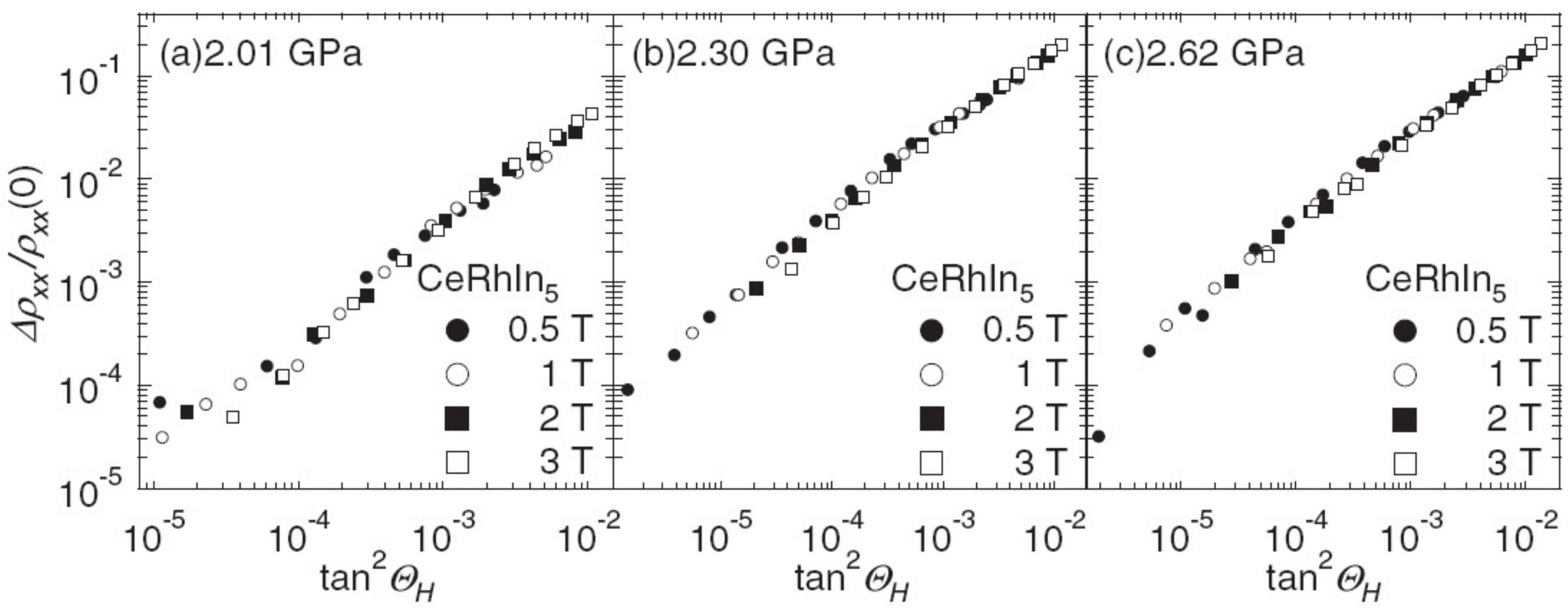}
\caption{\label{naka4} Modified Kohler's plots equating the
magnetoresistance with the Hall angle as measured in CeRhIn$_5$ at
pressures above 2 GPa. Reprinted with permission from Y.\ Nakajima
{\it et al.}, Journal of the Physical Society of Japan {\bf 76}
(2007) 024703 \cite{nak07}. Copyright (2007) by the Physical
Society of Japan.}
\end{figure}
properties enumerated above are retrieved. Fig.~\ref{naka4} shows
the modified Kohler's plot as deduced in CeRhIn$_5$ for a range of
pressures above 2 GPa where superconductivity is observed.

Though this similarity between the heavy fermion metals and the
superconducting cuprates is surprising, this common ground is
possibly brought about by the ingredient which influences the
electronic properties of both these systems, viz., the presence of
antiferromagnetic fluctuations. In magnetotransport, this is
reflected in the form described above, with the Hall and
resistivity being dictated by two different scattering rates.
Unfortunately, an analysis of this nature mandates the
simultaneous measurements of the Hall effect and the resistivity
which at least in the case of heavy fermion systems has been
scarce. However, it is interesting to note that for all the
systems where such measurements have been reported---namely, the
members of the Ce-115 family and YbRh$_2$Si$_2$ \cite{pas-nat}---a
$T^2$-dependence of the Hall angle has been found along with a
quasi-linear temperature dependence of resistivity. The fact that
this behaviour is observed in spite of the wide span of low
temperature electronic ground states observed in these materials
implies that the scenario of two scattering times could be a
feature of many heavy fermion systems.

\subsubsection{Hall effect in the oxy-pnictides and related
systems} \label{sec:pnic} A valuable addition to the existing
assemblage of strongly correlated electron systems was the
discovery of high temperature superconductivity in an iron based
system by Kamihara and co workers \cite{kam08}. The consequent
flurry of activity (for a review see, {\it e.g.}, \cite{john10})
has revealed that emergence of superconductivity in the proximity
to a magnetic instability could be a factor which binds these
systems with the heavy fermion systems and the high temperature
cuprate superconductors in spite of the different band structures
of these systems. Though analysis of Hall effect measurements on
these systems have not been extensive till date, there is enough
indication at the time of writing this review to indicate the
anomalous nature of the magnetotransport in this class of
materials. For instance, Cheng and co-workers reported on the Hall
effect and magnetoresistance in single crystals of
NdFeAsO$_{1-x}$F$_x$ and observed a pronounced temperature
dependence of the Hall effect \cite{che08}. Moreover, analysis of
magnetoresistance revealed a breakdown of Kohler's scaling in this
system. The temperature evolution and, more importantly, the sign
change of $R_H$ in Ba(Fe$_{1-x}$Co$_x$)$_2$As$_2$ was used to
conjecture on the nature of the electron-hole asymmetry in this
system \cite{fan09}. Using a two-band Fermi liquid model it was
inferred that the electrons and holes in this multiband system
have disparate scattering rates. Similar inferences were also
drawn by Rullier-Albenque and coworkers \cite{rul09}. The use of
the two-band Fermi liquid formalism in both these works meant,
however, that the utility of non-Fermi liquid models were not
explicitly considered.

An interesting pointer in this regard is a recent use of the Hall
effect in monitoring the doping-induced evolution from a Fermi
liquid to non-Fermi liquid regime in BaFe$_2$(As$_{1-x}$P$_x$)$_2$
\cite{dai09,kas10}. In this series of samples, increasing P
substitution suppressed the spin-density-wave transition near the
\begin{figure}
\centering\includegraphics[width=6.8cm,clip]{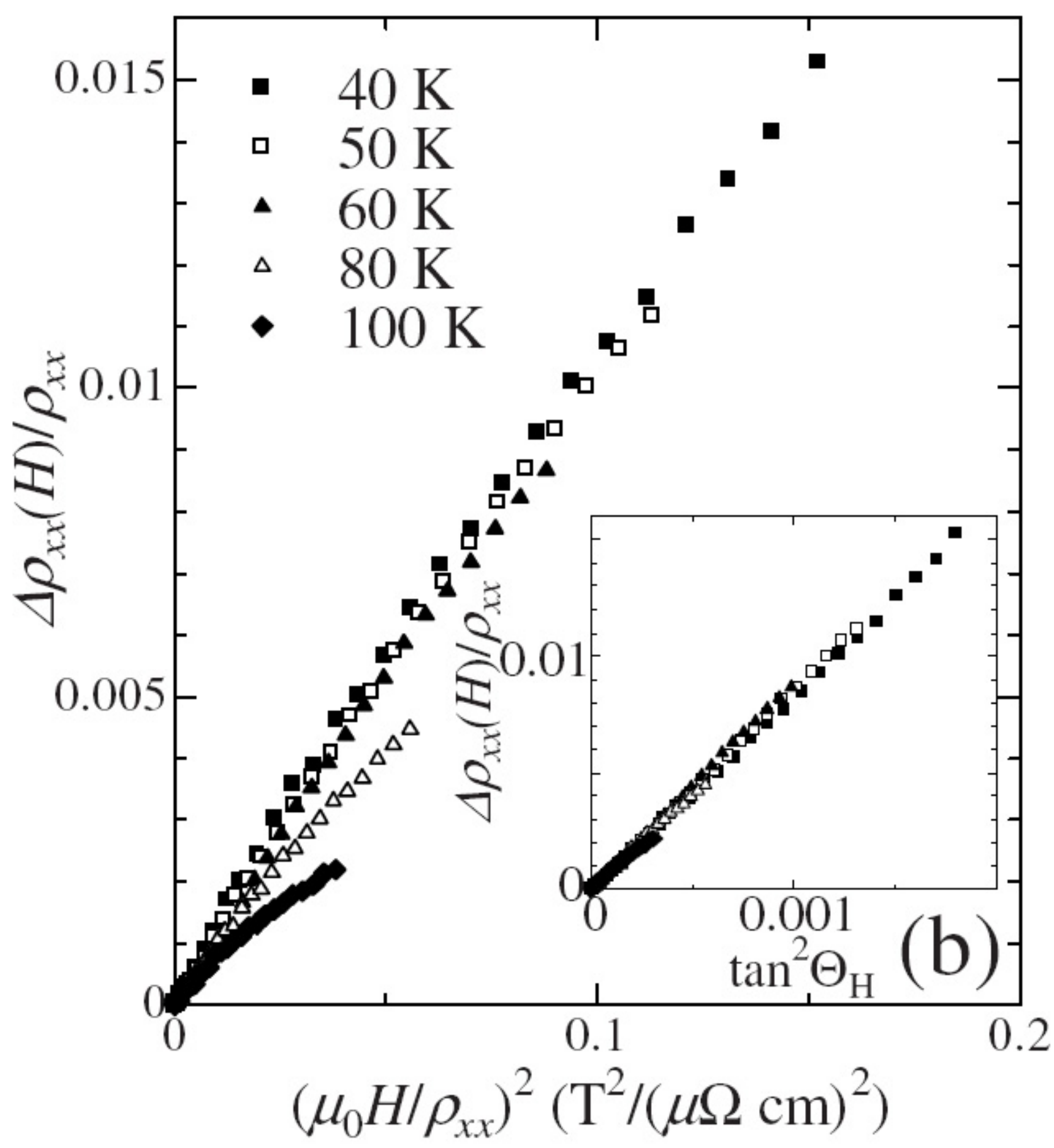}
\caption{\label{kasa} Violation of conventional Kohler's scaling
in BaFe$_2$(As$_{0.67}$P$_{0.33}$)$_2$. Inset shows the modified
Kohler's plot with the magnetoresistance equated in terms with the
Hall angle. Reprinted figure with permission from S. Kasahara {\it
et al.}, Physical Review B {\bf 81}, 184519 (2010) \cite{kas10}.
Copyright \copyright\,(2010) by the American Physical Society.}
\end{figure}
end point of which a superconducting regime was uncovered. For the
optimally doped system (with $x = 0.33$), pronounced non-Fermi
liquid characteristics like a linear $T$-dependence of the
resistivity and a violation of Kohler's scaling were observed.
Moreover, a modified Kohler's scaling which again relates the
magnetoresistance with the Hall angle was observed, as shown in
Fig.~\ref{kasa}. This observation indicates that the anomalous
transport properties in at least some of these multiband systems
could mirror those observed in the cuprates and the heavy fermion
systems.

\subsection{Other systems of related interest}
\label{sec:relat} The possibility of the Hall effect and the
magnetoresistance having different scattering rates has also been
discussed in a few other strongly correlated systems. For
instance, in the prototypical Mott Hubbard system V$_{2-y}$O$_3$ a
pronounced temperature dependence of $R_H$ was observed
\cite{ros98}. Moreover, the temperature dependence of the Hall
effect and the magnetoresistance in the paramagnetic regime was
different; with the Hall angle $\theta_H$ and the resistivity
$\rho_{xx}$ varying as $T^2$ and $T^{1.5}$, respectively. This was
ascribed to the possibility of a zero temperature metal insulator
transition, which is driven by hole doping in the 3$d$ band as a
consequence of increasing vanadium deficiency. Another celebrated
example is metallic chromium in which the long-range magnetic
order can be driven to absolute zero by alloying with small
amounts of vanadium, with the putative QCP being approached at
critical concentrations of about $x = 0.035$ in the
Cr$_{1-x}$V$_x$ system. Magnetotransport in specimens of this
concentration revealed a temperature dependent $R_H$ in
conjunction with a $T^3$-dependence of $\rho_{xx}$ and a
$T^2$-dependence of the Hall angle $\theta_H$ \cite{yeh02}.

The organics have emerged as in interesting playground in which
strong electronic correlations drive a host of novel electronic
ground states in close proximity to each other in phase space.
Though analysis of the Hall effect in conjunction with the
magnetoresistance has been relatively scarce in this class of
materials, a few examples do exist in current literature. For
instance, measurements of magnetotransport in the superconducting
salt $\kappa$-(BEDT-TTF)$_2$Cu[N(CN)$_2$]Cl in the pressure range
between 4 to 10 kbar reveal a $T^2$ dependence of the Hall angle
$\theta_H$ whereas the resistivity appears to be sub-linear
\cite{sus97}. Similar behavior was also observed in the system
$\Theta$-(DIETS)$_2$[Au(CN)$_4$] where measurements of the Hall
effect and magnetoresistance under pressures between 5 to 20 kbar
along the $c$-axis revealed a linear $T$-dependence of the
resistivity, along with a $T^2$-dependence of $\theta_H$ over a
reasonable temperature span \cite{taj03}. Considering the fact
that these anomalous transport properties appear to transcend
material classes and is observed in systems as diverse as the
cuprates, the heavy fermions, the organics, and other correlated
metals, we suggest that this disparity between the Hall effect and
the magnetoresistance could be a generic feature of many
correlated systems in the presence of strong (antiferro-)magnetic
fluctuations.

\subsection{Colossal magnetoresistive manganites}
\label{sec:cmr} A recently highly active field of research in
condensed matter physics is concerned with the manganites,
$R$MnO$_3$ with $R$ being a rare earth, and their doped variants,
$R_{1-x}$$A_x$MnO$_3$ where $A$ denotes a divalent or tetravalent
cation. Also in this class of materials the correlations between
the electrons play a crucial role \cite{coe}. These correlations
involve structural, charge, orbital as well as spin degrees of
freedom leading to a vast variety of different types of order that
may result in competition, or even coexistence, of different
phases. The balance between these phases can be subtle such that
small changes of a tuning parameter (again, {\it e.g.}, chemical
composition or magnetic field) prompt a significant response of
the material's properties. Typical examples here are the
metal-insulator transition (MIT, at a temperature $T_{\rm C}$) and
the so-called colossal magnetoresistance (CMR, an unusually strong
change of resistance with applied magnetic field) \cite{toku00}.
Even though the CMR effect with its potential technological
applications \cite{lem07} has originally drawn the research
interest to the manganites, their rich phase diagrams due to the
above-mentioned four degrees of freedom which gives rise to the
possibility for intrinsic inhomogeneity appears to be of more
contemporary interest \cite{mat07}. The phase coexistence with
{\em inhomogeneous} electronic or/and magnetic material properties
makes the manganites also prime candidates for {\em local} probes
like scanning tunneling microscopy and spectroscopy (STM/S) or
magnetic force microscopy (MFM).

Although the manganites seem to be quite different from the
heavy-fermion metals there are a number of similarities beyond the
mere statement of both being strongly correlated electron systems
\cite{alex00}. The above-mentioned Kondo effect was originally
developed for single magnetic impurities in metals \cite{kon64}.
In the heavy fermion metals, however, the magnetic ions are
periodically arranged on the lattice and may actually interact.
Therefore, a variant of the Kondo model, often referred to as the
``Kondo lattice model'' \cite{tsun,kondo05}, needs to be
considered in which the (often antiferro-) magnetic interaction
between localized moments and itinerant electrons at each relevant
lattice site is taken into account. On the other hand, the famous
double exchange \cite{zener} mediates the ferromagnetic coupling
between Mn$^{3+}$ and Mn$^{4+}$ ions through two simultaneous
hopping events of the $e_g$ conduction electrons via the
intermediate O$^{2-}$ ion in the manganites. Consequently, the
double exchange model can be considered as similar to the Kondo
lattice model just with ferromagnetic coupling of local moments
and conduction electrons \cite{dag03}. The combination of
spin---resulting from localized electrons---and charge---carried
by conducting electrons---degrees of freedom is also at the base
of spintronics \cite{wolf}, and the dilute magnetic semiconductors
\cite{hideo} bear some resemblance to the Kondo impurity problem.
Moreover, the phenomenon of phase separation (in the sense of
locally inhomogeneous electronic or/and magnetic properties) is
certainly not only discussed for magnetic oxides (see {\it e.g.}
\cite{khom97}) but was applied early on to doped Eu chalcogenides
\cite{svm67,svm68}, to the thorium phosphide-type structures
\cite{svm83} and to EuB$_6$ \cite{svm09}. Also, in its ``stripe''
version it has been considered for cuprate superconductors
\cite{emery,tran} and may even be at play within the coexistence
range of antiferromagnetism and superconductivity in
CeCu$_2$Si$_2$ \cite{osto06}.

Hall effect measurements in the manganites were used to infer the
Fermi surface of these materials \cite{jak01,gor00}. LaMnO$_3$ is
an insulating A-type antiferromagnet whose magnetic properties are
brought about by superexchange. The gap in the density of states
(DOS) right at $E_F$ is nicely reproduced by band structure
calculations within the local density approximation (LDA), {\it
e.g.} in Ref.\ \cite{satpathy}. Considering a simplified picture
of a rigid band structure it is obvious that either removing or
adding electrons, {\it i.e.} lowering or raising $E_F$ with
respect to the bands, may result in a conducting material at low
temperature. The usual hole doping ({\it e.g.} by Ca or Sr)
results in a corresponding mixture of Mn$^{4+}$ and Mn$^{3+}$
whereas a less well-known electron doping ({\it e.g.} by Ce) gives
rise to Mn$^{2+}$ and Mn$^{3+}$. Both types of mixed valencies
allow (within a certain doping range) for Zener double exchange
with a ferromagnetic metallic ground state of the material. The
metallic state prevails up to the ferromagnetic transition at
$T_{\rm C}$ where the mobility of the electrons decreases due to
spin disorder, and the carriers are localized via the formation of
Jahn-Teller polarons. Hence, the polaronic insulating state
results from the local lattice deformation around Mn$^{3+}$, and
these materials can be considered as polaronic semiconductors
above $T_{\rm C}$ \cite{jai97,hart06}.

Hall measurements on the manganites are notoriously difficult
because of often large contributions from the (negative) anomalous
Hall effect and the magnetoresistance \cite{coe}. Nonetheless,
some results of Hall measurements in the more common hole doped
manganites were reported, see {\it e.g.} \cite{lya01} for an
excellent overview. For example, charge carrier densities of $n
\approx 10^{29}$ m$^{-3}$ for La$_{0.65}$(PbCa)$_{0.35}$MnO$_3$ at
$T = 5$ K were obtained \cite{liu95} as expected for a typical
metal. At higher temperatures, sign reversals of $R_{\rm H}$ were
observed \cite{cao99,mal03} which could be a result of different
bands contributing to the Hall response, see below. In films there
could be additional localization effects of charge carriers due to
strain or disorder \cite{gran04}.

\begin{figure}
\centering\includegraphics[width=7.8cm,clip]{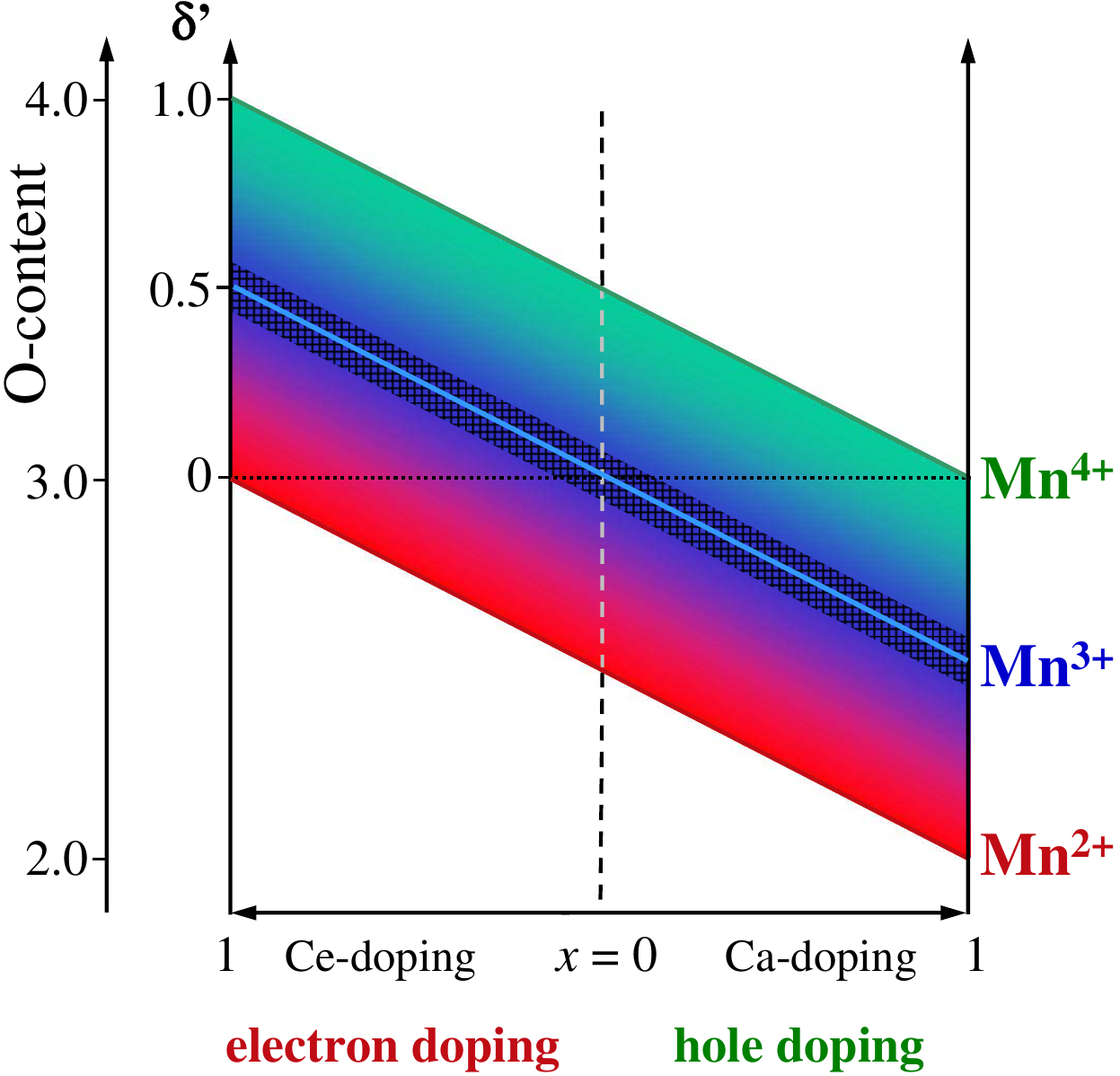}
\caption{Illustration of the Mn valence resulting from both, the
doping on the La site (abscissa) as well as the oxygen content
(ordinate) in La$_{1-x}$Ca$_{x}$MnO$_{3+\delta'}$ or
La$_{1-x}$Ce$_{x}$MnO$_{3+\delta'}$. The Mn valence is color-coded
in red, blue and green presenting Mn$^{2+}$, Mn$^{3+}$ and
Mn$^{4+}$, respectively. Sufficiently high O-excess can easily
outweigh an electron doping intended by Ce substitution of La, and
may even result in an unwanted Mn$^{3+}$/Mn$^{4+}$ mixture.}
\label{o-cont}
\end{figure}

Another issue becomes specifically obvious if nominally electron
doped manganites are considered: the resulting valency of the Mn
ions does not only depend upon the $A$-site doping level $x$ but
also on the oxygen stoichiometry $\delta'$ where
La$_{1-x}$A$_x$MnO$_{3+\delta'}$ \cite{wang06}. Electron doping in
films La$_{0.7}$Ce$_{0.3}$MnO$_3$ grown by pulsed laser deposition
(PLD) at low oxygen partial pressure ($\lesssim 10$ Pa) was
evidenced not only by x-ray absorption spectroscopy
\cite{mit03,han04} but was discernible also in Hall effect
measurements \cite{ray03}. However, with increasing oxygen content
the material is driven more and more into hole doping
\cite{wer09}. This is sketched in Fig.\ \ref{o-cont}: While oxygen
excess (going upward in Fig.\ \ref{o-cont}) is of no harmful
consequence to the nominally hole ({\it e.g.} Ca) doped manganites
it can easily outweigh or even overcome the intended electron
doping through Ce substitution in
La$_{1-x}$Ce$_{x}$MnO$_{3+\delta'}$ resulting in an effective
Mn$^{3+}$/Mn$^{4+}$ mixture \cite{yan04,sti07}. In addition,
unwanted phases ({\it e.g.} CeO$_2$) are very likely to be found
in the electron-doped manganites \cite{sti07,wer09}. The formation
of additional phases is certainly also supported by the small
tolerance factor ($t \approx 0.91$ for
La$_{0.7}$Ce$_{0.3}$MnO$_3$) resulting from the small size of the
tetravalent ions \cite{jol02,tan03}. The small $t$ may also
explain the fact that mostly thin films of
La$_{0.7}$Ce$_{0.3}$MnO$_3$ or other \cite{chang04,zhang04}
electron doped manganites could be synthesized; with the exception
of La$_{0.9}$Te$_{0.1}$MnO$_3$ for which electron doping in bulk
samples was shown by Hall effect measurements \cite{yang06}. All
these difficulties might easily explain the contradicting results
found in the literature as much as they render the electron doped
manganites less attractive for further applications.

As already discussed in section \ref{sec:anom} an anomalous
contribution to the Hall effect is expected in magnetic materials,
{\it cf.} eq.~(\ref{rhoAHE}). For several manganites it was found
that the AHE is negligible at low temperature $T \ll T_{\rm C}$
but contributes significantly to $\varrho_{xy}$ as $T$ is raised
towards $T_{\rm C}$ \cite{matl98,chun00,lya01,ray03,kuw03,beb08}.
For La$_{2/3}$Sr$_{1/3}$MnO$_3$ and within the metallic phase a
linear relationship between $R_{\rm S}(T)$ and $\varrho_{xx}(T)$
as well as the temperature driven change of magnetization, $M(0) -
M(T)$, was reported \cite{chen99}. A similar relation, $R_{\rm S}
\propto \varrho_{xx}^{\alpha}$, holds in La$_{1-x}$Ca$_x$MnO$_3$
with $\alpha = 1.38$ for $x = 0.3$ and $\alpha = 0.6$ for $x =
0.67$ where electron-type carriers were observed \cite{gor00}.
Below $T_{\rm C}$ the metallic manganites behave as bad metals
with the transport being dominated by holes of low mobility
\cite{coe,lya01,beb08}. For comparison we wish to add that well
above $T_{\rm C}$ the bandwidth of the $e_g$ electrons is narrowed
by the Jahn-Teller distortion along with strong electron-phonon
interactions which causes the formation of small polarons
\cite{coe,jaim96} whereas manganites in the insulating regime are
governed by variable-range hopping.

Certainly the most interesting temperature regime is the one
around the MIT. An early Hall effect study \cite{matl98} on
La$_{1-x}$Ca$_x$MnO$_3$ thin films focusing on this temperature
range again reported $R_{\rm S} \propto \varrho_{xx}$ as well as
the Hall angle (see eq.~(\ref{HallAngle})) $\tan \theta_{\rm H}
\sim M$ for $T > T_{\rm C}$. It was pointed out that the large
value of $R_{\rm S}$ in the hopping regime of the manganites is
different from the anomalous Hall effect in typical ferromagnetic
metals ({\it cf.} sections \ref{sec:anom} and \ref{sec:Hallmech}).
While hopping the electrons acquire a phase that reflects their
induced alignment to the core spins and is consequently governed
by the magnetization \cite{matl98}. Based on such ideas a model
for the AHE was developed \cite{chun00,lya01} that takes into
account the quantal phase due to strong Hund's-rule coupling and
the spin-orbit interactions.

We note that an additional (positive) contribution to the AHE is
observed in polycrystalline thin films due to grain boundaries
\cite{tani03}.

In layered manganites La$_{2-x}$Sr$_{1+x}$Mn$_2$O$_7$ an
additional contribution to the AHE is observed possibly related to
the intrinsic inhomogeneity resulting from the spin glass state
\cite{chun01,hir06}. In contrast to the manganites, the pyrochlore
Tl$_2$Mn$_2$O$_7$ exhibits a negligible AHE and its CMR near
$T_{\rm C}$ is governed by a change in charge carrier density
\cite{imai00}. Similarly, the CMR observed in
Eu$_{0.6}$Ca$_{0.4}$B$_6$ appears to be related to a crossover
from hole- to electron-dominated transport \cite{glu09}.

\section{Summary}
\label{sec:sum} The quest of shedding light on such highly complex
materials as the strongly correlated electron systems calls for a
comprehensive study and sophisticated experimental approaches.
Amongst the plethora of laboratory based tools routinely employed
in the investigation of strongly correlated electron systems, Hall
effect measurements have proven to be an exemplary one. With other
measurement techniques often not feasible in the temperature and
field range of interest for heavy fermion compounds, Hall effect
measurements can provide insight into the evolution of their
renomalized Fermi surfaces, even across a quantum critical point.
As such, it remains unparalleled in tackling some of the
outstanding problems at the frontiers of current condensed matter
physics. Though demonstrated in the case of a prototypical heavy
fermion system, YbRh$_2$Si$_2$ in section \ref{sec:YRS}, the
inferences drawn therein are crucial for understanding the physics
of systems like heavy fermion superconductors (in section
\ref{sec:fluc} we focused on those with 115-type of structure) or
even the cuprates, in which magnetism and superconductivity appear
to be inexorably intertwined. The possibility that a similar Fermi
surface reconstruction occurs at the crossover from the overdoped
to the underdoped regime in the high temperature superconducting
cuprates continues to be at the focus of extensive investigations.
A related development pertains to the possibility of two
scattering times in the heavy fermion systems as deduced from
recent Hall measurements in some Ce-based heavy fermions---very
much in similarity to that observed earlier in the cuprates. The
recently discovered Fe-based superconductors also appear to share
some of these intriguing properties. Considering current issues
related to sample purity in these systems, it is not unrealistic
to predict that they would continue to be extensively investigated
for several years to come.

In this review we focused on the investigation of Fermi surface
effects in heavy fermion metals. Consequently, we concentrated on
the {\em normal} contribution to the Hall effect, leaving the
anomalous Hall effect aside, even though it also may yield
valuable information \cite{nag10}. Fortunately in this context,
the anomalous contribution is often negligibly small for heavy
fermion compounds at low temperatures, see section
\ref{sec:AHE-HF}. Our material basis also implies that the active
research areas of the quantized Hall effect \cite{klitzing86}, its
fractional counterpart \cite{stormer99} as well as the related,
topical spin Hall effect \cite{schlie06} and topological phases
\cite{wen95} are not touched upon.

Hall effect measurements probe the electronic states directly at
the Fermi energy $E_{\rm F}$. As a major advantage, such
measurements can be conducted over a wide parameter space, {\it
e.g.} in temperature, magnetic field (including pulsed fields) or
pressure, making this an invaluable probe. Moreover, the results
of Hall measurements can be compared to theoretical predictions,
or could serve as input for theoretical considerations. Here, Hall
effect measurements on YbRh$_2$Si$_2$ constitute a good example,
as outlined in section \ref{sec:YRS}. The theoretical prediction
and interpretation have taken advantage of the simplification the
zero-temperature limit brings about to the evolution of the Hall
coefficient across a quantum critical point. At the same time, we
have emphasized that for quantitative understandings as well as
for the description of the Hall coefficient at finite
temperatures, it is often important to incorporate the
complexities in the band structure, such as the multiplicity of
the bands, or the various interaction effects on the electronic
excitations near the Fermi surfaces. Experimentally, it can be
challenging to get the samples into the thin, plate-like shape
shown in Fig.\ \ref{fig:Hall}. As many of these materials are good
metals the thickness $d$ often needs to be as small as a few tens
of micrometers to give reasonably measurable Hall voltages of
several ten nV. In providing detailed information about the
measurement technique (section \ref{sec:exp}) as well as some
basic theoretical background (section \ref{sec:theory}) we wish to
inspire more research within the exciting field of heavy fermion
physics and, hopefully, beyond.

\section{Appendix} \label{sec:append}
The following list provides references in which results of Hall
effect measurements are presented for numerous heavy fermion and
related materials. It is meant to serve as a guide and is by
no means exhaustive.\\[0.4cm]
\hspace*{1cm} \begin{tabular}{p{3.6cm}l} \hline Compound & Reference \\
\hline CeCu$_2$Si$_2$ &
\cite{cat85,adria88,stewa83,aliev83,catta86b}\\ 
(Ce$_{1-x}$La$_x$)Cu$_2$Si$_2$ & \cite{aliev83}\\ 
CeRu$_2$Si$_2$ & \cite{had86,Daou2006}\\ 
CeRh$_2$Si$_2$ & \cite{bou06}\\ 
CeNi$_2$Ge$_2$ & \cite{sato98}\\ 
CeNiGe$_3$ & \cite{pikul03}\\ 
CeAl$_2$ & \cite{diekm92,chris77,catta86b}\\ 
Ce(Al$_{2-x}$Co$_x$) & \cite{bogac07}\\ 
CeSn$_3$ & \cite{catta86b}\\ \hline 
\end{tabular} \newpage
\hspace*{1cm} \begin{tabular}{p{3.6cm}l} \hline Compound &
Reference \\ \hline
CePtSi & \cite{ham88} \\ 
CeCu$_6$ & \cite{pen86a,pen86b,adria88,winze86,sato85} \\ 
(Ce$_{1-x}$La$_x$)Cu$_6$ & \cite{sato85} \\ 
CeB$_6$ & \cite{kobay02} \\ 
(Ce$_{1-x}$La$_x$)B$_6$ & \cite{dreye83,kobay02} \\ 
CePd$_3$ & \cite{cat85,catta86b} \\ 
(Ce$_{1-x}$Y$_x$)Pd$_3$ & \cite{fer85} \\ 
Ce(Pd$_{0.88}$Ag$_{0.12}$)$_3$ & \cite{cat85,catta86b} \\ 
CeBe$_{13}$ & \cite{cat85,catta86b} \\ 
CeRh$_{3}$ & \cite{catta86b} \\ 
CeIn$_{3}$ & \cite{gao85} \\ 
CeAl$_{3}$ & \cite{had86,brand85} \\ 
(Ce$_{1-x}$La$_x$)Al$_{3}$ & \cite{brand85} \\ 
CeCoIn$_5$ & \cite{nak07,hun04,sin07,ono07,nak04a,nak06} \\ 
CeCo(In$_{1-x}$Cd$_x$)$_5$ & \cite{nai09c} \\ 
CeRhIn$_5$ & \cite{nak07,hun04,nak04a,nak06} \\ 
CeIrIn$_5$ & \cite{hun04,nai09b,nai08,nak08a} \\ 
Ce$_3$Pd$_{20}$Si$_6$ & \cite{paschen12} \\ 
Ce$_2$RhIn$_8$ & \cite{sakam03} \\ 
Ce$_2$IrIn$_8$ & \cite{sakam03} \\ 
Ce$_2$CoIn$_8$ & \cite{gchen03} \\ 
Ce$_2$PdIn$_8$ & \cite{gnida12} \\ 
CePt$_2$In$_7$ & \cite{tobas12} \\ 
YbRh$_2$Si$_2$ &
\cite{Friedemann2010e,Friedemann2010b,pas-nat,Paschen2005,pfau12un} \\ 
YbCuAl & \cite{cat85,catta86a} \\ 
YbCu$_2$Al$_2$ & \cite{cat85,catta86a} \\ 
YbNi$_2$B$_2$C & \cite{budko05} \\ 
YbAgCu$_4$ & \cite{sarra99} \\ 
YbInAu$_2$ & \cite{cat85,catta86a} \\ 
YbAgGe & \cite{Budko2005a} \\ 
YbPd & \cite{catta86a} \\ 
YbAl$_2$ & \cite{catta86a} \\ 
YbAl$_3$ & \cite{catta86a,jung02} \\ 
U$_2$Zn$_{17}$ & \cite{sie86} \\ 
URu$_2$Si$_2$ & \cite{sch87,oh07,lev09} \\ 
U(Ru$_{0.96}$Rh$_{0.04}$)$_2$Si$_2$ & \cite{oh06} \\ 
UPt$_3$ & \cite{Kambe1999,had86} \\ 
U$_2$PtC$_2$ & \cite{penne87} \\ 
UAl$_{2}$ & \cite{had86} \\ 
UGe$_{2}$ & \cite{tran04} \\ 
UBe$_{13}$ & \cite{pen86a,penne87,hetti89} \\ 
UPd$_{2}$Al$_3$ & \cite{huth94} \\ 
NpPd$_{5}$Al$_2$ & \cite{grive08,matsu10} \\ 
\hline
\end{tabular}

\section*{Acknowledgments} \label{sec:ackn}
We thank E.~Abrahams, P.~Coleman, P.~Gegenwart, C.~Geibel,
S.~Kirchner and S.~Paschen for insightful discussions. Work at
Dresden was partly supported by the German Research Foundation
through DFG Forschergruppe 960. S.N. and S.F. acknowledge support
by the Alexander von Humboldt Foundation, Germany. Q.S.
acknowledges the support of NSF Grant No.~DMR-1006985 and the
Robert A.~Welch Foundation Grant No.~C-1411.

For all figures reprinted from American Physical Society material
readers may view, browse, and/or download material for temporary
copying purpose only, provided these uses are for noncommercial
personal purposes. Except as provided by law, this material may
not be further reproduced, distributed, transmitted, modified,
adapted, preformed, displayed, published, or sold in whole or
part, without prior written permission from the American Physical
Society.

\bibliographystyle{tADP}
\label{sec:bib}
\bibliography{Hallrev}
\end{document}